\documentclass[prb,twocolumn,showpacs,preprintnumbers,amsmath,amssymb,superscriptaddress,nofootinbib]{revtex4}
\pdfoutput=1

\usepackage{bbm,bm,slashed}
\usepackage{graphicx,graphics,psfrag,epsfig}
\usepackage{color,array,tabularx}
\usepackage{hyperref}

\newcommand{\be}{\begin{eqnarray}}
\newcommand{\ee}{\end{eqnarray}}
\newcommand{\txi}{\tilde{\xi}}
\newcommand{\teta}{\tilde{\eta}}
\newcommand{\JK}{J_{\mathrm{K}}}
\newcommand{\JH}{J_{\mathrm{H}}}
\newcommand{\nn}{\nonumber } 
\newcommand{\Eqref}[1]{Eq.~\eqref{#1}}

\begin{document}

\author{Daniel D. Scherer}
\affiliation{Institut f\"ur Theoretische Physik, Universit\"at
Leipzig, D-04103 Leipzig, Germany}
\email{daniel.scherer@physik.uni-leipzig.de}
\author{Michael M. Scherer}
\affiliation{Institut f\"ur Theoretische Physik, Universit\"at
Heidelberg, D-69120 Heidelberg, Germany}
\author{Giniyat Khaliullin}
\affiliation{Max-Planck-Institut f\"ur Festk\"orperforschung, D-70569 Stuttgart, Germany}
\author{Carsten Honerkamp}
\affiliation{Institute for Theoretical Solid State Physics, RWTH Aachen University, D-52056 Aachen, Germany
and JARA - FIT Fundamentals of Future Information Technology, Germany}
\author{Bernd Rosenow}
\affiliation{Institut f\"ur Theoretische Physik, Universit\"at
Leipzig, D-04103 Leipzig, Germany}
\affiliation{Physics Department, Harvard University, Cambridge, Massachusetts 02138, USA}

\title{Unconventional pairing and electronic dimerization instabilities in the doped
Kitaev-Heisenberg model}

\begin{abstract} 
We study the quantum many-body instabilities of the $t -\JK - \JH$ Kitaev-Heisenberg Hamiltonian on the honeycomb lattice as a minimal model for a doped spin-orbit Mott insulator. This spin-$1/2$ model is believed to describe the magnetic properties of the layered transition-metal oxide Na$_2$IrO$_3$. We determine the ground-state of the system with finite charge-carrier density from the functional renormalization group (fRG) for correlated fermionic systems. To this end, we derive fRG flow-equations adapted to the lack of full spin-rotational invariance in the fermionic interactions, here represented by the highly frustrated and anisotropic Kitaev exchange term. Additionally employing a set of Ward identities for the Kitaev-Heisenberg model, the numerical solution of the flow equations suggests a rich phase diagram emerging upon doping charge carriers into the ground-state manifold ($\mathbbm{Z}_{2}$ quantum spin liquids and magnetically ordered phases). We corroborate superconducting triplet $p$-wave instabilities driven by ferromagnetic exchange and various singlet pairing phases. For filling $\delta > 1/4$, the $p$-wave pairing gives rise to a topological state with protected Majorana edge-modes. For antiferromagnetic Kitaev and ferromagnetic Heisenberg exchange we obtain bond-order instabilities at van Hove filling supported by nesting and density-of-states enhancement, yielding dimerization patterns of the electronic degrees of freedom on the honeycomb lattice. Further, our flow equations are applicable to a wider class of model Hamiltonians.
\end{abstract}

\maketitle

\section{Introduction}
\label{sec:intro}

The postulation of new topological states of matter and the quest for unraveling their properties
has spurred a huge amount of research activity over the last decade, building on the groundbreaking
insight that the appearance of the integer quantum Hall effect (IQHE)~\cite{klitzing1980,laughlin1981} in two-dimensional electron gases subject to quantizing perpendicular magnetic fields is intimately related to the topology of wavefunctions~\cite{niu1982,niu1985,kohmoto1985}. The prediction of the existence of a topological state generated by the presence of spin-orbit coupling in HgTe/CdTe quantum wells~\cite{bernevig2006a,bernevig2006b} effectively opened out in the creation of the field of toplogical insulators and superconductors~\cite{hasan2010,qi2009,qi2010,schnyder2008,ryu2010}. While the IQHE still serves as a time-honored prototype to the field, many different models have been identified featuring the same or similar kinds of topological non-triviality in their wavefunctions~\cite{hasan2010,qi2009,qi2010,schnyder2008,ryu2010}.

Another sub-field of condensed matter physics with strong connections to topology is the search
for quantum spin liquids~\cite{balents2010} -- non-magnetic phases with typically exotic excitations and topological order~\cite{wen1990,wen1995}, a concept first introduced in the context of the fractional quantum Hall effect (FQHE)~\cite{tsui1982,haldane1983,laughlin1983,halperin1984}. On the theoretical side, the honeycomb Kitaev model~\cite{kitaev2005} provided the first exactly solvable model with a quantum spin liquid ground-state in two spatial dimensions. Though a pure spin model, its excitations are Majorana fermions obeying non-Abelian exchange statistics~\cite{kitaev2005} .

A route to explore Kitaev physics in a transition-metal oxide solid-state system, however, was suggested only recently. 
The naturally large spin-orbit and crystal-field energy scales in layered transition-metal oxides and the rather strong correlation effects in $5d$ orbitals lead to highly anisotropic exchange interactions, entangling spin and orbital degrees of freedom. As a candidate compound, the layered honeycomb iridate Na$_2$IrO$_3$ was proposed~\cite{jackeli2009,chaloupka2010}. It turned out to order magnetically below $T_{\mathrm{N}} \simeq 15\,\mathrm{K}$ in a so-called zigzag pattern different from the N$\mathrm{\acute e}$el state on the bipartite honeycomb lattice~\cite{singh2010,liu2011,bhatt2012,singh2012,chaloupka2013,ye2012,kargarian2012,reuther2012,schaffer2012}.
Such a low ordering temperature was further taken as a sign of strongly frustrated exchange - a hallmark of Kitaev physics. Indeed, an
effective (iso-)spin $1/2$ model for the two states in the $j_{\mathrm{eff}} = 1/2$ part of the spin-orbit split manifold of $t_{2g}$ electrons was suggested to capture the magnetic properties of the Mott-insulating ground-state. It is simply given by Kitaev exchange and isotropic Heisenberg exchange
for (iso-)spin $1/2$ degrees of freedom on nearest-neighbor sites in the honeycomb lattice.
As the character of exchange interactions is varied among the possibilities of ferromagnetic and antiferromagnetic coupling, the model features a ground-state manifold with several interesting magnetic orderings with characteristic imprints on the spin-wave excitation spectrum. The $\mathbbm{Z}_{2}$ quantum spin liquid ground state occurs for either ferromagnetic or antiferromagnetic Kitaev exchange, as long as the perturbation due to the Heisenberg coupling remains sufficiently small. 

While the adequacy of the Kitaev-Heisenberg Hamiltonian as a minimal model
for the magnetic properties of the honeycomb iridate Na$_2$IrO$_3$ is still subject to debate~\cite{foyevtsova2013}, the idea of studying the unconventional pairing states of the doped system has attracted additional attention from theory~\cite{hyart2012,you2012,okamoto2013,kimme2013}. The resulting Hamiltonian can be understood as a paradigmatic model of a spin-orbit coupled, frustrated and doped Mott insulator. Besides singlet pairing phases, mean-field studies revealed $p$-wave triplet pairing phases~\cite{hyart2012,you2012,okamoto2013}, similar to the $B$-phase in $^{3}\mathrm{He}$. Finally, when doping
the system beyond quarter filling, the $p$-wave (mean-field) states are guaranteed~\cite{qi2009,sato2009a,sato2009b,qi2010,sato2010} to undergo a transition to a topological $p$-wave triplet phase~\cite{hyart2012,you2012,okamoto2013}. This argument applies at least within weak-pairing theory. Almost in reminiscence of the
Majorana excitations of the $\mathbbm{Z}_{2}$ quantum spin liquid of the pure Kitaev model, the 
topological $p$-wave phase would posses Majorana states at vortex cores and Majorana edge modes
propagating along the boundaries of the system. Amazingly, this line of investigation unveils the possible unification of the two different branches of topological insulator/superconductors and topologically ordered phases in the phase diagram of a paradigmatic single model Hamiltonian.

Regarding the possibility to experimentally realize doped states of the honeycomb iridate Na$_2$IrO$_3$, we would like to mention that electron doping of this material has been achieved recently by covering the sample with a potassium layer~\cite{comin2012}. Further, doping of Sr$_2$IrO$_4$ iridium compounds became possible by the same technique and pseudogap physics as in cuprates was observed~\cite{kim2014}. Studying minimal models of doped spin-orbit Mott insulators therefore seems a worthwhile enterprise with possible connections to future experiments.

In this work, we investigate the phase diagram of the doped Kitaev-Heisenberg model beyond
mean-field theory. This aim is accomplished within the flexible flow equation approach provided by
the functional renormalization group~\cite{metzner2011} (fRG). 

\section{Model, Results and Outline}
\label{sec:summary}

In the following, we introduce the Hamiltonian of the doped Kitaev-Heisenberg model, see Subsect~\ref{sec:model}. Our overall goal is to draw
the ground-state phase diagram of the system, paying attention to different, competing ordering tendencies. We summarize and describe the gist of our results
in Subsect.~\ref{sec:mainresults}, demonstrating the richness of the phase diagram borne by doping of the Kitaev-Heisenberg system. We then provide an outline of this work in Subsect.~\ref{sec:outline}. Details are then provided in subsequent sections.

\subsection{The doped Kitaev-Heisenberg Model}
\label{sec:model}

We study the Hamiltonian Eq.~(\ref{eq:hamiltonian}) on the honeycomb lattice as a minimal model for a doped spin-orbit Mott insulator. The Hamiltonian includes a kinetic part, which describes the hopping of the electrons, a Kitaev coupling $\JK$, describing a bond-dependent Ising-like spin exchange, and a Heisenberg exchange, including a nearest-neighbor density-density interaction of magnitude $\JH$ that occurs upon doping.
\begin{figure}[t!]
\centering
\includegraphics[height=.4\columnwidth]{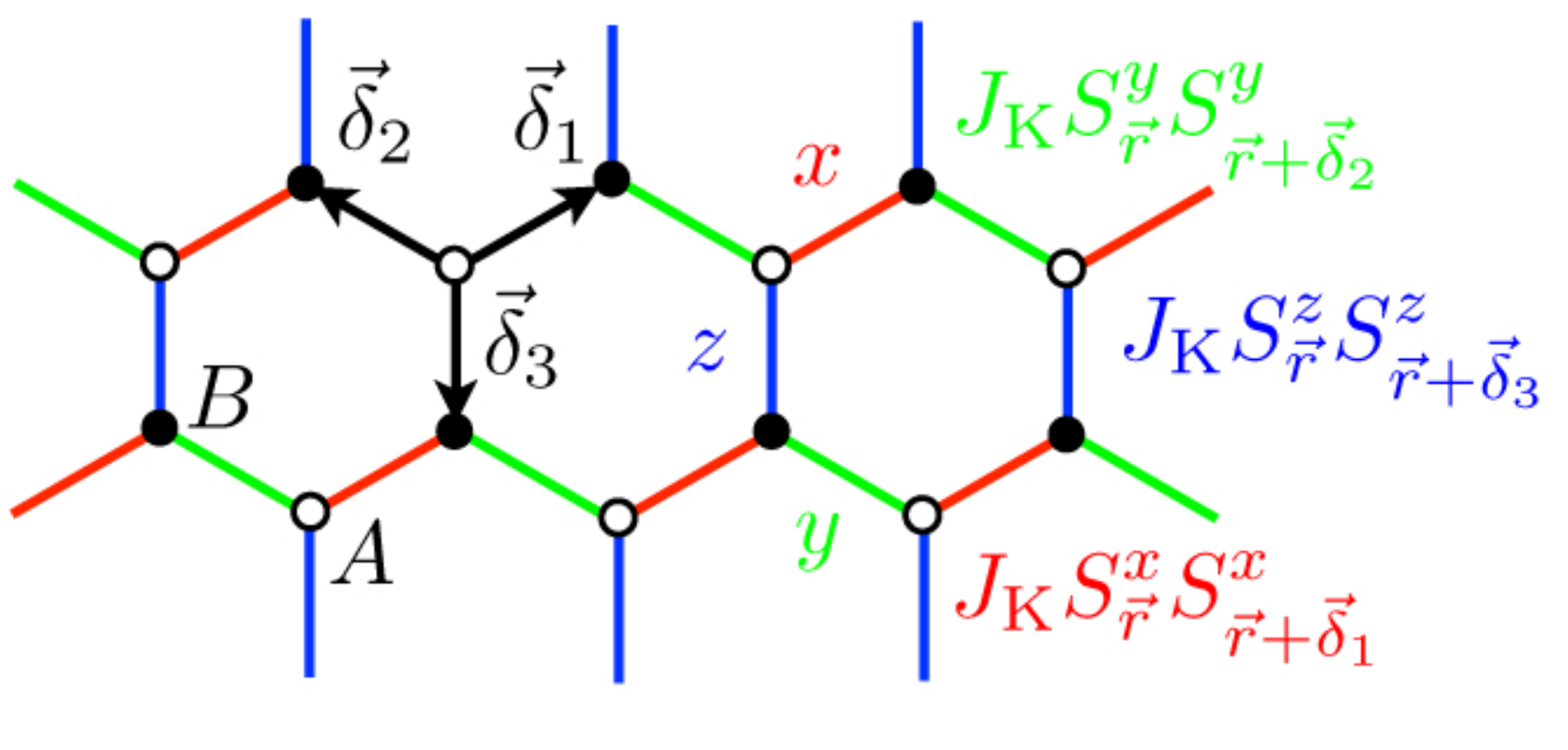}
\caption{Three plaquettes on the honeycomb lattice. White disks mark sites of the $A$-sublattice, while black disks those residing in the $B$-sublattice. Shown are
also the three nearest-neighbor vectors $\vec{\delta}_{i}$, $i=1,2,3$ pointing from a site in the $A$-sublattice to its nearest neighbors in the
$B$-sublattice. The different colors of the bonds linking the sites of the honeycomb encode the bond-specific nearest-neighbor interaction of the Ising-like Kitaev exchange entering
the Kitaev-Heisenberg Hamiltonian. Red bonds are called $x$-bonds and correspondingly, along an $x$-bond the $x$-components of neighboring spin operators
are exchange-coupled. The conventions for $y$- and $z$-bonds are analogous.}
\label{fig:model}
\end{figure}
The Hamiltonian for the doped spin-orbit Mott insulator thus reads
\be\label{eq:hamiltonian}
H &=& H_{\mathrm{kin}} + H_{\mathrm{K}} + H_{\mathrm{H}},
\ee
where~\footnote{Here, we defined the exchange couplings $\JH$ and $\JK$ differently compared to Ref.~\cite{hyart2012}.}
\be
\label{eq:hamiltoniancontd}
H_{\mathrm{kin}} & = & -t_{0}\,\mathcal{P}\sum_{\sigma,\vec{r}\in A,\vec{\delta}_{i}}\bigl[c_{A,\sigma,\vec{r}}^{\dagger}c_{B,\sigma,\vec{r}+\vec{\delta}_{i}}
+ \mathrm{h.c.} \bigr]\mathcal{P} \\  
& & - \mu\sum_{\sigma,o\in A,B,\vec{r}} c_{o,\sigma,\vec{r}}^{\dagger}c_{o,\sigma,\vec{r}} , \nn \\ 
H_{\mathrm{K}} & = &  \JK \sum_{\vec{r}\in A,\vec{\delta}_{i}}
S^{\gamma}_{\vec{r}}\, S^{\gamma}_{\vec{r}+\vec{\delta}_{i}}, \label{eq:hamiltoniancontdK} \\
H_{\mathrm{H}} & = & \JH
\sum_{\vec{r}\in A,\vec{\delta}{i}}\left(\vec{S}_{\vec{r}}\cdot\vec{S}_{\vec{r}+\vec{\delta}_{i}}-\frac{1}{4}n_{\vec{r}}n_{\vec{r}+\vec{\delta}_{i}}\right). \label{eq:hamiltoniancontdH}
\ee
The kinetic part of the Hamiltonian $H_{\mathrm{kin}}$ describes spin-independent nearest-neighbor hopping of electrons with hopping amplitude $t_{0}$, while the chemical potential $\mu$
is adjusted to yield a charge concentration corresponding to the doping level. The Gutzwiller projection $\mathcal{P}$ enforces the constraint
of no doubly occupied sites, incorporating the strong
correlation effects of the Mott insulating state. 
\begin{figure}[t!]
\centering
\includegraphics[height=.44\columnwidth]{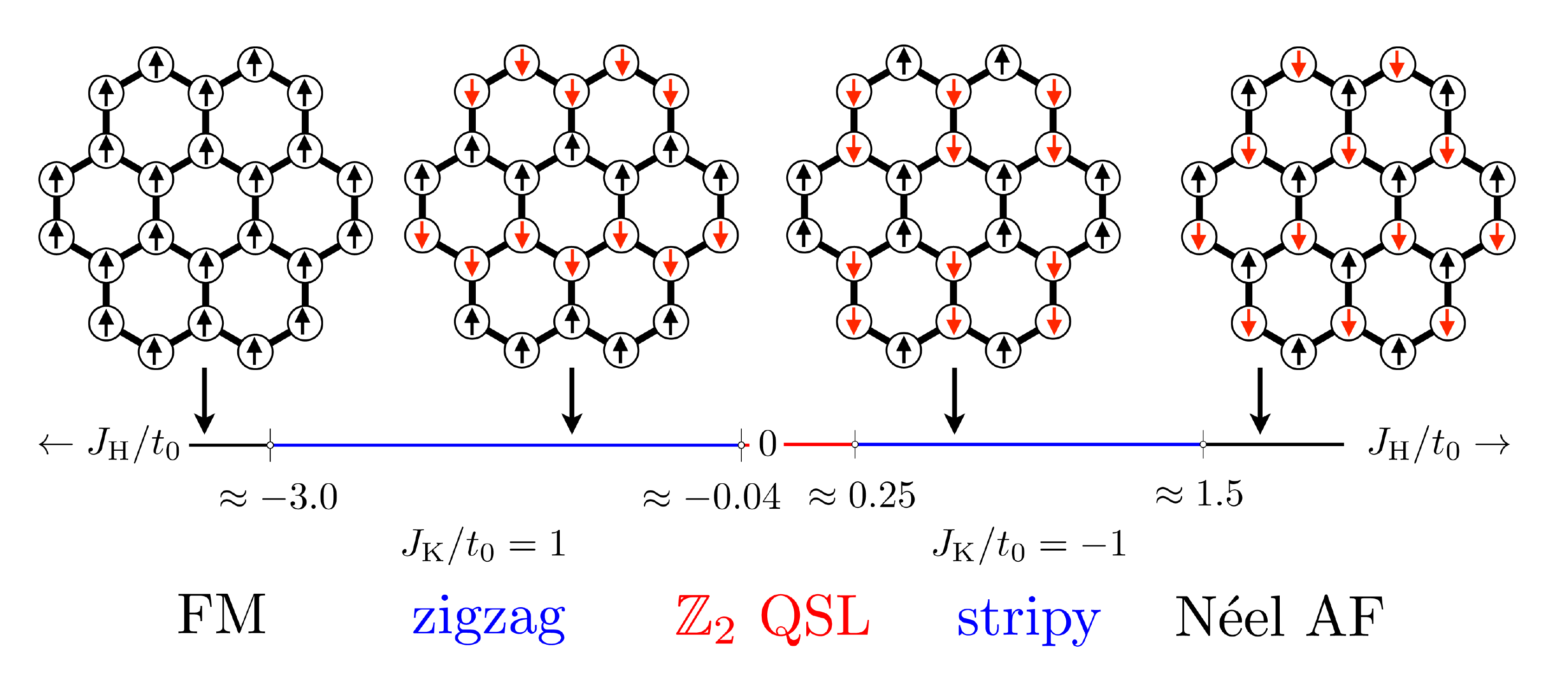}
\caption{Cuts in the ground-state manifold of the undoped Kitaev-Heisenberg model in the two-dimensional parameter space spanned by Kitaev ($\JK$) and Heisenberg ($\JH$) exchange couplings. In our work on the doped case, we restrict our attention to fixed $\JK$. To facilitate a comparison to the undoped system, we chose the bare hopping as the overall energy scale and plot the ground-states for fixed Kitaev exchange as the magnitude of Heisenberg exchange is varied. For the full phase diagram in the $(\JK,\JH)$-plane, we refer to Ref.~\onlinecite{chaloupka2010}. The first row shows the magnetic ground-state patterns on the honeycomb lattice, where red and black arrows denote spin up and spin down with respect to a given quantization axis. The second row gives the corresponding parameter ranges. The numbers below indicate the critical coupling strengths, where transitions
from one phase to another occur. The $\mathbbm{Z}_{2}$ quantum spin liquid (which occurs for both $\JK >0$ and $\JK < 0$) is destabilized by some amount of Heisenberg exchange and turns into either a magnetically ordered zigzag pattern or into an ordered stripy phase. For large ferromagnetic Heisenberg exchange, the ground-state is rendered ferromagnetic, while for large antiferromagnetic Heisenberg exchange, one finds a N$\mathrm{\acute e}$el antiferromagnet.}
\label{fig:undoped}
\end{figure}
Due to the two-atom unit cell of the honeycomb lattice, the kinetic term leads
to a two-band description for mobile charge degrees
of freedom. The sites within the
two-dimensional bipartite honeycomb lattice are labeled by
$\vec{r}$. For fixed $\vec{r}$ in sublattice $A$, there are three nearest neighbor sites within
the $B$-sublattice whose position is given by $\vec{r}+\vec{\delta}_i$ with $i \in \{1,2,3\}$, cf. Fig~\ref{fig:model}.
The nearest-neighbor vectors $\vec{\delta}_i$ are given by 
$\vec{\delta}_1= \frac{\sqrt{3}a}{2}
\hat{e}_x+\frac{a}{2}\hat{e}_y$, $\vec{\delta}_2=
-\frac{\sqrt{3}a}{2} \hat{e}_x+\frac{a}{2}\hat{e}_y$ and
$\vec{\delta}_3= -a\hat{e}_y$,
with $a$ being the distance between two neighboring lattice
sites and the vectors point from the $A$-sublattice to the
$B$-sublattice. The operators $c_{o,\sigma,\vec{r}}$ and $c_{o,\sigma,\vec{r}}^\dagger$
describe annihilation and creation of an electron at
position $\vec{r}$ in sublattice $o \in \{A,B\}$ with $\sigma=
\uparrow,\downarrow$ the isospin polarization, respectively. For simplicity,
we will refer to $\sigma$ as `spin' in the remainder of the paper.
We note that generally summation runs over one sublattice only (thus counting every nearest neighbor bond only once), while the sum in the local chemical-potential term runs over either sublattice $A$ or $B$, depending on whether $o=A$ or $o=B$.

Starting from a model with local interactions, within a strong-coupling expansion~\cite{lee2006} virtual charge excitations
above the Mott-Hubbard gap create effective spin-spin interactions. Due 
to large spin-orbit effects and low-symmetry crystal fields as in the iridates, the
exchange interactions are highly anisotropic. The so-called Kitaev
interaction $H_{\mathrm{K}}$ describes a bond-dependent Ising-like spin exchange, see the rightmost plaquette in Fig~\ref{fig:model}. Its strength is
described by the Kitaev coupling $\JK$. 
For $i\in\{1,2,3\}$ running over the adjacent nearest-neighbor sites residing in the
$B$-sublattice, $\gamma\equiv \gamma(i)$ takes on the values
$\gamma \in \{x,y,z\}$. 
Besides the Kitaev term, the model contains an additional Heisenberg exchange
$H_{\mathrm{H}}$ with Heisenberg coupling constant $\JH$, where a nearest-neighbor density-density interaction due to doping is included. We note that the presence of the density-density term allows to re-formulate the Heisenberg exchange solely in terms of bond-singlet operators.

The Hamiltonian Eq.~(\ref{eq:hamiltonian}) gives rise to a $t -\JK - \JH$ model description of a doped
spin-orbit Mott insulator.

\subsection{Main results}
\label{sec:mainresults}

In this work, we consider the case of ferromagnetic Kitaev ($\JK <  0$) and
antiferromagnetic ($\JH > 0$) Heisenberg exchange, as well as antiferromagnetic Kitaev ($\JK >  0$) and
ferromagnetic ($\JH < 0$) Heisenberg exchange. While the former realizes a magnetic phase with
stripy antiferromagnetic order (alternating ferromagnetic stripes that are coupled antiferromagnetically) at $\delta = 0$ in the pure spin model, the latter case brings about
magnetic order in a zigzag pattern (ferromagnetic zigzag chains that are coupled antiferromagnetically)~\cite{chaloupka2010,singh2010,liu2011,bhatt2012,singh2012,chaloupka2013,ye2012,kargarian2012,reuther2012,schaffer2012}, see Fig.~\ref{fig:undoped}. 
The $\mathbbm{Z}_{2}$ quantum spin liquid exists for both ferromagnetic and antiferromagnetic Kitaev exchange 
for sufficiently small Heisenberg exchange strength. When both exchange couplings come with the same sign, the magnetic ordering pattern is either of ferromagnetic or  N$\mathrm{\acute e}$el type.

Motivated by an estimate of the energy scale of the Kitaev exchange relative to the spin-independent nearest neighbor hopping~\cite{chaloupka2013}, and in order to reduce the number of parameters in the model, we restrict our attention to $|\JK|/t_{0}=1$. As noted previously~\cite{hyart2012}, at fixed doping the ratio $|\JK|/t_{0}$ largely controls the overall scale
for critical temperatures, while the ratio $|\JK|/|\JH|$ determines the ground-state of the doped model.

The numerical solutions to the fRG equations adapted to the Kitaev-Heisenberg model allow us to identify the leading Fermi surface instabilities of the auxiliary fermion system. We introduce
the fRG equations in some detail in Sect.~\ref{sec:method}. Additionally, we can read off estimates of ordering scales or critical temperatures from the fRG flows through the critical scale $\Lambda_{c}$, at which an instability manifests as a divergence in the scale-dependent effective interaction.
In the following, we will present the phase diagrams of the doped Kitaev-Heisenberg model for $\JK < 0$, $\JH > 0$ and $\JK > 0$, $\JH < 0$ as our main results. A detailed discussion
of our results can be found in Sect.~\ref{sec:results}.

\subsubsection{Ferromagnetic Kitaev $\JK < 0$, antiferromagnetic Heisenberg $\JH > 0$ exchange}
\label{sec:mainresultsA}
%
\begin{figure}[t!]
\centering
\includegraphics[height=.85\columnwidth]{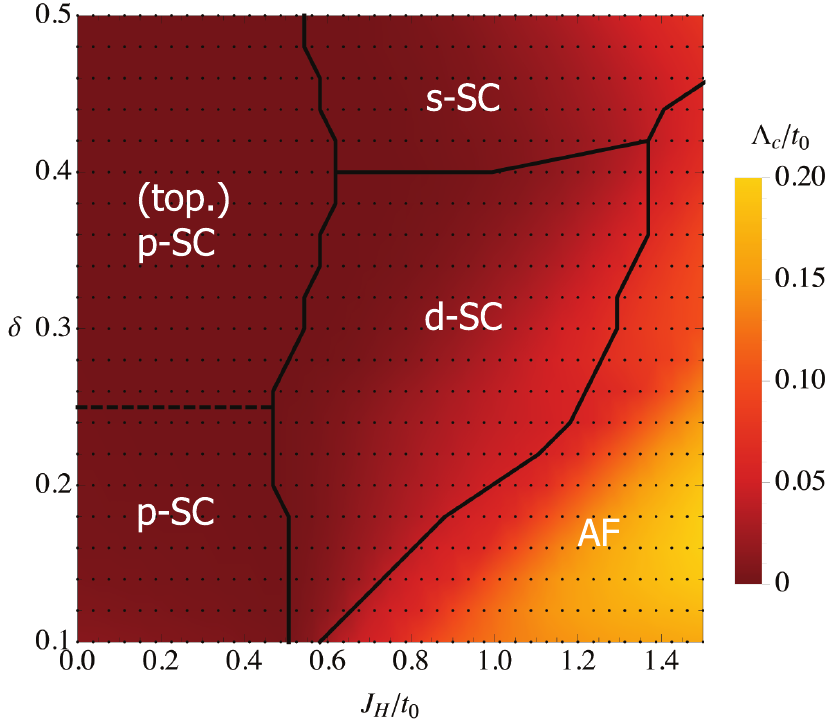}
\caption{The phase diagram as obtained from the numerical solution of a $(N=24)$-patching scheme with
full fRG with $\JK/t_{0} = -1$. Black dots mark the parameters for which fRG-flows were evaluated. The horizontal axis gives the strength
of the antiferromagnetic Heisenberg coupling $\JH > 0$ in units of the bare hopping amplitude $t_{0}$, while the doping level $\delta$ is given on the vertical axis. The colorcode describes the magnitude of the critical scale $\Lambda_c$ across the phase diagram. The dashed black line
marks the transition to a topological odd-parity pairing state across the van Hove singularity. The inclusion of particle-hole fluctuations only affects the singlet instability. The $d$-wave instability ($d$-SC) is turned into a magnetic instability with a momentum structure corresponding to a  N$\mathrm{\acute e}$el antiferromagnet (AF). For superconducting instabilities,
labels in the phase diagram refer to intraband pairing symmetries. See Sect.~\ref{sec:stripy} for a detailed description.}
\label{fig:pd1}
\end{figure}
We start by considering the Kitaev-Heisenberg model with ferromagnetic Kitaev, $\JK < 0$, and antiferromagnetic Heisenberg coupling, $\JH > 0$. Fig.~\ref{fig:pd1} shows the phase diagram obtained from our fRG instability analysis in the parameter space spanned by doping $\delta$ and Heisenberg exchange $\JH$ in units of the bare hopping amplitude $t_{0}$. The Kitaev coupling is fixed to $\JK/t_{0}=-1$ in units of the bare hopping. The general structure of the phase diagram becomes apparent already at small doping. A $p$-wave triplet superconductor is supported by ferromagnetic Kitaev exchange. The superconducting instability switches from triplet to singlet as the
antiferromagnetic Heisenberg coupling increases and finally gives way to a N$\mathrm{\acute e}$el antiferromagnet, expected for large $\JH$. Even at low
doping, we find no hint of the stripy phase, one of the ground-states encountered in the undoped system, cf. Fig.~\ref{fig:undoped}. Although
our method becomes unreliable in the limit $\delta \to 0$ due to truncation and additional approximations (see beginning of Sect.~\ref{sec:results} and Sect.~\ref{sec:conclusions}), we interpret this observation as a destabilization of the stripy phase due to finite doping. For doping level $\delta > 1/4$, we confirm topological $p$-wave phases in the doped Kitaev-Heisenberg model. Further, the mean-field prediction of a topological triplet $p$-wave superconductor with topologically protected Majorana modes is left untouched~\cite{hyart2012} by our findings. Our results further suggest that the generation of triplet pairing instabilities in this parameter regime hinges solely on a finite ferromagnetic Kitaev exchange. 

Since in the quest for realizing accessible $p$-wave superconductors, a high transition temperature into the superconducting state is desirable,
let us note that we identify a window of critical scales corresponding to $p$-wave instabilities in the range $10^{-2} t_{0}$ to $10^{-4} t_{0}$ in units of the bare hopping. 

\subsubsection{Antiferromagnetic Kitaev $\JK > 0$, ferromagnetic Heisenberg $\JH < 0$ exchange}
\label{sec:mainresultsB}
%
\begin{figure}[t!]
\centering
\includegraphics[height=.85\columnwidth]{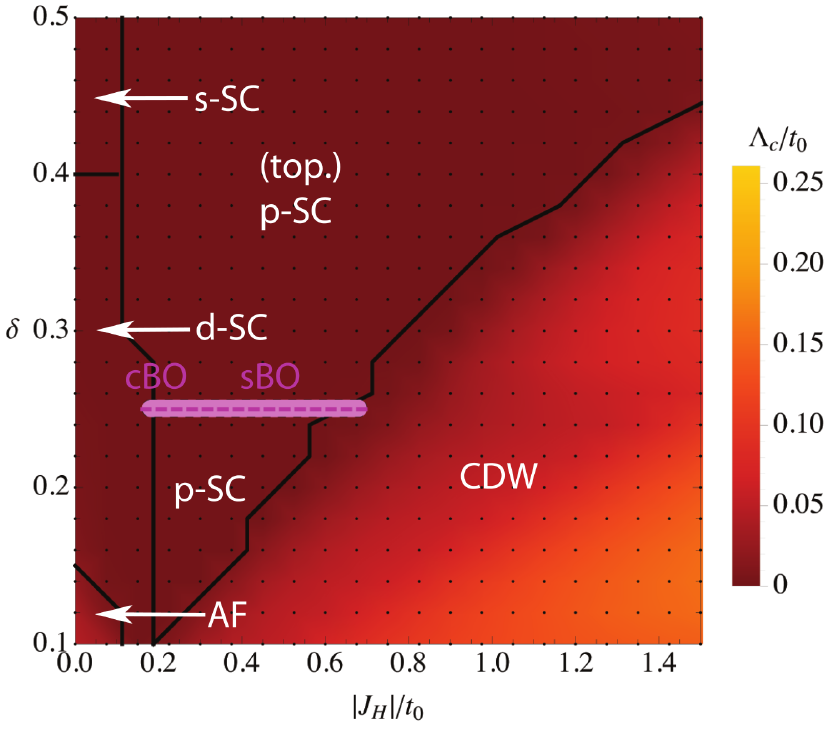}
\caption{The phase diagram as obtained from the numerical solution of a $(N=24)$-patching scheme with
full fRG with $\JK/t_{0} = 1$. The phase boundaries were checked with a $(N=96)$-patching scheme, and only
the singlet SC to triplet SC phase-boundary was mildly revised. Black dots mark the parameters for which fRG-flows were evaluated. The horizontal axis gives the strength
of the ferromagnetic Heisenberg coupling $\JH < 0$ in units of the bare hopping amplitude $t_{0}$, while the doping level $\delta$ is given on the vertical axis. The colorcode describes the magnitude of the critical scale $\Lambda_c$ across the phase diagram. The dashed magenta line
marks the van Hove singularity. The magenta shading represents the formation of charge (cBO) and spin bond-order (sBO) instabilities at van Hove filling, as obtained
from a $(N=96)$-patching scheme. For the critical scales at van Hove filling, see Fig.~\ref{fig:critscalevHF}. For details on bond-order, see Sect.~\ref{sec:bondorder}. The $p$-wave triplet phase appears in a parameter range where ferromagnetic exchange interactions dominate. Labels for superconducting instabilities refer to intraband pairing symmetries. For large Heisenberg exchange, the instability is of CDW type due to partice-hole fluctuations. See Sect.~\ref{sec:zigzag} for a detailed description.}
\label{fig:pd2}
\end{figure}
In the case of antiferromagnetic Kitaev, $\JK > 0$, and ferromagnetic Heisenberg exchange, $\JH < 0$, we again find triplet $p$-wave solutions, which are supported by both Kitaev and Heisenberg exchange. While ferromagnetic Heisenberg exchange polarizes the electronic states and eases the formation of Cooper pairs in the triplet channel, the antiferromagnetic Kitaev interaction turns out to still play a vital role in forming the instability. Indeed, for $\JK = 0$ and below a doping-dependent critical Heisenberg coupling $\JH$, we observe no ordering tendencies down to the lowest scales accessible within our approach in neither singlet nor triplet channels with $\delta \neq 0$ (an exception is the special filling $\delta = 1/4$, see below). The phase diagram we obtained in the plane of doping $\delta$ and Heisenberg exchange $|\JH|$ is shown in Fig.~\ref{fig:pd2} for the case of a constant Kitaev interaction $J_K/t_0 = 1$. 

At small doping and Heisenberg exchange, we find the ground-state to realize a N$\mathrm{\acute e}$el antiferromagnet. Increasing the strength of the Heisenberg interaction while keeping the doping level low, a narrow superconducting window is sandwiched between the N$\mathrm{\acute e}$el state and a charge density wave stabilized by the density-density contribution to Eq.~(\ref{eq:hamiltonian}). This superconducting region increases in size as the doping grows. The ferromagnetic Heisenberg exchange naturally favors the $p$-wave triplet over singlet pairing, and we observe an extended region of triplet $p$-wave pairing throughout the phase diagram. A topological $p$-wave state hosting Majorana edge-excitations is formed for doping $\delta > 1/4$. The range of critical scales that serves as an estimate of transition temperatures, is
here given as $10^{-1} t_{0}$ to $10^{-8} t_{0}$.

For the special case of van Hove filling $\delta = 1/4$, however, we encounter a family of instabilities not present for ferromagnetic Kitaev and antiferromagnetic Heisenberg exchange, see
dashed magenta line in Fig.~\ref{fig:pd2}. At $\delta = 1/4$, the Fermi surface becomes straight, leading to nesting and enhanced DOS effects. 
As expected, these conditions strongly enhance the tendency for particle-hole instabilities. For antiferromagnetic Kitaev and ferromagnetic Heisenberg exchange, we observe bond-order instabilities beyond a rather small critical Heisenberg coupling $|\JH|/t_{0}\simeq 0.2$. In fact,
even for $\JK = 0$, we observe the formation of bond-order instabilities at van Hove filling.
The bond-order instability is leading until we hit the triplet $p$-wave/charge-density wave phase-boundary. Once the pairing neighborhood in the phase diagram has disappeared, 
bond-order signatures in the vertex function are rendered subleading. No leading bond-order instability is found for ferromagnetic Kitaev and antiferromagnetic Heisenberg exchange.
Our results add to the current surge of unconventional bond-order instabilities with concomitant electronic dimerization in triangular lattice systems with non-trivial orbital or sublattice
structure and longer-ranged interactions. We find, however, only nearest-neighbor dimerization, which rules out the possibility of a dynamically generated 
topological Mott insulator, that has previously been found for extended single- and multilayer honeycomb Hubbard models. 
Exotic properties of previously identified bond-order phases include charge fractionalization due to vortices in the Kekul$\mathrm{\acute e}$ order~\cite{hou2007,grushin2013}, valence-bond crystal states~\cite{lang2013}, spontaneously generated current patterns~\cite{grushin2013} and interaction-driven emergent topological states~\cite{raghu2008,wen2010,daghofer2014,grushin2013,martinez2013,duric2014}. One other remarkable feature of the dimerized phase with spin bond-order of the Kitaev-Heisenberg model is the dynamical re-generation of spin-orbit coupling. A similar observation was made in the Kagome Hubbard model with fRG methods~\cite{kiesel2013}. Interestingly, in the present case it does not lead to a gapped state, but instead we obtain a metallic state in a downfolded Brillouin zone. We note that bond order is not exclusive to systems with an underlying triangular lattice or complex orbital structure, but is also observed in square lattice systems~\cite{sachdev2013,sau2013,sachdev2014a,sachdev2014b}.

\subsection{Outline of this Paper}
\label{sec:outline}

In Sect.~\ref{sec:mott} we derive an auxiliary fermion Hamiltonian
on the honeycomb lattice that will form the input to our fRG calculations. To account for doping effects in the Kitaev-Heisenberg system, we start with a $t - \JK - \JH$ model. The kinetic energy will be minimally described by spin-independent hopping, assuming that high-energy spin-orbit effects are accounted for by the Kitaev exchange term. The exclusion of doubly occupied sites due to the strong interactions in the Mott insulating state, i.e., a Gutzwiller projection, is dealt with by the slave-boson method. 
In Sect.~\ref{sec:slave} we discuss the slave-boson treatment of the Mott insulating state and the mean-field approximations in the bosonic sector to map the problem onto a metal of auxiliary fermions with renormalized Fermi surface coupled by exchange interactions. The resulting Hamiltonian is then split into singlet and triplet bond-operators. We then briefly recapitulate slave-boson mean-field results obtained previously by various groups in Sect.~\ref{sec:mean}. 
In Sect.~\ref{sec:method} we provide the necessary background of the functional renormalization group method. It is based on the idea of the Wilsonian renormalization procedure of treating quantum corrections to the classical sector of a given theory by successively integrating out high-energy momentum shells, renormalizing the vertices of the remaining low-energy degrees of freedom. It relies on an exact hierarchy of coupled renormalization group equations for the 1-particle irreducible vertex functions. A solution of our flow equations corresponds to an unbiased resummation of 1-loop diagrams in both particle-particle and particle-hole channels.
Our results are presented in detail in Sect.~\ref{sec:results}. The phase diagram for ferromagnetic Kitaev and antiferromagnetic Heisenberg exchange is 
discussed in Sect.~\ref{sec:stripy}. We first demonstrate the capability of the fRG method to reproduce mean-field results for the pairing channel in Sect.~\ref{subsec:pp} by studying the effect of exclusive particle-particle fluctuations. This is done both at the level of the phase diagram and the form-factors of the superconducting order-parameters. We then move on to include particle-hole fluctuations and discuss the resulting modifications of the phase diagram in Sect.~\ref{subsec:ppph} as compared to the pure particle-particle resummation in the pairing channel. 
In Sect.~\ref{sec:zigzag} we flip the signs of exchange interactions and discuss the phase diagram for
the doped Kitaev-Heisenberg model with antiferromagnetic Kitaev and ferromagnetic Heisenberg exchange.
Sect.~\ref{sec:bondorder} is devoted to the rather special filling condition $\delta = 1/4$ with coinciding van Hove singularity and perfect Fermi surface nesting. 
Finally, in Sect.~\ref{sec:conclusions} we briefly discuss our findings and discuss the validity of applying
the fRG method to the $t- \JK - \JH$ model with strong interactions.

\section{Slave-boson formulation for a doped and frustrated Mott insulator}
\label{sec:mott}

In the following, we briefly describe the $U(1)$ slave-boson construction~\cite{hyart2012} to deal with the Gutzwiller projection
onto the Hilbert space of no doubly occupied sites. This
yields an effective theory in terms of auxiliary fermionic
degrees of freedom, where the bosonic charge excitations (holons) have condensed by assumption. The
background holon-condensate will then give rise to an effective renormalized dispersion for the auxiliary fermions. This dispersive fermionic model
will form the starting point for our fRG analysis of the doped Kitaev-Heisenberg model.

\subsection{From Kitaev-Heisenberg
$t-\JK-\JH$ model to a $U(1)$ slave-boson model}
\label{sec:slave}

The local electron Hilbert-space corresponding to
$c_{o,\sigma,\vec{r}}$ and $c_{o,\sigma,\vec{r}}^{\dagger}$ can be
represented by fermionic $f_{o,\sigma,\vec{r}}$, bosonic holon
$b_{o,\vec{r}}$ and doublon $d_{o,\vec{r}}$ degrees of freedom.
This description, however, introduces unphysical states. 
To project the artificially enlarged Hilbert space down
onto the physical states, the local constraint 
\be
\label{eq:constraint}
\sum_{\sigma}f_{o,\sigma,\vec{r}}^{\dagger}f_{o,\sigma,\vec{r}} + b_{o,\vec{r}}^{\dagger} b_{o,\vec{r}} + d_{o,\vec{r}}^{\dagger}d_{o,\vec{r}} = 1
\ee
needs to be enforced on the operator level. The Gutzwiller projection is now
taken into account by deleting doublon operators and states
from the theory.

The electron creation and annihilation operators are now to be
replaced by
\be
c_{o,\sigma,\vec{r}} \rightarrow b_{o,\vec{r}}^{\dagger} \, f_{o,\sigma,\vec{r}}, \quad
c_{o,\sigma,\vec{r}}^{\dagger} \rightarrow f_{o,\sigma,\vec{r}}^{\dagger} \, b_{o,\vec{r}} .
\ee
Importantly, the spin operator can be written in terms of
auxiliary fermions only,
$\vec{S}_{\vec{r}}=f_{\vec{r},\sigma}^{\dagger}\vec{\sigma}_{\sigma \sigma^{\prime}}f_{\vec{r},\sigma^{\prime}}$, with
$\vec{\sigma}=(\sigma_{x},\sigma_{y},\sigma_{z})^{T}$ the vector of Pauli matrices.
Assuming the holons to Bose-condense into a collective state,
$\langle b^{\dagger}b\rangle = \delta$,
the local constraint for the fermions becomes
$\sum_{\sigma}f_{o,\sigma,\vec{r}}^{\dagger}f_{o,\sigma,\vec{r}}=1-\delta$.

The kinetic part of the Hamiltonian can now be cast in
the form
\be\label{eq:kinetic}
H_{\mathrm{kin}} & = & -t
\sum_{\sigma,\vec{r}\in A,\vec{\delta}_{i}}\left[f_{A,\sigma,\vec{r}}^{\dagger}f_{B,\sigma,\vec{r}+\vec{\delta}_{i}} +
\mathrm{h.c.}\right]  \\
& & - \mu_{f}\sum_{\sigma,o\in A,B,\vec{r}} f_{o,\sigma,\vec{r}}^{\dagger}f_{o,\sigma,\vec{r}} \nn
\ee
with the renormalized hopping amplitude $t=t_{0}\,\delta$ and $\mu_{f}=\delta\, \mu$
the chemical potential which we adjust such that $\sum_{\sigma}\langle f_{o,\sigma,\vec{r}}^{\dagger}f_{o,\sigma,\vec{r}} \rangle=1-\delta$ is fulfilled. 
The constraint eliminating unphysical states is thus only included on average in this approach.

In going from the tight-binding
to the Bloch representation by introducing the operators
\be
f_{o,\sigma,\vec{k}} = 
\frac{1}{\sqrt{\mathcal{N}}}\sum_{\vec{r}}\mathrm{e}^{\mathrm{i}\vec{k}\cdot\vec{r}}\,f_{o,\sigma,\vec{r}}\,,\,
\ee
where $\mathcal{N}$ is the number of unit cells of the honeycomb lattice, we obtain the Bloch Hamiltonian
\be
H_{\mathrm{kin}} =
-\sum_{\sigma,\vec{k}}(f_{A,\sigma,\vec{k}}^{\dagger},f_{B,\sigma,\vec{k}}^{\dagger})
\begin{pmatrix}
\mu_{f} &  t d_{\vec{k}}^{\ast}\\
 t d_{\vec{k}} & \mu_{f}
\end{pmatrix}
\begin{pmatrix}
f_{A,\sigma,\vec{k}} \\
f_{B,\sigma,\vec{k}}
\end{pmatrix}
\ee
with
$d_{\vec{k}}=\sum_{\vec{\delta}_{i}}\mathrm{e}^{\mathrm{i}\vec{k}\cdot\vec{\delta}_{i}}$.
The dispersion is analogous to electrons moving in a graphene monolayer, with the concomitant $\vec{K}$ and $\vec{K}^{\prime}$
points in the Brillouin zone.

The interaction terms quartic in the fermion operators can
be expressed in terms of singlet and triplet contributions,
which yields a very convenient starting point for the analysis of 
superconducting instabilities.

The spin-singlet operator defined on a bond
connecting site $\vec{r}\in A$ to $\vec{r}+\vec{\delta}_{i}\in
B$ is defined as
\be
s_{\vec{r},\vec{\delta}_{i}} 
& = & f_{A,\sigma,\vec{r}}\, [\Gamma_{0}]_{\sigma\sigma^{\prime}}\, f_{B,\sigma^{\prime},\vec{r}+\vec{\delta}_{i}} \\
& = & \frac{1}{\sqrt{2}}\left(f_{A,\uparrow,\vec{r}}f_{B,\downarrow,\vec{r}+\vec{\delta}_{i}}-f_{A,\downarrow,\vec{r}}f_{B,\uparrow,\vec{r}+\vec{\delta}_{i}}\right) \nn
\ee
and correspondingly the $x$, $y$, $z$ triplet operators read
\be
t_{\vec{r},\vec{\delta}_{i};x} 
& = & f_{A,\sigma,\vec{r}}\, [\Gamma_{x}]_{\sigma\sigma^{\prime}}\, f_{B,\sigma^{\prime},\vec{r}+\vec{\delta}_{i}},
\\
& = & 
\frac{1}{\sqrt{2}}\left(f_{A,\downarrow,\vec{r}}f_{B,\downarrow,\vec{r}+\vec{\delta}_{i}}-f_{A,\uparrow,\vec{r}}f_{B,\uparrow,\vec{r}+\vec{\delta}_{i}}\right), \nn \\ 
t_{\vec{r},\vec{\delta}_{i};y} 
& = & f_{A,\sigma,\vec{r}}\, [\Gamma_{y}]_{\sigma\sigma^{\prime}}\, f_{B,\sigma^{\prime},\vec{r}+\vec{\delta}_{i}}
\\
& = & 
\frac{\mathrm{i}}{\sqrt{2}}\left(f_{A,\uparrow,\vec{r}}f_{B,\uparrow,\vec{r}+\vec{\delta}_{i}}+f_{A,\downarrow,\vec{r}}f_{B,\downarrow,\vec{R}+\vec{\delta}_{i}}\right), \nn \\ 
t_{\vec{r},\vec{\delta}_{i};z} 
& = & f_{A,\sigma,\vec{r}}\, [\Gamma_{z}]_{\sigma\sigma^{\prime}}\, f_{B,\sigma^{\prime},\vec{r}+\vec{\delta}_{i}}
\\
& = & 
\frac{1}{\sqrt{2}}\left(f_{B,\uparrow,\vec{r}}f_{A,\downarrow,\vec{r}+\vec{\delta}_{i}}+f_{B,\downarrow,\vec{r}}f_{A,\uparrow,\vec{r}+\vec{\delta}_{i}}\right) \nn.
\ee
Here, we used the $\Gamma$ matrices $\Gamma_{0}=\frac{1}{\sqrt{2}}\sigma_{0}\mathrm{i}\sigma_{y}$, $\Gamma_{x}=\frac{1}{\sqrt{2}}\sigma_{x}\mathrm{i}\sigma_{y}$,
 $\Gamma_{y}=\frac{1}{\sqrt{2}}\sigma_{y}\mathrm{i}\sigma_{y}$ and $\Gamma_{z}=\frac{1}{\sqrt{2}}\sigma_{z}\mathrm{i}\sigma_{y}$. 
See Appendix~\ref{app:matrix} for explicit expressions for these matrices and their relation to superconducting order parameters. 
The interaction part of the slave-boson Hamiltonian
\be\label{eq:slaveboson}
H_{\mathrm{slave}} = H_{\mathrm{kin}} + H_{\mathrm{int}}^{(\mathrm{s})} + H_{\mathrm{int}}^{(\mathrm{t})} 
\ee
can now be re-cast as a sum of a singlet interaction
\be\label{eq:initial_singlet}
H_{\mathrm{int}}^{(\mathrm{s})}=-\left(\JH
+ \frac{\JK}{4}\right)\sum_{\vec{r}\in A,\vec{\delta}_{i}}s_{\vec{r},\vec{\delta}_{i}}^{\dagger}s_{\vec{r},\vec{\delta}_{i}}
\ee
and a triplet interaction
\be\label{eq:initial_triplet}
H_{\mathrm{int}}^{(\mathrm{t})}=-
\frac{\JK}{4}\sum_{l\in\{x,y,z\}}\sum_{\vec{r}\in A,\vec{\delta}_{i}}\,\zeta_{\vec{r},\delta_{i}}^{l}
t_{\vec{r},\vec{\delta}_{i};l}^{\dagger}t_{\vec{r},\vec{\delta}_{i};l}\, .
\ee
Here, the index $l$ runs over the triplet components $x$, $y$ and $z$. The pre-factor $\zeta_{\vec{r},\delta_{i}}^{l}$ describes a bond-dependent sign modulation of the interaction term for each triplet component: we have $\zeta_{\vec{r},\delta_{i}}^{l}=+1$
if the bond-type ($x$, $y$ or $z$) from site $\vec{r}$ to $\vec{r}+\vec{\delta}_{i}$ coincides with the triplet component $l$.
Otherwise,  $\zeta_{\vec{r},\delta_{i}}^{l}=-1$. The highly frustrated Kitaev term contains a singlet contribution, which 
renormalizes the singlet interaction coming from Heisenberg exchange. The contribution to the triplet channel from
the Kitaev term is irreducible in the sense that it does not contain any `hidden' singlet contributions. Thus, the singlet-triplet
decomposition is unique.

We note that the interaction does not contain terms that describe the decay of
a singlet state into a triplet state or vice versa. It is worth emphasizing
that we model the kinetic term for the auxiliary fermions with a simple nearest-neighbor
hopping, which furthermore preserves spin. Thus the only $SU(2)_{\mathrm{spin}}$ violating contribution
to the Hamiltonian comes from Kitaev exchange.

In a solid-state system with strong spin-orbit coupling, the $SU(2)_{\mathrm{spin}}$ symmetry is
locked to the point group of the lattice~\cite{sigrist1991,you2012}, i.e., only the \textit{simultaneous}
application of point-group transformations and the corresponding representation of
the point-group operations on the spin degrees of freedom leave the Hamiltonian invariant.
In, e.g., the iridates the Kitaev term arises precisely due to the presence of strong spin-orbit coupling.
As noted previously~\cite{you2012}, it is thus natural that the symmetry of the Kitaev term
involves simultaneous transformation of both spin and lattice (or wavevector). The relevant symmetry
transformations in the case of the Kitaev term (with the same coupling $\JK$ on nearest-neighbor bonds) acting on the lattice can be understood as a $2\pi/3$ rotation around the center of a honeycomb 
hexagon. To maintain invariance of the Kitaev Hamiltonian, spins have to be rotated
by the same angle around the axis $\hat{n}=\frac{1}{\sqrt{3}}(1,1,1)^{T}$, where the coordinate system corresponds
to an embedding of a honeycomb layer into a 3D cubic lattice~\cite{you2012}. See Appendix~\ref{app:ward} for
further details.

\subsection{Comparison with mean-field theory} 
\label{sec:mean}

The phase diagram of the doped Kitaev-Heisenberg model has previously been studied
within different slave-boson formulations. If quantum fluctuations were treated exactly in 
these different formulations, they should yield equivalent results. In practice, however,
the emergent gauge fields in slave-boson approaches for strongly correlated lattice fermions
are treated in a mean-field approximation. The local Hilbert-space constraint is realized only on average. While this is
also true in our slave-boson formulation of the $t-\JK-\JH$ model, where the holon-condensation
is built into the theory in mean-field fashion, the fermionic fRG takes into account the fermionic fluctuations in
all channels in an otherwise unbiased way. To allow for a systematic and self-contained comparison of our fRG results
to mean-field studies reported in the literature, we briefly summarize recent findings.

\subsubsection{$U(1)$ Slave-boson mean-field theory}

A previous self-consistent mean-field study~\cite{hyart2012} of the doped Kitaev-Heisenberg model taking into
account only superconducting order-parameters demonstrated a rich phase diagram upon doping charge carriers 
into the $\mathbbm{Z}_{2}$ quantum spin liquid (QSL) and the Mott insulating stripy phase. For ferromagnetic Kitaev coupling, $\JK < 0$,
the QSL state at zero doping is stable with respect to Heisenberg-type perturbations for $|\JH| < |\JK|/8$, cf. Ref.~\onlinecite{chaloupka2013}.
Keeping the value of the Kitaev exchange coupling fixed for the undoped system, the stripy phase is realized for antiferromagnetic Heisenberg
exchange coupling, $\JH > 0$, with $\JH \simeq 0.25, \dots, 1.5$. At finite doping, $\delta \gtrsim 0.1$, for small antiferromagnetic Heisenberg coupling and dominating
ferromagnetic Kitaev coupling, $|\JK|/2 > \JH$, a time-reversal invariant $p$-wave state ($p$-SC) was found. Upon increasing doping in this regime, a topological phase transition 
to a topological pairing state with $p$-wave symmetry in the triplet channel was obtained~\cite{hyart2012} in the vicinity of van Hove filling. This state
is stable to interband pairing correlations and was recently shown to be robust also against weak non-magnetic disorder~\cite{kimme2013}.
Its topological properties can be characterized by a non-trivial topological $\mathbbm{Z}_{2}$-invariant~\cite{hyart2012,kimme2013} from which we can infer that it falls into the DIII symmetry class of the Altland-Zirnbauer classification~\cite{schnyder2008,ryu2010}.

For antiferromagnetic Heisenberg coupling $\JH \gtrsim |\JK|/2$ at small doping and ferromagnetic Kitaev coupling, the leading pairing correlations
occur in the singlet channel with intraband $d$-wave symmetry. In the large $\JH$ regime, an extended singlet $s$-wave state was reported.

\subsubsection{$SU(2)$ Slave-boson mean-field theory} 

Within an $SU(2)$ formulation~\cite{you2012} and a specific mean-field \textit{Ansatz} that reproduces the
$\mathbbm{Z}_{2}$ quantum spin liquid in the $\delta = 0$ Kitaev limit $\JH=0$, a time-reversal symmetry breaking
$p$-wave state in the triplet channel ($p$-SC$_1$) was found upon doping the stripy phase. Interestingly, it supports chiral edge modes and localized Majorana states in the low-doping
regime. The physics of this state was argued to be dominated by its vicinity to the QSL at $\JH = 0$, while
for increased doping a first order transition to a BCS like state ($p$-SC$_2$) occurs. 

A further extensive mean-field study~\cite{okamoto2013} for the doped Kitaev-Heisenberg model
utilizing the $SU(2)$ formulation with an \textit{Ansatz} that includes pairing and magnetic order-parameters
supports the previous findings. In both mean-field studies for ferromagnetic $\JK$ and antiferromagnetic $\JH$
couplings, however, time-reversal breaking and time-reversal invariant $p$-wave states are reported as (almost)
energetically degenerate for the $p$-SC$_2$ state. In the time-reversal invariant case, 
the $p$-SC$_2$ state obtained from the $SU(2)$ formulation coincides with the $p$-SC state from
the $U(1)$ slave-boson theory. For large $|\JH|$, again singlet instabilities are obtained, with pairing symmetry changing
from $d$- to $s$-wave upon increasing the doping level. 

 For antiferromagnetic $\JK$ and ferromagnetic $\JH$, a $p$-SC$_1$ phase 
 that extends to large doping was reported. Roughly, for dominating antiferromagnetic Kitaev coupling, 
 a $d$-wave singlet solution was obtained, with a transition to a $p$-SC$_2$ phase upon increasing $\JH$ and/or doping. For 
 $\JH \gg |\JK|$, all pairing correlations disappear and a ferromagnetic ordering emerges. Mean-field \textit{Ans\"atze} other
 than superconducting and ferromagnetic were not considered.

\section{fRG method}
\label{sec:method}

We employ a functional renormalization group (fRG) approach for
the one-particle-irreducible (1PI) vertices with a momentum cutoff.
For a recent review on the fRG method, see Refs.~\onlinecite{metzner2011,platt2013}.
The actual fRG calculation is performed in the band-basis in
which the quadratic part of the fermion Hamiltonian is diagonal. The
free Hamiltonian can be diagonalized by a unitary transformation
of the form
\be\label{eq:unitary}
f_{b,\sigma,\vec{k}} = \sum_o u_{bo,\vec{k}}\, f_{o,\sigma,\vec{k}}\,,\quad
f^{\dagger}_{b,\sigma,\vec{k}} = \sum_o u^{*}_{bo,\vec{k}}\, f^{\dagger}_{o,\sigma,\vec{k}}\,,
\ee
where $o\in\{A,B\}$ labels the two sublattices and the index $b$
denotes the corresponding bands. This transformation also affects the bare
interaction vertex and leads to so-called orbital make-up. Since we do not consider mixing of spin states by the kinetic term of the Hamiltonian, this unitary transformation does not
involve spin projection. In band representation, the propagator is also diagonal,
with diagonal entries encoding the dispersion $\epsilon(\vec{k},b)$ of the various bands labeled by $b$.
In the standard 1PI fRG-scheme we employ here, an infrared regulator with energy scale
$\Lambda$ is introduced into the bare propagator function in band representation
$G_{0}(\xi,\xi^{\prime}) \sim \delta_{\xi,\xi^{\prime}}$, where the label $\xi=(\sigma,b,\omega,\vec{k})$ 
collects spin projection $\sigma$, band index $b$, frequency $\omega$ and Bloch momentum $\vec{k}$.
We thus replace
\begin{equation}
G_0(b,\omega,\vec{k})\rightarrow
G_0^\Lambda(b,\omega,\vec{k})=\frac{C^\Lambda
[\epsilon(\vec{k},b)] }{i \omega -\epsilon(\vec{k},b) }.
\end{equation}
As the spin quantum number $\sigma$
carried by the auxiliary fermion degrees of freedom is conserved by the
kinetic part of the Hamiltonian
\Eqref{eq:slaveboson} the free propagator is diagonal also in spin indices. The
cutoff function is chosen to enforce an energy cutoff, which
regularizes the free Green's function by suppressing the modes
with band energy below the scale $\Lambda$,
\begin{equation}
C^\Lambda [\epsilon(\vec{k},b)] \approx \Theta \left(
|\epsilon(\vec{k},b)| - \Lambda \right)\,.
\end{equation}
For better numerical feasibility the step function is slightly
softened in the actual implementation. With this modified
scale-dependent propagator, we can define the scale-dependent
effective action $\Gamma^\Lambda$ as the Legendre transform of
the generating functional $\mathcal G^\Lambda$ for correlation
functions, cf. Refs.~\onlinecite{metzner2011,negorl}.
The RG flow of $\Gamma^\Lambda$ is generated upon variation of
$\Lambda$. By integrating the flow down from an initial scale
$\Lambda_0$ one smoothly interpolates between the bare action of the system and the
effective action at low energy.

\subsection{Flow equations for $SU(2)_{\mathrm{spin}}$ non-invariant systems}

While the $U(1)$ slave-boson model~\Eqref{eq:slaveboson} is equipped with
global $U(1)$ symmetry
in the fermion sector even after the bosonic holons have
condensed, there is no $SU(2)_{\mathrm{spin}}$ symmetry present as for electronic
systems \textit{without} spin-orbit coupling. Since the
interactions do not conserve the spin quantum number carried by
the fermions, the four-point vertex will also depend on the
specific spin configuration of incoming and outgoing states.
Without any further symmetry constraints, this leads to a total
of $16 = 2^{4}$ independent coupling functions, one for each possible spin configuration. 
For the Hamiltonian in \Eqref{eq:slaveboson}, however, it suffices to consider
only four vertex functions. These are simply given by singlet-singlet and
triplet-triplet interactions. The renormalization group flow
preserves this property, i.e., if it is present in the initial
condition, singlet-triplet mixing terms will never be generated during the
RG evolution.\footnote{This property hinges on the $SU(2)_{\mathrm{spin}}$ invariance
of the kinetic Hamiltonian. Generally, the fRG-flow is expected to leave the manifold spanned by the singlet and triplet vertex functions, if we were to include spin dependent hopping processes. Then the full space of spin configurations needs to be resolved.}

This allows us to factor out the spin-indices from the vertex functions. To
this end we introduce the short-hand notation  $\xi=(\sigma,\txi)$ for the
set of quantum numbers, where $\txi=(o,\omega,\vec{k})$ in orbital/sublattice
representation and $\txi=(b,\omega,\vec{k})$ in band representation. After factorization,
the flow equations can be formulated for only four vertex functions depending
on the quantum numbers $\tilde{\xi}$, exclusively.
With these premises, the scale-dependent coupling function $V^{\Lambda}=V^{\Lambda}(\xi_1,\xi_2,\xi_3,\xi_4)$
can be expanded in terms of the singlet vertex function, $V^{(\mathrm{s})}=V^{(\mathrm{s})}(\txi_1,\txi_2,\txi_3,\txi_4)$, and the triplet vertex functions, $V^{(\mathrm{t})}_{l}=V^{(\mathrm{t})}_{l}(\txi_1,\txi_2,\txi_3,\txi_4)$, as
\be
V^{\Lambda} &=& 
-V^{(\mathrm{s})}
[\Gamma_{0}^{\dagger}]_{\sigma_1\sigma_2}[\Gamma_{0}]_{\sigma_3\sigma_4}  
\nn \\ &{}&\quad
+\sum_{l\in\{x,y,z\}}V^{(\mathrm{t})}_{l}
[\Gamma_{l}^{\dagger}]_{\sigma_1\sigma_2}[\Gamma_{l}]_{\sigma_3\sigma_4}, 
\ee
The projection onto scale-dependent singlet $V^{(\mathrm{s})}$
and triplet $V^{(\mathrm{t})}_{l}$ vertex functions can be
facilitated with the orthogonality properties of the $\Gamma$
matrices, see Appendix \ref{app:matrix}, effectively tracing out spin
quantum numbers $\sigma_{i}$, $i=1,\dots,4$.

The singlet vertex function $V^{(\mathrm{s})}$ is fully symmetric with 
respect to exchanging $\txi_1 \leftrightarrow \txi_2$ and $\txi_3 \leftrightarrow \txi_4$.
The antisymmetric spin matrix $\Gamma_{0}$ ensures overall antisymmetry under these
exchange operations, as required by a fermionic 4-point function. Correspondingly,
the triplet vertex functions $V_{l}^{(\mathrm{t})}$ are antisymmetric under these exchange operations with
symmetric spin matrices $\Gamma_{l}$.

\begin{figure}[t!]
\centering
\includegraphics[height=1.2\columnwidth]{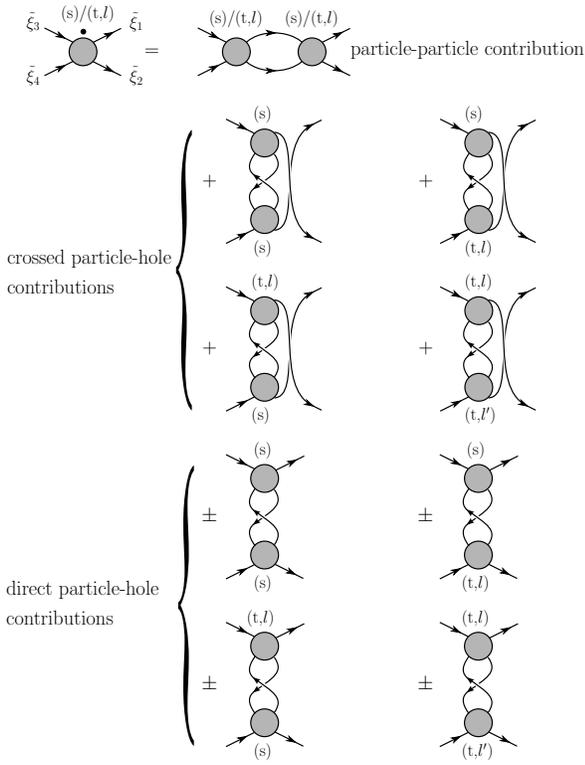}
\caption{Diagrammatic representation of flow \Eqref{eq:singletflow} and \Eqref{eq:tripletflow}. 
The dot on the l.h.s of the equation denotes the scale derivative $d/d\Lambda$. In the loops, one line always corresponds to a bare
regularized propagator $G_{0}^{\Lambda}$, while the other one corresponds to a so-called single scale
propagator $S^{\Lambda} = d/d\Lambda\, G_{0}^{\Lambda}$. The labels $(\mathrm{s})$ and $(\mathrm{t},l)$
distinguish singlet and triplet vertices. The first contribution to the flow is the respective particle-particle diagram $\phi_{\mathrm{pp}}^{(\mathrm{s})}$, $\phi_{\mathrm{pp};l}^{(\mathrm{t})}$
that does not couple singlet and triplet channels. The remaining two sets of diagrams correspond to
the crossed ($\phi_{\mathrm{ph,cr}}^{(\mathrm{s})}$, $\phi_{\mathrm{ph,cr};l}^{(\mathrm{t})}$) and direct particle-hole bubbles ($\phi_{\mathrm{ph,d}}^{(\mathrm{s})}$, $\phi_{\mathrm{ph,d};l}^{(\mathrm{t})}$). In the direct particle-hole diagrams the positive (negative) signs refer to the contributions to the flow of the singlet (triplet) vertex. See also Appendix~\ref{app:flow} for details on 
the different sign structures for singlet and triplet flow equations.}
\label{fig:diagrams}
\end{figure}

Employing the flow equations appropriate for global $U(1)$ symmetry~\cite{salmhofer2001},
we find for the singlet evolution
\be\label{eq:singletflow}
\frac{d}{d\Lambda}V^{(\mathrm{s})} =
\phi_{\mathrm{pp}}^{(\mathrm{s})} + \phi_{\mathrm{ph,cr}}^{(\mathrm{s})} +
\phi_{\mathrm{ph,d}}^{(\mathrm{s})},
\ee
where $\phi_{\mathrm{pp}}^{(\mathrm{s})}$, $\phi_{\mathrm{ph,d}}^{(\mathrm{s})}$ 
and $\phi_{\mathrm{ph,cr}}^{(\mathrm{s})}$ denote the particle-particle,
direct particle-hole and crossed particle-hole RG contributions
to the singlet channel. The triplet evolutions are found along the same lines as
\be\label{eq:tripletflow}
\frac{d}{d\Lambda}V^{(\mathrm{t})}_{l} =
\phi_{\mathrm{pp};l}^{(\mathrm{t})} + \phi_{\mathrm{ph,cr};l}^{(\mathrm{t})} -
\phi_{\mathrm{ph,d};l}^{(\mathrm{t})},\quad l\in\{x,y,z\},
\ee
with the corresponding RG contributions for the three triplet channels. The scale-dependent 
bubble contributions appearing on the right hand sides of \Eqref{eq:singletflow} and \Eqref{eq:tripletflow} are in turn quadratic functionals
of $V^{(\mathrm{s})}$ and $V^{(\mathrm{t})}_{l}$. The explicit
expressions are summarized in Appendix~\ref{app:flow}. From these it follows, that singlet and
triplet channel exert a mutual influence only through particle-hole fluctuations. 
For a diagrammatic representation of the flow equations~\Eqref{eq:singletflow} and~\Eqref{eq:tripletflow} for singlet
and triplet vertices, see Fig.~\ref{fig:diagrams}.
Neglecting both direct and crossed particle-hole contributions to the RG evolutions of singlet and triplet
coupling functions, \Eqref{eq:singletflow} and \Eqref{eq:tripletflow} decouple so that singlet and triplet vertex functions
evolve independently. This decoupling of singlet and triplet channels is also found
in BCS-like mean-field theory from a linearized gap-equation. We thus expect that the inclusion
of only the particle-particle bubbles $\phi_{\mathrm{pp}}^{(\mathrm{s})}$ and $\phi_{\mathrm{pp};l}^{(\mathrm{t})}$
as driving forces in the flow equations yields a resummation that reproduces mean-field results.
Details are discussed in Sect.~\ref{sec:stripy}.

We note there is a difference in sign between crossed and direct particle-hole contributions
in the singlet~\Eqref{eq:singletflow} and triplet~\Eqref{eq:tripletflow} flow equations. 
This sign reflects the exchange symmetries of the various vertex functions, such that the incremental changes $dV^{(\mathrm{s})}$ and
$dV_{l}^{(\mathrm{t})}$ come with the correct (anti-)symmetrization, since neither crossed nor direct particle-hole bubbles
have the symmetry property on their own.

\subsection{Approximations and numerical implementation}

\begin{figure}[t!]
\centering
\includegraphics[height=0.45\columnwidth]{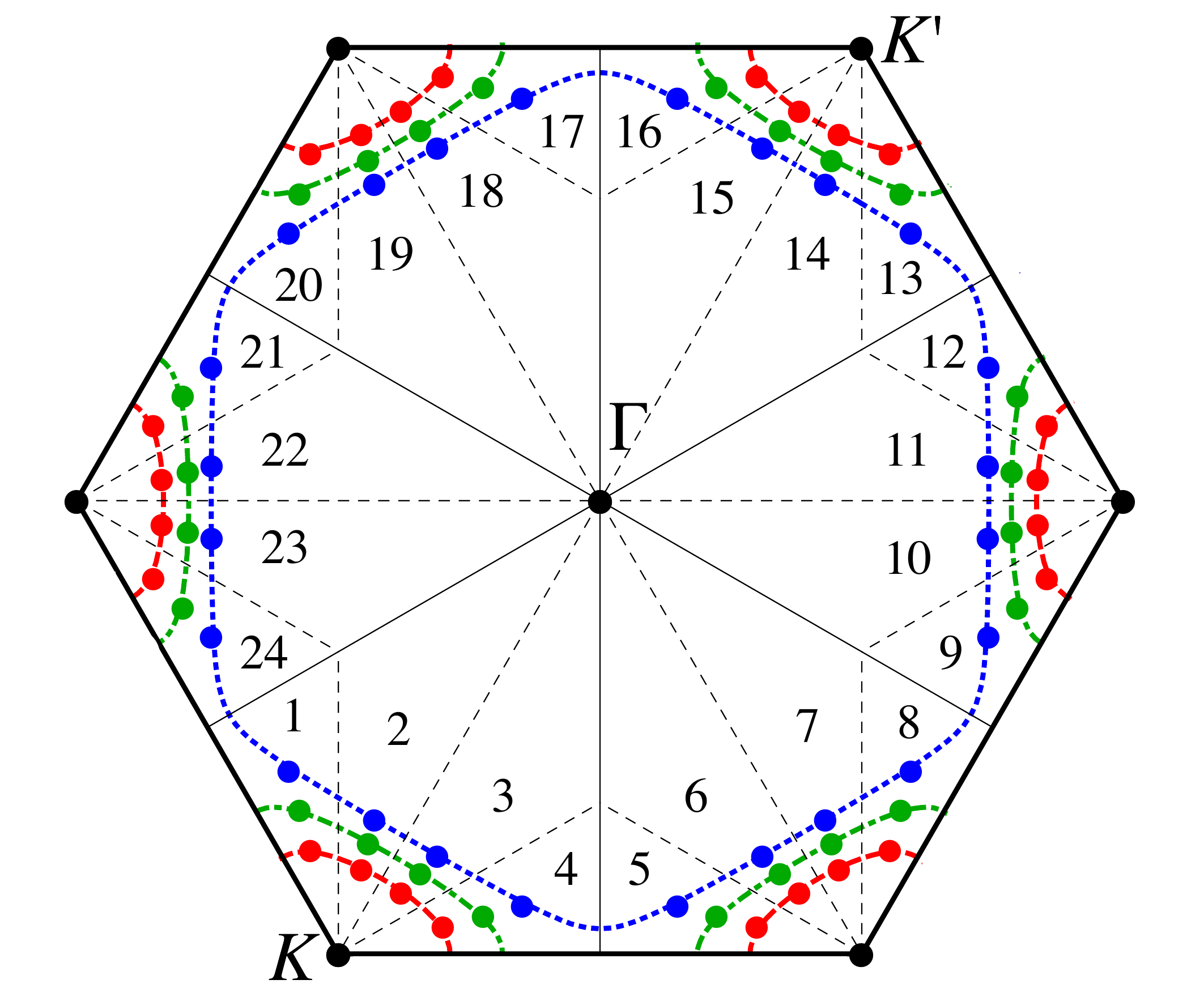}
\caption{Patching of the Brillouin zone (BZ) with $N=24$ for different doping levels. For $\mu \neq 0$ the representative patch momenta (colored dots)
move toward the $\Gamma$-point in the BZ center. At $|\mu| = t_{0}$ the Fermi surface segments become `straight', i.e., perfect nesting is realized, which
is reflected in a van Hove singularity in the density of states. For $|\mu| \gtrsim t_{0}$ 
the actual implementation of the patching needs to be slightly modified. The red, green and blue lines correspond
to the Fermi surfaces for doping $\delta\approx0.08, 0.14, 0.27$, respectively.}
\label{fig:patching}
\end{figure}

In order to limit the numerical effort we employ a number of approximations. First, the
hierarchy of flowing vertex functions is truncated after the
four-point (two-particle interaction) vertex. Second, we employ the static approximation,
i.e., we neglect the frequency dependence of vertex functions, by setting all external
frequencies to zero, as we are interested in ground-state
properties. Third, self-energy corrections are neglected. This
approximate fRG scheme then amounts to an infinite-order summation of one-loop
particle-particle and particle-hole terms of second order in the
effective interactions.
It allows for an unbiased investigation of the competition between various correlations, by analyzing
the components of $V^{(\mathrm{s})}$ and $V^{(\mathrm{t})}_{l}$ that create instabilities by growing large at a
critical scale $\Lambda_c$~\cite{metzner2011}. From the evolving pronounced momentum
structure one can then infer the leading ordering tendencies.
With the approximations mentioned above, this procedure is
well-controlled for small interactions. At intermediate
interaction strengths we still expect to obtain reasonable
results and it was recently shown that the fRG-flow produces sensible
results even in proximity to the singularity~\cite{eberlein2013}. In any case, the fRG takes into account effects beyond mean-field and random phase approximations. This way, the fRG represents an alternative to the inclusion of gauge fluctuations in the mean field theories.
The wavevector dependence of the interaction vertices is
simplified by a discretization -- the $N$-patch scheme -- that resolves the angular dependence
along the Fermi surface for a given chemical potential. The Brillouin zone (BZ) is divided into $N$ patches
with constant wavevector dependence within one patch, so that
the coupling function has to be calculated for only one
representative momentum in each patch. The representative
momenta for the patches are chosen to lie close to the Fermi
level. The patching scheme is shown in Fig.~\ref{fig:patching}
for $N=24$. Calculations were performed for different but fixed angular
resolution with $N=24$ and $N=48$ as well as $N=96$ to check the reliability of the results with respect to higher resolution. The vertex functions further
depend on sublattice or band labels. Since overall momentum conservation leaves
only three independent wavevectors in the BZ, a single vertex function is approximated by a $2^4\times N^3$
component object. In total, we thus obtain $4\times2^4\times N^3$ coupled differential equations 
for the approximated singlet and triplet vertex functions. Exploiting the implications of residual rotational symmetry as
outlined above (see also Appendix~\ref{app:ward}) for vertex reconstruction in the triplet channel, this number can be reduced by a factor of $2$. 
%

\section{Ordering tendencies from functional RG flows}
\label{sec:results}

We start the fRG flow at the initial scale $\Lambda_0$ which we choose as the largest distance
in energy from the location of the Fermi surface to the lower and top band edges of valence and conduction bands, respectively. 
By solving the flow equations numerically, we successively integrate out all modes of these bands in 
energy shells with support peaked around the RG-scale $\Lambda$.
In the case of an instability, some components of the scale-dependent effective interaction
vertices become large and eventually diverge at a critical scale $\Lambda_c >
0$. Since the flow needs to be stopped at a scale $\Lambda^{\ast}\lesssim\Lambda_c$, we take as a stopping criterion the condition that the absolute value of one of the vertex functions 
exceeds a value of the order of 100 times the bare bandwidth $t_{0}$. We
further assume that $\Lambda^{\ast}\simeq \Lambda_{c}$. The precise choice for the stopping criterion
affects the extracted value for the critical scale very mildly,
as the couplings grow very fast in the vicinity of the
divergence. 

In our analysis, we kept the value of the bare hopping $t_{0}$ fixed, and also
fixed the value for the Kitaev coupling $\JK$, while doping $\delta$ and Heisenberg coupling $\JH$
are varied. 

To elucidate the role of holon-condensation on the system parameters, cf. \Eqref{eq:kinetic}, we consider the auxiliary fermion Hamiltonian \Eqref{eq:slaveboson},
and its corresponding partition function $Z(\beta) = \mathrm{Tr}\,\mathrm{e}^{-\beta H_{\mathrm{slave}}}$. While doping levels $\delta < 1$
reduce the bandwidth of the fermion system, we can equivalently view this as a renormalization of the bare exchange interactions. 
Since the partition function $Z(\beta)$ is invariant upon rescaling temperature as $\beta \to \beta\delta$ and at the same
time rescaling the Hamiltonian $H_{\mathrm{slave}} \to H_{\mathrm{slave}}^{\prime} = H_{\mathrm{slave}}/\delta$,
the ground-state properties of $H_{\mathrm{slave}}$ in the limit $\beta\to \infty$ can be extracted from $H_{\mathrm{slave}}^{\prime}$.

In total, this corresponds to rescaling hopping amplitude, chemical potential and couplings as
\be
t \rightarrow \frac{t}{\delta}=t_{0},\,\, \mu_{f}\to\frac{\mu_{f}}{\delta}=\mu,\,\, \JH \rightarrow \frac{\JH}{\delta},\,\, \JK \rightarrow \frac{\JK}{\delta}.
\ee
The rescaling entails large absolute values for the vertex functions already in the initial condition at least
in the low-doping regime $\delta \simeq 0.1$. While we can conveniently keep the
kinetic energy scale at $t_{0}$, we cannot set the bar for the stopping criterion too low
on the vertex functions to allow $-$ at least $-$ for a sizable evolution along the $\Lambda$ direction. From these
considerations it is also immediately obvious that within our slave-boson approach and the employed approximations to
the exact hierarchy of flow equations we cannot describe
the magnetic instabilities of the Mott insulator as $\delta\rightarrow 0$. Including
the flowing self-energy $\Sigma^{\Lambda}$ and keeping frequency dependence in the flows for
self-energy and effective interaction vertex, we could expect to bridge the description to the Mott insulating state, cf. Sect.~\ref{sec:conclusions} and Refs.~\onlinecite{reuther,reuther2012,singh2012}. 

A divergence in the interaction vertex can be considered
as an artefact of our truncation, which as it stands completely neglects
self-energy feedback. Here, we thus restrict ourselves to an analysis of leading ordering tendencies at finite doping.
The pronounced momentum structure of a vertex function close to the
critical scale $\Lambda_{c}$ can be used to extract an effective Hamiltonian
for the low-energy degrees of freedom, which can in principle be decoupled by a suitable Hubbard-Stratonovich field that is then dealt with on a mean-field level. This is used to determine
the order parameter corresponding to the leading instability.
Furthermore, the scale $\Lambda_c$ can be interpreted as an
estimate for ordering temperatures, if ordering is allowed, or at least as the temperature below
which the dominant correlations should be clearly observable.

\subsection{Doping QSL and stripy phase - \\ FM Kitaev and AF Heisenberg exchange}
\label{sec:stripy}

We first consider the case of a ferromagnetic Kitaev, $\JK < 0$, and antiferromagnetic
Heisenberg coupling, $\JH > 0$. At doping level $\delta = 0$, there exists an extended region
in the space of couplings $\JH$, $\JK$ where the stripy phase is realized as the magnetically
ordered ground-state of the strongly correlated spin-orbit Mott insulator. In the following, we will fix
the ferromagnetic Kitaev coupling as $\JK/t_{0} = -1$. The range of Heisenberg couplings we focus on is given by $\JH/t_{0} \in [0,1.5]$. This parameter range includes the full extent of the stripy phase at $\delta = 0$, and also for small $\JH$, the dominating Kitaev term is responsible for the realization of the quantum spin liquid phase (QSL). However, here we
focus on a doping regime $\delta\in[0.1,0.5]$, where the effects of the proximity to the QSL phase are not visible any more.
We thus cannot observe a signature of the $p$-SC$_1$ state~\cite{you2012,okamoto2013}, cf. Sect.~\ref{sec:mean}. 
Slave-boson mean-field studies connecting to the Kitaev-limit as $\delta\to 0$ suggest~\cite{you2012,okamoto2013} that holon-condensation sets in rapidly at $\JH\neq 0$ as $\delta$ is increased from 0 to a small but finite value. Although this seems to render our Hamiltonian a sensible starting point from the point of view of mean-field theory,
the small renormalized band-width $\sim t$ yields huge rescaled couplings, turning the fRG-flow unreliable. In the doping regime $\delta < 0.1$,
a pure particle-particle resummation detects no sign of a first order transition between two different triplet $p$-wave phases. While upon the inclusion of partice-hole fluctuations
the superconductivity seems to disappear, the divergent vertex functions do not yield a clear picture as to what kind of instability is actually realized.
\begin{figure}[t!]
\centering
 \includegraphics[height=.8\columnwidth]{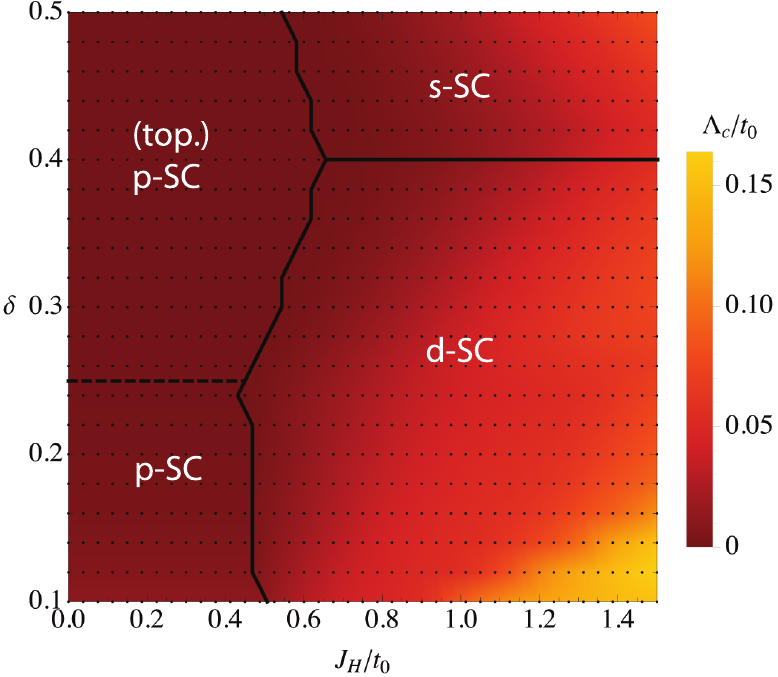}
\caption{The phase diagram as obtained from the numerical solution of a $(N=24)$-patching scheme with
only particle-particle bubbles with $\JK/t_{0}=-1$. Black dots mark the parameters for which fRG-flows were evaluated. The horizontal axis gives the strength
of the antiferromagnetic Heisenberg coupling $\JH > 0$ in units of the bare hopping amplitude $t_{0}$, while the doping level $\delta$ is given on the vertical axis. The colorcode describes the magnitude of the critical scale $\Lambda_c$ in units of the bare hopping $t_{0}$ across the phase diagram. The dashed black line
marks the transition to a topological odd-parity pairing state across the van Hove singularity. In this approximation we obtain
several superconducting instabilities, where labels refer to intraband pairing symmetries. See the main text for a detailed description.}
\label{fig:pdmf}
\end{figure}
As a consistency check on our flow equations, we first completely neglect crossed and direct particle-hole bubbles in the flow. From a diagrammatic perspective, the numerical solution of the flow equations is expected to reproduce the results of a mean-field analysis of the slave-boson Hamiltonian with only
superconducting order parameters.
Second, we obtain the phase diagram for the doped stripy phase with particle-particle
and particle-hole bubbles included on equal footing in the flow equations.

\subsubsection{Resummation in particle-particle channel}
\label{subsec:pp}

As already mentioned in Sect.~\ref{sec:method}, only particle-hole fluctuations couple singlet and
triplet vertex functions among each other. Neglecting particle-hole bubbles amounts to taking into
account only the first diagram in Fig.~\ref{fig:diagrams} in the flow equation. We find, that the leading instability
is essentially determined by the structure of the initial condition for the flow equation. For 
a dominating singlet vertex, the flow leads to an increase in the singlet channel, where also the
dominating pairing symmetry of the initial condition is enhanced. The amplitude of the subleading 
pairing solutions is not substantially increased by the fRG-evolution.
\begin{figure}[t!]
\centering
\includegraphics[height=.325\columnwidth]{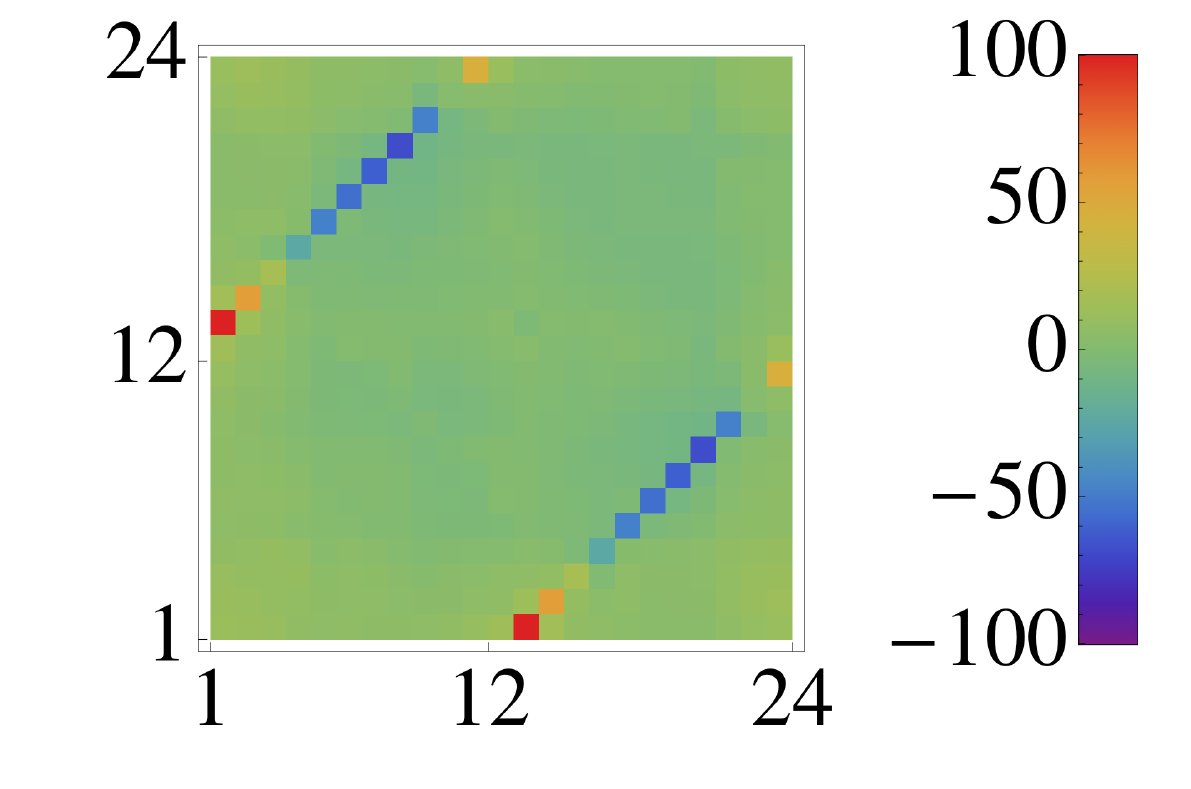}
\includegraphics[height=.325\columnwidth]{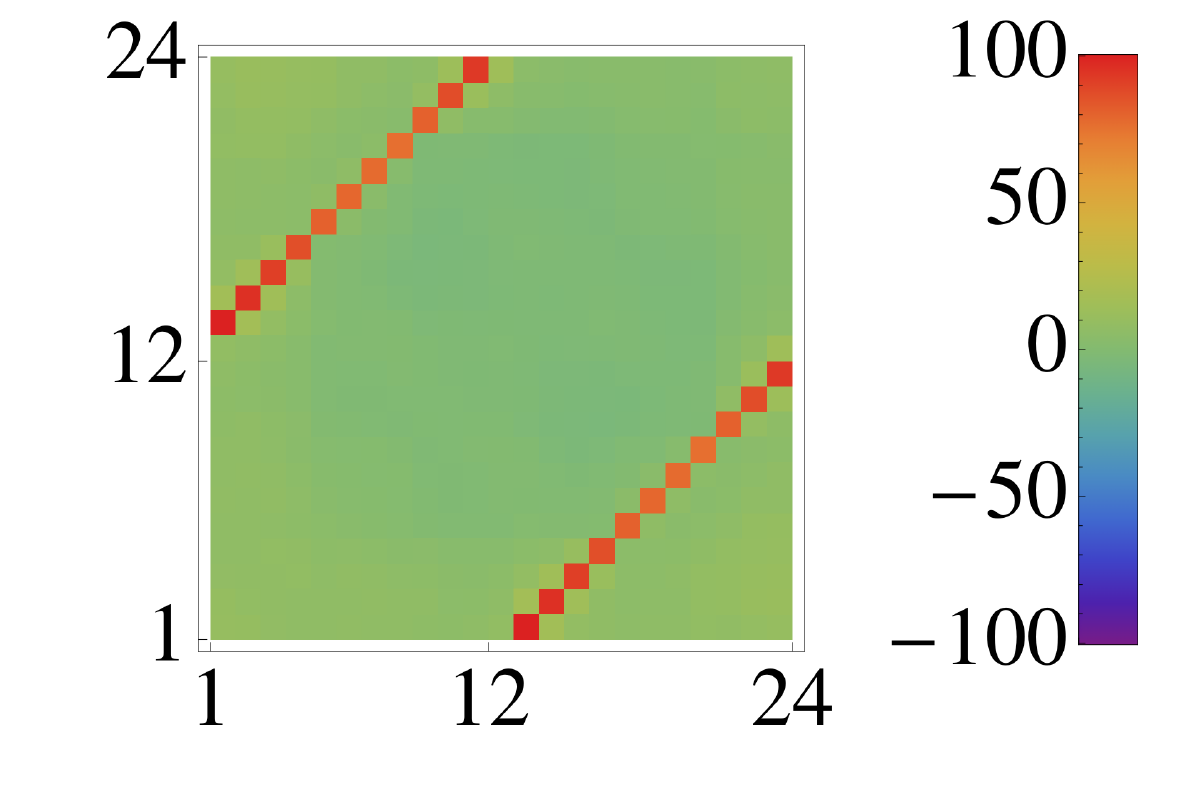}\\
\hspace{-0.5em}\includegraphics[height=.325\columnwidth]{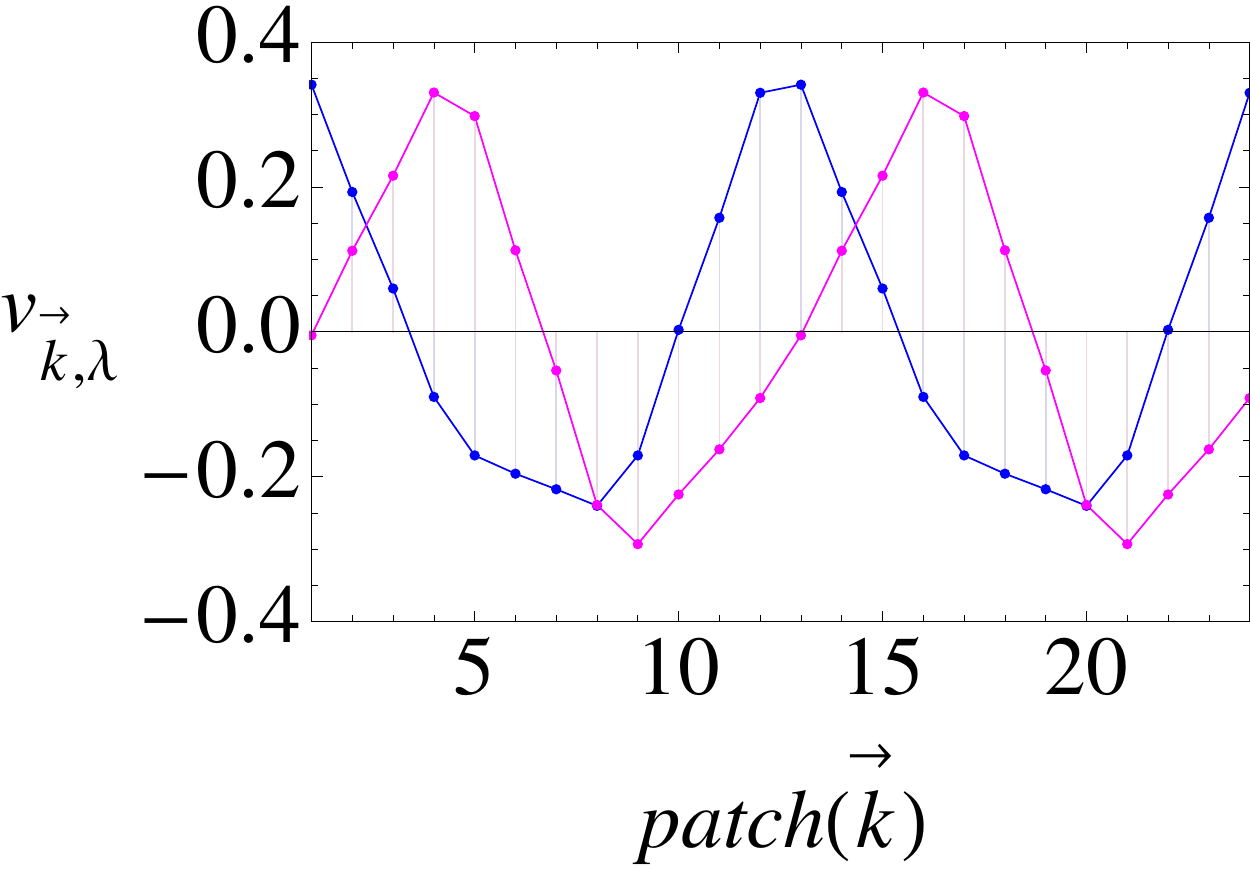}\hspace{0.5em}
\includegraphics[height=.325\columnwidth]{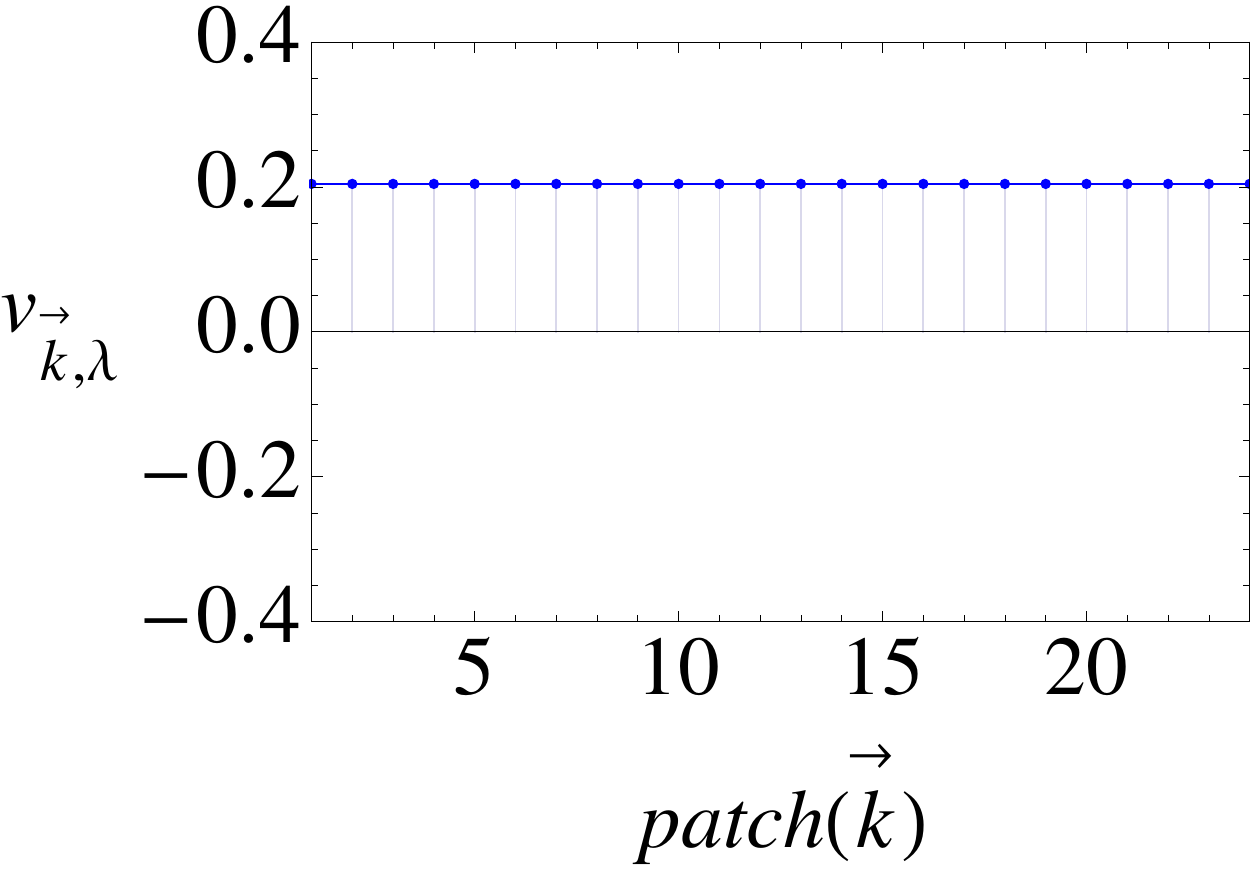}
\caption{The upper left panel shows the
intraband component of the divergent singlet vertex function for $\delta=0.2$ and $\JH/t_{0}=0.9$, where the patch numbers corresponding to patch momentum $\vec{k}_{1}$ are given on the ordinate and $\vec{k}_{2}$ on the abscissa. The remaining free momentum $\vec{k}_{3}$ is fixed to the first patch, cf. Fig.~\ref{fig:patching}. The divergent momentum structure 
corresponds to a $d$-wave instability. The upper right panel shows the intraband component of the divergent singlet vertex function for $\delta=0.44$ and $\JH/t_{0}=1.31$, with a momentum structure yielding an $s$-wave instability. In the lower left panel we display the normalized amplitude of the two degenerate eigenvectors with $d$-wave symmetry of the singlet pair-scattering amplitude along the Fermi surface. The patch number is enumerated on the abscissa. The lower right panel shows the corresponding eigenvector with $s$-wave symmetry.}
\label{fig:ff1}
\end{figure}
For a superconducting instability, the associated spin-structure of the leading pairing correlations
can be immediately inferred from whether the singlet vertex $V^{(\mathrm{s})}$ or the triplet vertices
 $V_{l}^{(\mathrm{t})}$ diverge. Due to the residual rotational symmetry of the Kitaev-Heisenberg model (see Sect.~\ref{sec:method}
 and Appendix~\ref{app:ward}), the three triplet vertex functions are bound to diverge simultaneously. This also leads
 to a degeneracy for the pairing solutions for the $\vec{d}$-vector describing the structure of 
 the corresponding Cooper pair. In order to obtain such information from the vertex functions, we 
 extract the pair scattering amplitudes in singlet and triplet channels as
\be
 V^{(\mathrm{s})}(\vec{k},\vec{k^{\prime}}) & \equiv & V^{(\mathrm{s})}(\vec{k},-\vec{k},\vec{k}^{\prime},-\vec{k}^{\prime}), \\
 V_{l}^{(\mathrm{t})}(\vec{k},\vec{k^{\prime}}) & \equiv & V_{l}^{(\mathrm{t})}(\vec{k},-\vec{k},\vec{k}^{\prime},-\vec{k}^{\prime}), 
\ee
where for brevity we suppressed sublattice or band labels. Since our discretization of the Brillouin zone
employs a total of $N$ representative patch momenta, the pair-scattering amplitudes can be treated as $N\times N$ matrices.
Diagonalization of $V^{(\mathrm{s})}(\vec{k},\vec{k^{\prime}})$ or $V_{l}^{(\mathrm{t})}(\vec{k},\vec{k^{\prime}})$
and determination of the eigenvectors $v_{\vec{k},\lambda}$ corresponding to the eigenvalues $\lambda$ unveils the leading and subleading pairing instabilities~\cite{honerkamp2008,zhai2009,wang2012,kiesel2012}.
The 2-particle contributions to the effective action $\Gamma^{\Lambda_{c}}$ at the critical scale $\Lambda_{c}$ determined by the leading instability (i.e. the eigenvalue $\lambda$ with largest absolute value) in e.g. the singlet channel becomes
\be
H_{\mathrm{SC}}^{(\mathrm{s})} \propto - \lambda \sum_{\vec{k},\vec{k}^{\prime}} \left[ v_{\vec{k},\lambda}^{\ast} f_{\vec{k}}^{\dagger}\Gamma_{0}^{\dagger}f_{-\vec{k}}^{\dagger}\right] \left[v_{\vec{k}^{\prime},\lambda} f_{\vec{k}^{\prime}}\Gamma_{0} f_{-\vec{k}^{\prime}}\right], 
\ee
where sublattice/band and spin labels were again suppressed for clarity.
The Hamiltonian $H_{\mathrm{SC}}^{(\mathrm{s})}$ can be decoupled by a Hubbard-Stratonovich transformation with 
a singlet order-parameter field $\psi_{\vec{k}}\sim v_{\vec{k},\lambda} \langle f_{\vec{k}}\Gamma_{0} f_{-\vec{k}} \rangle$. Analogous definitions hold for the triplet case, where the decoupling is performed with the vector order-parameter $\vec{d}_{\vec{k}}$, cf. Appendix~\ref{app:matrix}. The order parameter symmetry, i.e., the momentum-space Cooper pair structure, is obtained by projecting the eigenvectors $v_{\vec{k},\lambda}$ onto suitably defined form factors. These can be obtained from the irreducible representations of the point-group of the hexagonal lattice in real space~\cite{sigrist1991,wang2012}. From a neighbor-resolved Fourier transform one can obtain momentum-space
representations of form factors with well-defined parity. See Appendix~\ref{app:formfactors} for details.  
\begin{figure}[t!]
\centering
\includegraphics[height=.325\columnwidth]{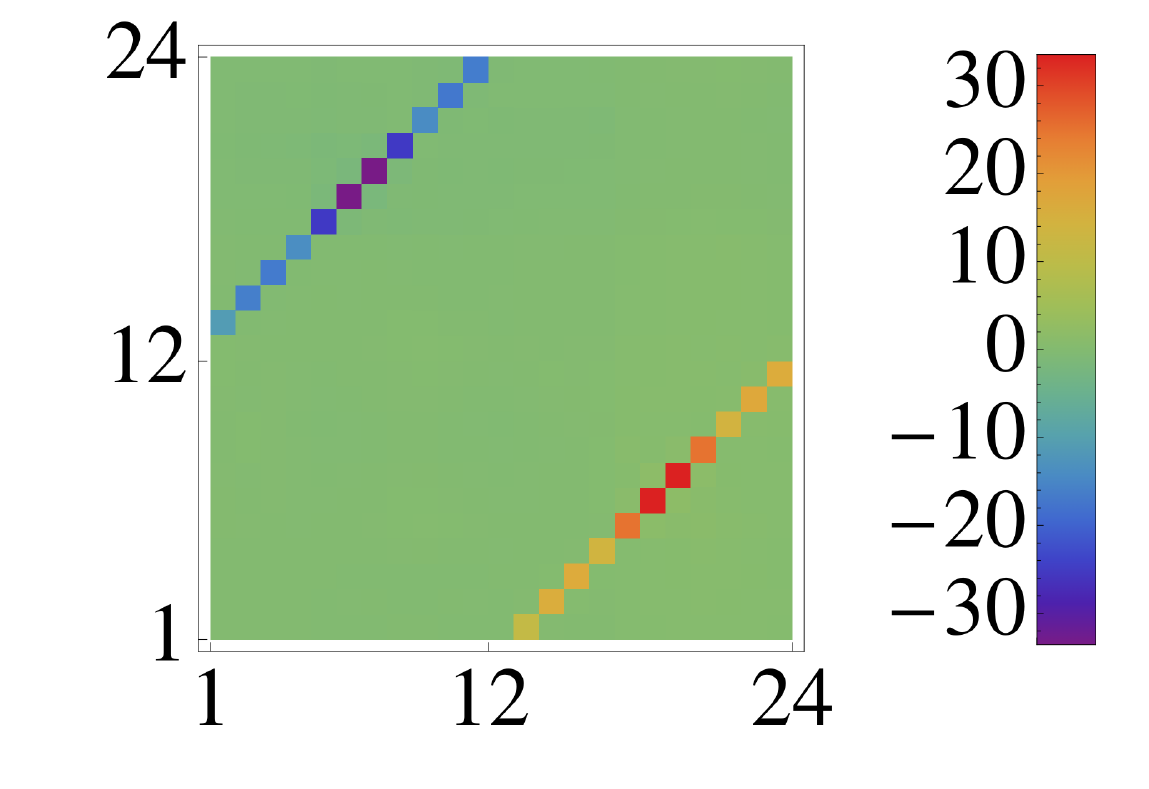}\hspace{0.5em}
\includegraphics[height=.325\columnwidth]{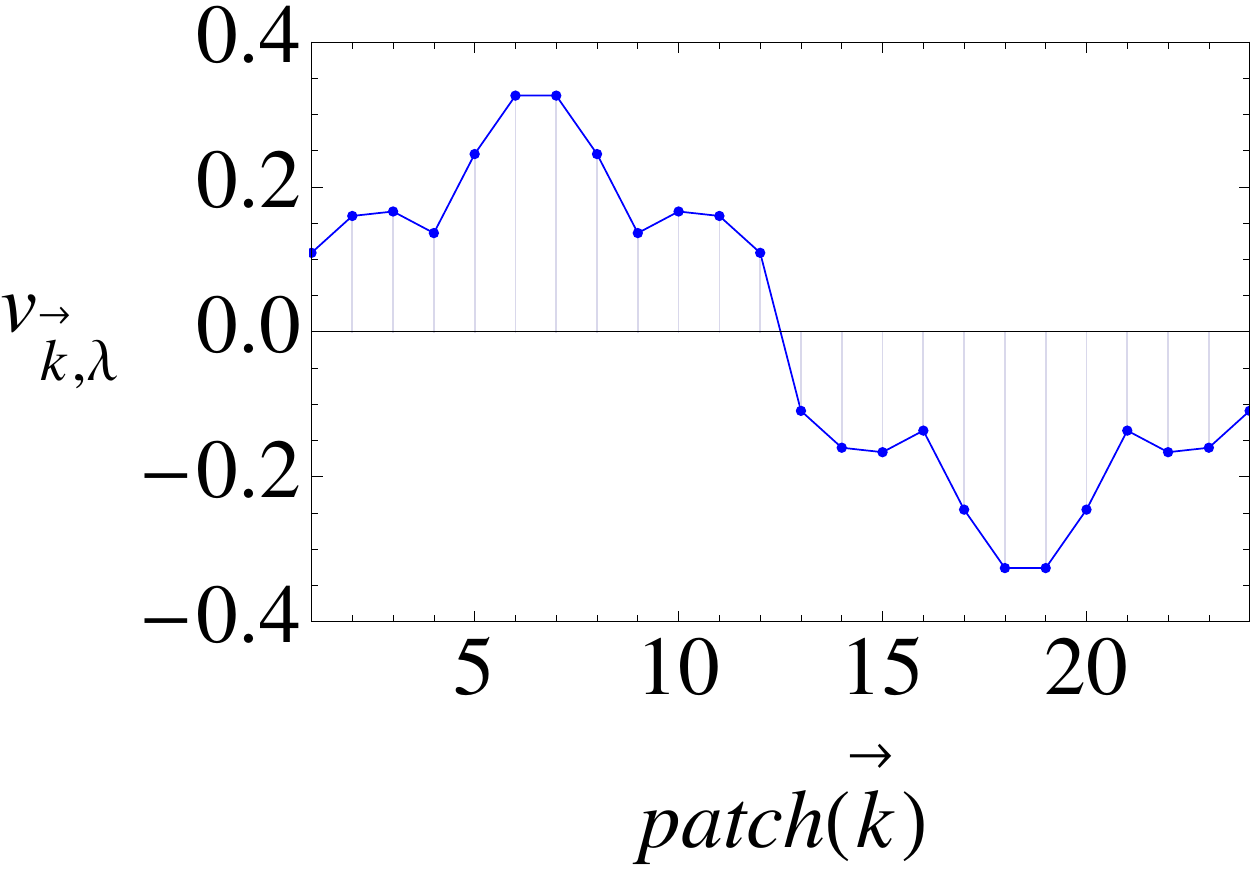}
\caption{The left panel shows the
intraband component of the divergent triplet vertex function for $\delta=0.3$ and $\JH/t_{0}=0.375$, where the patch numbers corresponding to patch momentum $\vec{k}_{1}$ are given on the ordinate and $\vec{k}_{2}$ on the abscissa. The remaining free momentum $\vec{k}_{3}$ is fixed to the first patch, cf. Fig.~\ref{fig:patching}. The divergent momentum structure 
corresponds to a $p$-wave instability. In the right panel we display the normalized amplitude of the $p$-wave eigenvectors ($p_{x}(\vec{k}) - \sqrt{3}\, p_{y}(\vec{k})$) of the triplet-$x$ pair-scattering amplitude along the Fermi surface. The patch number is enumerated on the abscissa.}
\label{fig:ff2}
\end{figure}
The phase diagram as obtained from an analysis of the leading instabilities is shown in Fig.~\ref{fig:pdmf}. The overall structure of the phase
boundary between intraband $p$-SC and singlet pairing phases agrees nicely with the findings in Ref.~\onlinecite{hyart2012}. While Ref.~\onlinecite{hyart2012} reports
a singlet $s$-wave regime that extends to low doping, the singlet $s$-wave instability appears only for $\delta > 0.4$ in our fRG calculations. This finding agrees with 
the phase diagram obtained in Ref.~\onlinecite{doniach2009}, where the triplet channel was neglected. We interpret this in favor of our present results.
A projection of the initial condition onto pair-scattering amplitudes in the singlet regime reveals a $d$-wave dominance for $\delta < 0.4$ and a subleading $s$-wave, while for $\delta > 0.4$ the situation is reversed and the $s$-wave form factor is dominating. The form factor with the largest weight is subsequently enhanced by the flow.
We also performed flows at finite temperature, which, however, showed that temperature does not exert an influence on the respective
$d$- or $s$-wave dominance in our flows. Rather, above the critical temperature the singular flow is smoothed, which signals the stability of a Fermi liquid ground-state.

Due to lattice symmetry, the intraband $d$-wave solution is doubly degenerate, i.e., the largest eigenvalue of the singlet pair-scattering comes with a two dimensional
eigenspace. Projecting onto the form factors given in Appendix~\ref{app:formfactors}, we find that each eigenvector has overlap with the two even-parity nearest-neighbor $d$-wave form factors $d_{xy}(\vec{k})$ and $d_{x^2-y^2}(\vec{k})$ as defined by ~\Eqref{eq:ffdxy} and~\Eqref{eq:ffdx2y2}. See Fig.~\ref{fig:ff1} for the momentum-space structure of the divergent singlet vertex function and the corresponding eigenvectors of the pair-scattering amplitudes. We note that due to the lack of particle-hole fluctuations in this reduced flow, no longer-ranged
intrasublattice pairing correlations develop.

The interband pair-scattering shows odd-parity $p$-wave correlations in the singlet regime, also with degenerate $p$-wave form factors 
$p_{x}(\vec{k})$ and $p_{y}(\vec{k})$ on nearest-neighbor bonds as defined by~\Eqref{eq:ffpx} and~\Eqref{eq:ffpy}. The interband correlations are in fact substantial close to $\Lambda_{c}$ and comparable in magnitude to intraband correlations. On the level of our fRG-flows this remains unchanged when turning on finite temperature.

At large doping $\delta > 0.4$, the leading intraband correlations change from $d$- to $s$-wave with even-parity nearest-neighbor form
factor, see ~\Eqref{eq:ffsp} corresponding to an extended $s$-wave pairing instability.

The degeneracies in the case of singlet instabilities are related to lattice symmetries~\cite{sigrist1991,doniach2009,kiesel2012}. Which linear combination
is finally realized in the superconducting state needs to be inferred from, e.g., a comparison of ground-state
energies~\cite{wang2009} as obtained from mean-field theory. Only when self-energy feedback or counter-terms~\cite{gersch2008,ossadnik2011,eberlein2013} are included in the fRG-flow,
symmetry breaking can be accounted for.

The triplet instability for $\JH \lesssim |\JK|/2$ is manifested by diverging triplet vertex functions.
As noted in Sect.~\ref{sec:method}, the discrete symmetry of the Kitaev interaction relates the
triplet functions among each other. Since the flow stays in the
symmetric regime, the triplet vertex functions diverge simultaneously. Moreover, symmetry ensures that the eigenvalues obtained from diagonalizing the triplet pair-scattering are also degenerate. From the current fRG scheme we can thus infer three degenerate $\vec{d}$-vectors, each one corresponding
to one of the degenerate triplet channels. As in the singlet case, the true ground state will pick a particular linear
combination, which is, however, inaccessible in the employed scheme. Since the Ward identity (see Sect.~\ref{app:ward})
derived from the discrete Kitaev symmetry allows reconstruction of two vertex functions from a given one,
we only keep the triplet vertex $V_{x}^{(\mathrm{t})}$ in the flow. The vertices $V_{y}^{(\mathrm{t})}$ and $V_{z}^{(\mathrm{t})}$ are
obtained from the final result for $V_{x}^{(\mathrm{t})}$. 

The triplet instability is dominated by intraband pairing, which in fact corresponds to the $p$-SC solution found in Ref.~\onlinecite{hyart2012}. From an analysis
of the pair-scattering amplitude we obtain with a high numerical accuracy the degenerate solutions (see also Fig.~\ref{fig:ff2})
\be\label{eq:dvec}
\vec{d}_{\vec{k},1} & = & \left[p_{x}(\vec{k}) - \sqrt{3}\, p_{y}(\vec{k})\right]\left(1, 0, 0 \right)^{T}, \nn \\
\vec{d}_{\vec{k},2} & = & \left[p_{x}(\vec{k}) + \sqrt{3}\, p_{y}(\vec{k})\right]\left(0,1 , 0 \right)^{T},  \\
\vec{d}_{\vec{k},3} & = & \left[2\, p_{x}(\vec{k})\right]\left(0, 0,  1 \right)^{T}. \nn
\ee
Expanding these form factors to leading order in $\vec{k}$ about the $\Gamma$-point in the BZ, we recover the results obtained
in Ref.~\cite{hyart2012}. There it was also shown, that such a $\vec{d}$-vector configuration realizes pairing between fermions with spin projections
aligned (with $\sim (k_{x} - \mathrm{i} k_{y})$-pairing) or anti-aligned (with $\sim (k_{x} + \mathrm{i} k_{y})$-pairing) along the $(1,1,1)^{T}$-axis in spin space.
\begin{figure}[t]
\centering
\includegraphics[height=.6\columnwidth]{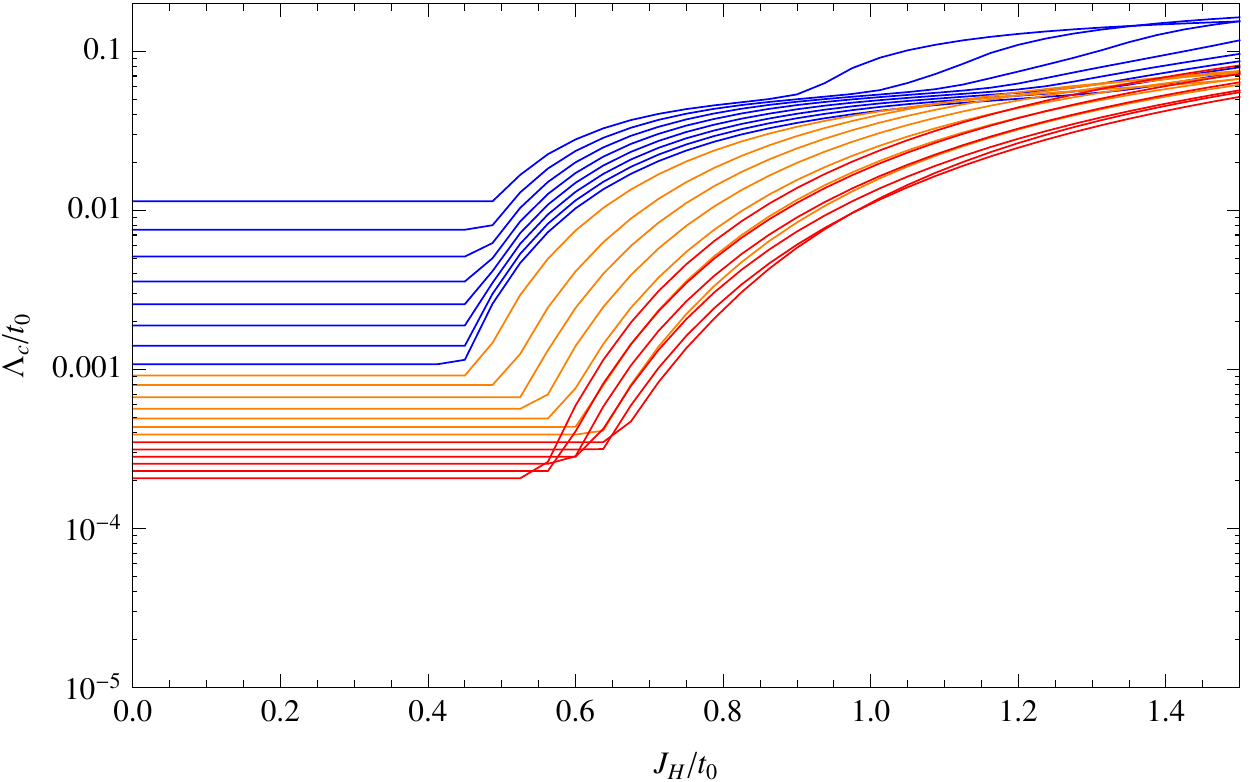}
\caption{The critical scale $\Lambda_{0}$ in units of the bare hopping $t_{0}$ as a function
of the antiferromagnetic Heisenberg coupling $\JH$, also in units of $t_{0}$ and for fixed $\JK/t_{0}=-1$. The plateaus of constant critical scales below critical Heisenberg coupling can be clearly identified. See main text for an explanation. $\Lambda/t_{0}$ is here given on a logarithmic scale. Starting with doping $\delta=0.1$,
the doping level decreases from the top to the bottom curve in steps of $\Delta\delta = 0.02$. Blue corresponds to
doping $\delta=0.1,\dots,0.24$, orange to $\delta=0.26,\dots,0.38$, and red to $\delta=0.4,\dots,0.5$.}
\label{fig:critscalePP}
\end{figure}
Further following mean-field arguments~\cite{hyart2012} and employing knowledge about the mechanism
for the creation of topological pairing, namely an odd number of time-reversal invariant points enclosed by the Fermi surface~\cite{qi2009,sato2009a,sato2009b,qi2010,sato2010},
for $\delta > 0.25$ (indicated by the black dashed line
in Fig.~\ref{fig:pdmf}) the triplet $p$-wave superconductor turns into a topological superconductor.   

In total, we obtain good agreement with results from mean-field theory~\cite{hyart2012,doniach2009} from the reduced pure particle-particle flows. 
Also when turning to the stability of the superconducting phases with respect to thermal excitations, we obtain estimates for critical temperatures from
the critical scale $\Lambda_{c}$ (see Fig.~\ref{fig:pdmf}) that are within the same orders of magnitude as reported in Ref.~\onlinecite{hyart2012}. Within the $p$-wave phase,
the critical temperature decreases from $k_{\mathrm{B}} T_{c} \sim 10^{-2}\, t_{0}$ at $\delta \gtrsim 0.1$ by two orders of magnitude to  $k_{\mathrm{B}} T_{c} \sim \, 10^{-4} t_{0}$ at $\delta \gtrsim 0.3$. For fixed doping level, the critical scale/temperature remains constant within the $p$-wave phase. This, however, can
be easily understood from the fixed Kitaev coupling $\JK / t_{0} = -1$. Since singlet and triplet vertices are decoupled
in the reduced flows, the channel that diverges first wins the race and determines the leading instability. The singlet vertex thus has
no influence on the critical scale within the $p$-wave regime determined by the leading triplet channel, see also Fig.~\ref{fig:critscalePP}. For parameters
in the singlet regime, the critical scale shows a prominent $\JH$-dependence. Increasing $\JH$, the critical scale/temperature
grows rather quickly to larger values $k_{\mathrm{B}} T_{c} \sim 10^{-1}\, t_{0}$ for $\delta \gtrsim 0.1$. A logarithmic plot of the
critical scale as a function of $\JH/t_{0}$ for various dopings is given in Fig.~\ref{fig:critscalePP}, where the plateaus for fixed doping within the
triplet regime can be clearly identified.

\subsubsection{Unbiased resummation of particle-particle and particle-hole bubbles}
\label{subsec:ppph}

Having established our method in the limit of exclusive particle-particle contributions to the flow of the
scale-dependent vertex functions, we now include the particle-hole fluctuations. These
lead to a coupling of singlet and triplet vertex functions. The particle-hole contributions are in fact
considerably more complicated than the particle-particle contributions alone. This originates from
our choice of channel decomposition of the initial condition, cf. \Eqref{eq:initial_singlet} and \Eqref{eq:initial_triplet}.

\begin{figure}[t]
\centering
\includegraphics[height=.6\columnwidth]{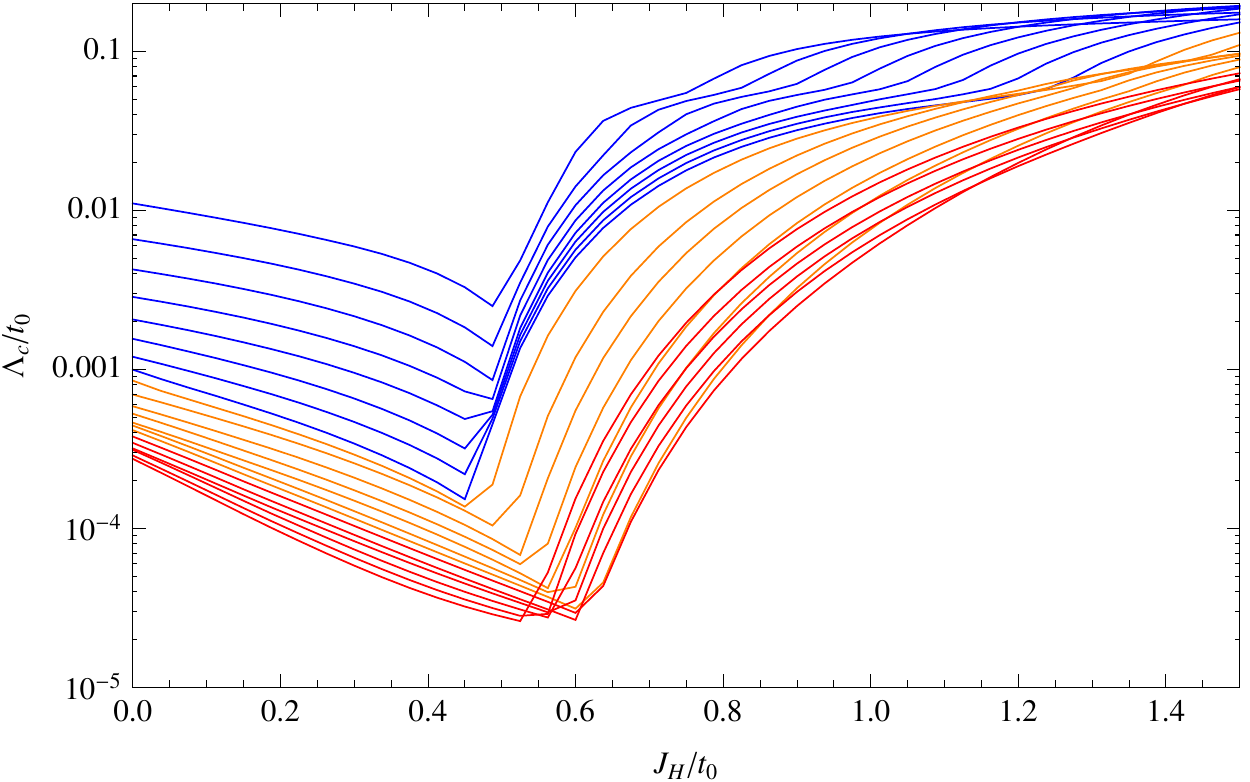}
\caption{The critical scale $\Lambda_{0}$ in units of the bare hopping $t_{0}$ as a function
of the antiferromagnetic Heisenberg coupling $\JH$, also in units of $t_{0}$ and for fixed $\JK/t_{0}=-1$. $\Lambda/t_{0}$ is here given on a logarithmic scale. Starting with doping $\delta=0.1$,
the doping level decreases from the top to the bottom curve in steps of $\Delta\delta = 0.02$. Blue corresponds to
doping $\delta=0.1,\dots,0.24$, orange to $\delta=0.26,\dots,0.38$, and red to $\delta=0.4,\dots,0.5$.}
\label{fig:critscalePPPH}
\end{figure}

\begin{figure}[t!]
\centering
\includegraphics[height=.45\columnwidth]{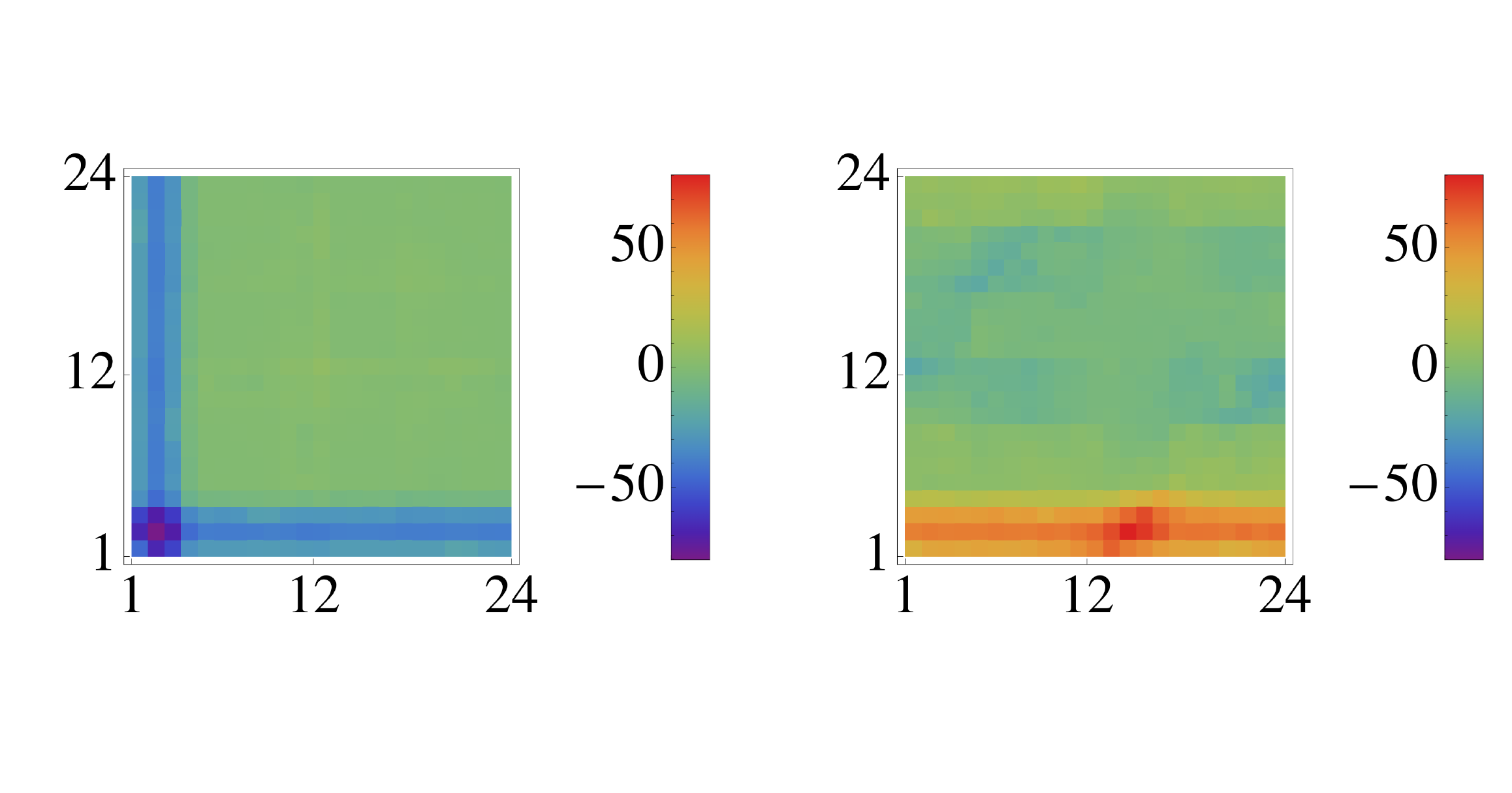}\\
\vspace{-2.5em}
\includegraphics[height=.45\columnwidth]{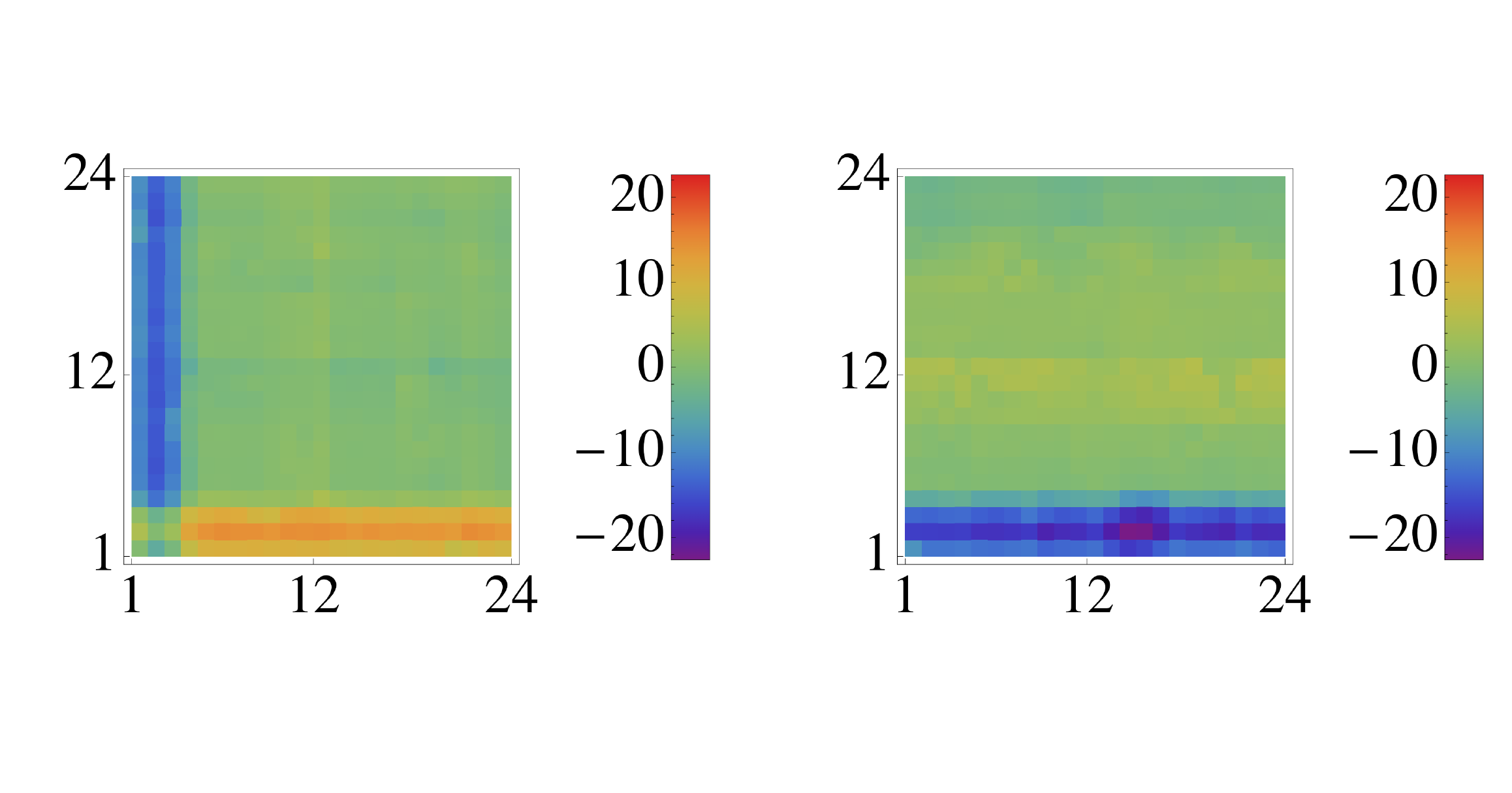}
\caption{The upper panel shows the
intrasublattice (left) and intersublattice (right) component of the divergent singlet vertex function for $\delta=0.12$ and $\JH/t_{0}=0.75$ with $\JK/t_{0}=-1$, where the patch numbers corresponding to patch momentum $\vec{k}_{1}$ are given on the vertical and $\vec{k}_{2}$ on the horizontal axis. The remaining free momentum $\vec{k}_{3}$ is here fixed to the second patch, cf. Fig.~\ref{fig:patching}. With the
same conventions, the lower panel shows the intrasublattice (left) and intersublattice (right) component of the divergent triplet-$x$ vertex function.
The divergent momentum structure corresponds to an antiferromagnetic  N$\mathrm{\acute e}$el instability. Both amplitude ratio of singlet to triplet vertex and the sign structure
conspire to re-combine singlet and triplet pairing interactions into a spin-spin interaction, cf. \Eqref{eq:af}.}
\label{fig:af}
\end{figure}

The resulting phase diagram is presented in Fig.~\ref{fig:pd1}.
The $p$-wave instability seems to be largely unaffected by the inclusion of particle-hole
fluctuations. As in the pure particle-particle case, symmetry guarantees degeneracy of the triplet
vertices. We even find that the $\vec{d}$-vector describing the triplet instability is
still rather well described by the form given in \Eqref{eq:dvec}. Particle-hole fluctuations, however, generate longer-ranged pairing correlations. 
In the triplet channel, these are subleading contributions compared to the leading nearest-neighbor
$p$-wave. For intermediate $\JH$ and $\delta$
the leading instability still occurs in the singlet channel with $d$-wave symmetry, and 
for larger doping $\delta \gtrsim 0.4$ the order-parameter symmetry switches to $s$-wave. 
The phase boundaries between the adjacent superconducting instabilities appear to be rather robust with 
respect to particle-hole fluctuations as compared to the previous pure particle-particle resummation.
Critical scales and temperatures are also only mildly affected. We plot the critical scale logarithmically in Fig.~\ref{fig:critscalePPPH} for various dopings
as a function of $\JH/t_{0}$. We do no longer find a constant $\Lambda_{c}$ for fixed
doping and $\JK/t_{0} = -1$ as $\JH$ is varied within the $p$-wave triplet regime. As expected, the particle-hole fluctuations
suppress the critical scale in the superconducting regimes. Quantitatively, the changes as compared to the pure particle-particle case reach up to an order of magnitude, cf. Fig.~\ref{fig:critscalePP}.
Finally, in the large-$\JH$ regime, the character of the instability changes from superconducting 
to magnetic. This can be read off from the singlet and triplet vertex functions as shown in Fig.~\ref{fig:af}.
In the case of spin or charge density wave (SDW, CDW) instabilities, the singlet and triplet vertex
functions encode the corresponding divergent momentum structure in a rather complicated way due to the
channel decomposition that is adapted to pairing instabilities. Nevertheless, 
the form of the full vertex function $V^{\Lambda}$ can in these cases be obtained essentially by
matrix algebra\footnote{The actual computations are most efficiently performed with the help of so-called Fierz identities,
which can be understood as `re-arrangement' formulas for the index structure of a quartic interaction term.} and the momentum structures corresponding to SDW and CDW instabilities can be obtained.
Using Fierz identities and re-combining singlet and triplet pairing channels, we recover a Hamiltonian
\be\label{eq:af}
H_{\mathrm{AF}} \propto - V \sum_{o,o^{\prime}\in A,B} \epsilon_{o,o^{\prime}}\, \vec{S}_{\vec{q}=0}^{o}\cdot\vec{S}_{\vec{q}=0}^{o^{\prime}},\quad V > 0,
\ee
that describes the low-energy degrees of freedom close to the critical scale $\Lambda_{c}$. Here,
$S_{\vec{q}}^{o} = \sum_{\vec{k}} f_{o,\vec{k},\sigma}^{\dagger} [\vec{\sigma}]_{\sigma\sigma^{\prime}} f_{o,\vec{k}-\vec{q},\sigma^{\prime}}$ is the $\vec{q}$-component of the fermion spin operator in sublattice $o\in A,B$. The pre-factor $\epsilon_{o,o^{\prime}}$ equals $-1$ for $o\neq o^{\prime}$
corresponding to antiferromagnetic correlations between the two sublattices. For $o = o^{\prime}$, $\epsilon_{o,o^{\prime}} = +1$,
which describes ferromagnetic correlations in a given sublattice. The long-range order corresponding to
such a Hamiltonian with infinitely ranged interaction is nothing but a two sublattice  N$\mathrm{\acute e}$el state, i.e., a commensurate antiferromagnet, where
the staggered magnetization is arranged over the two sublattices.

The momentum structure displayed in Fig.~\ref{fig:af} is rather broad and smeared out. We confirmed that these
features are also obtained from a Hubbard model in the large-$U$ regime on the honeycomb lattice, where the  N$\mathrm{\acute e}$el antiferromagnet
was established as the magnetically ordered ground-state.

\subsection{Doping QSL and zigzag phase - \\ AF Kitaev and FM Heisenberg exchange}
\label{sec:zigzag}

%
\begin{figure}[t]
\centering
\includegraphics[height=.6\columnwidth]{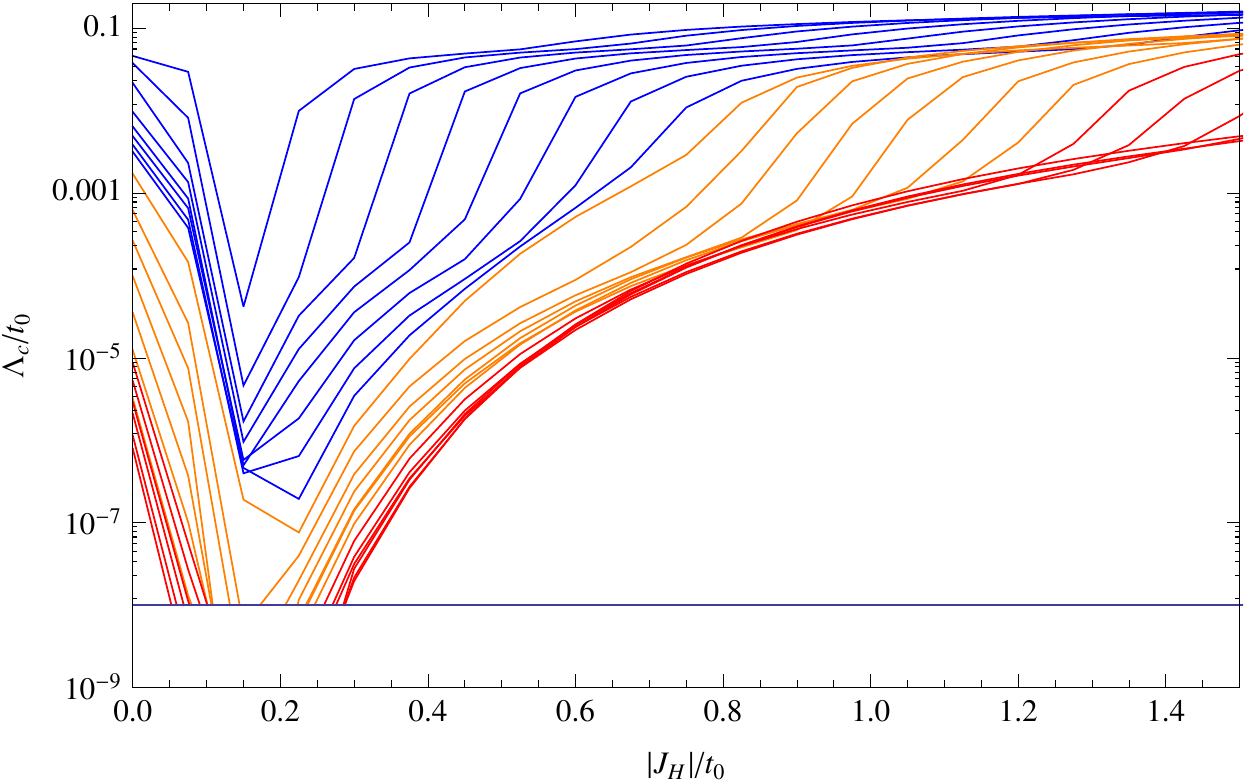}
\caption{The critical scale $\Lambda_{0}$ in units of the bare hopping $t_{0}$ as a function
of the ferromagnetic Heisenberg coupling strength $|\JH|$, also in units of $t_{0}$ and for fixed $\JK/t_{0}=1$. $\Lambda/t_{0}$ is here given on a logarithmic scale. Starting with doping $\delta=0.1$,
the doping level decreases from the top to the bottom curve in steps of $\Delta\delta = 0.02$. Blue corresponds to
doping $\delta=0.1,\dots,0.24$, orange to $\delta=0.26,\dots,0.38$, and red to $\delta=0.4,\dots,0.5$. For scales $\Lambda/t_{0}<10^{-8}$, marked
by the solid blue line, the fRG-flow could not be evaluated properly to even lower scales for the given choice of parameters used for numerical integration of the flow equation.}
\label{fig:critscalePPPH2}
\end{figure}
To analyze the effect of doping charge carriers into the QSL and the magnetically ordered zigzag phase,
we select the parameter range of the ferromagnetic ($\JH < 0$) Heisenberg coupling as $|\JH|/t_{0} \in [0,1.5]$, while
again keeping the now antiferromagnetic Kitaev coupling fixed $\JK/t_{0} = 1$. This parameter range
covers both QSL and zigzag phase at $\delta = 0$. The phase diagram extracted from fRG-flows with
both partice-particle and particle-hole bubbles is shown in Fig.~\ref{fig:pd2}. We again find singlet and triplet pairing instabilities,
where as in the case of doping the QSL/stripy phase, the singlet instability comes with different pairing symmetries 
depending on the doping level. 
Here, we find two different density-wave regimes (SDW, CDW). A further type on density wave state, a bond-order wave,
occurs at the special filling $\delta = 1/4$, i.e., van Hove filling. This type of instability will be discussed below in Sect.~\ref{sec:bondorder} after we presented our findings for superconducting and SDW/CDW regimes.

\begin{figure}[t!]
\centering
\includegraphics[height=.325\columnwidth]{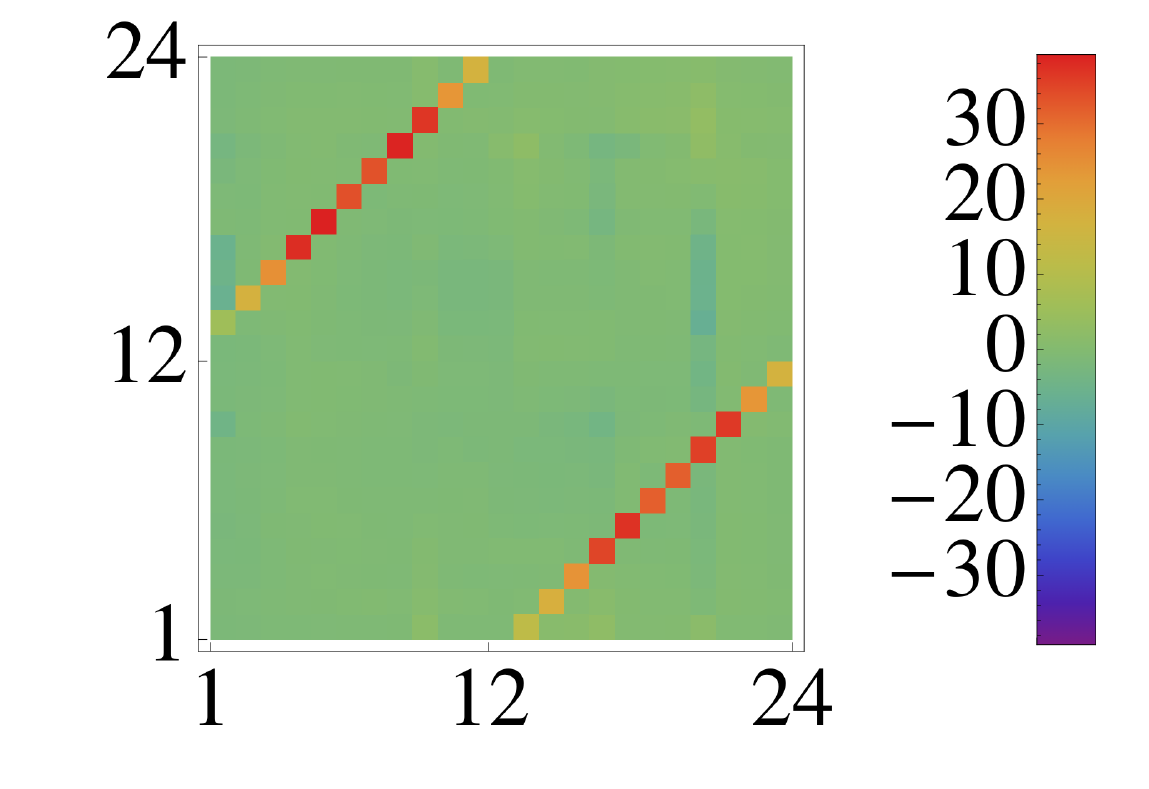}
\includegraphics[height=.325\columnwidth]{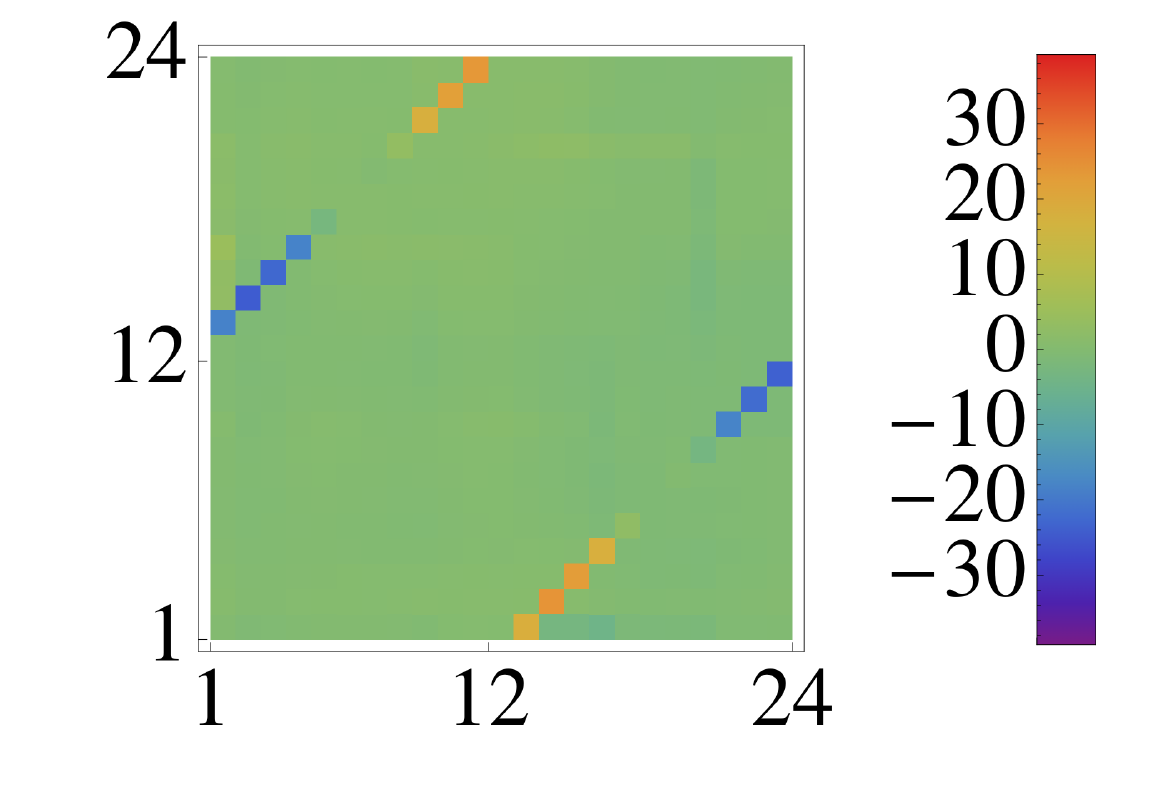}\\
\hspace{-0.5em}\includegraphics[height=.325\columnwidth]{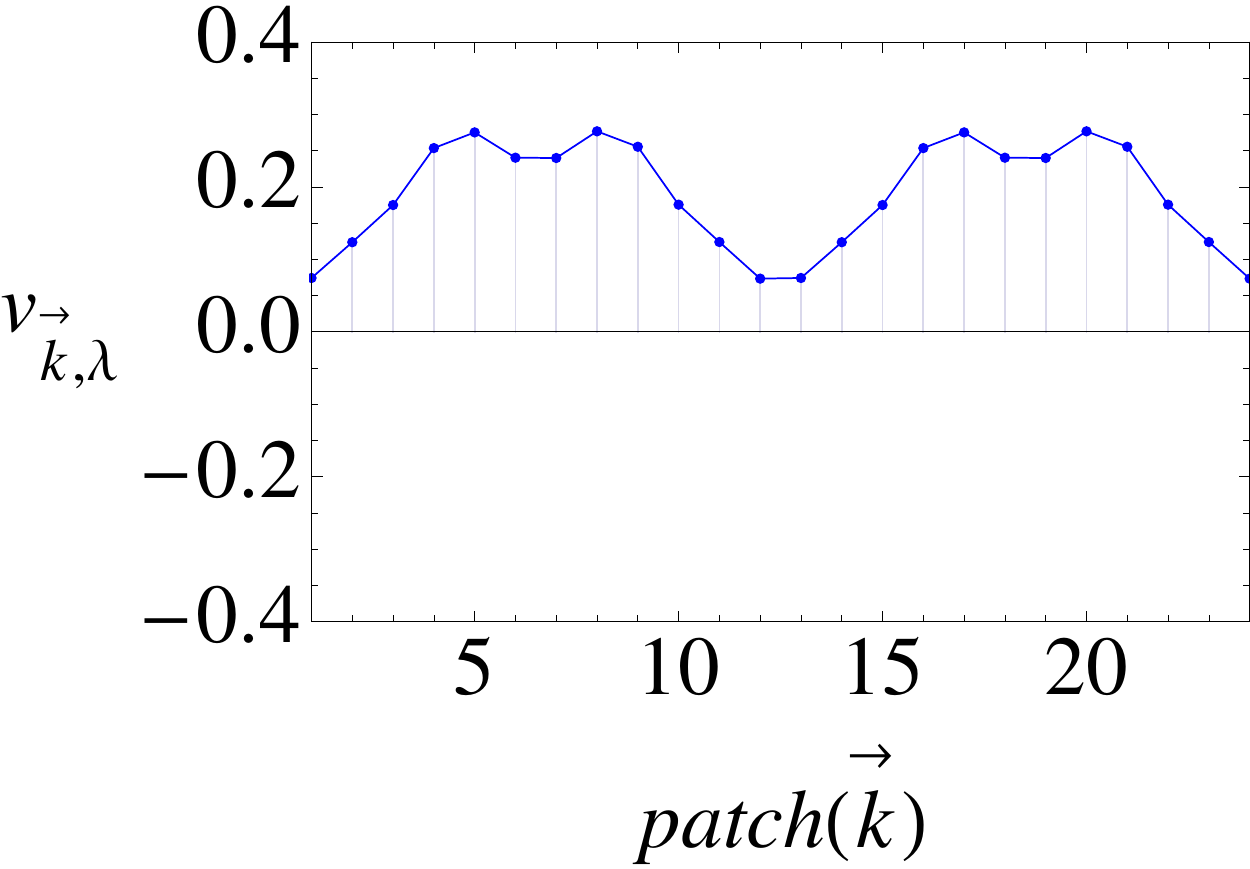}\hspace{0.5em}
\includegraphics[height=.325\columnwidth]{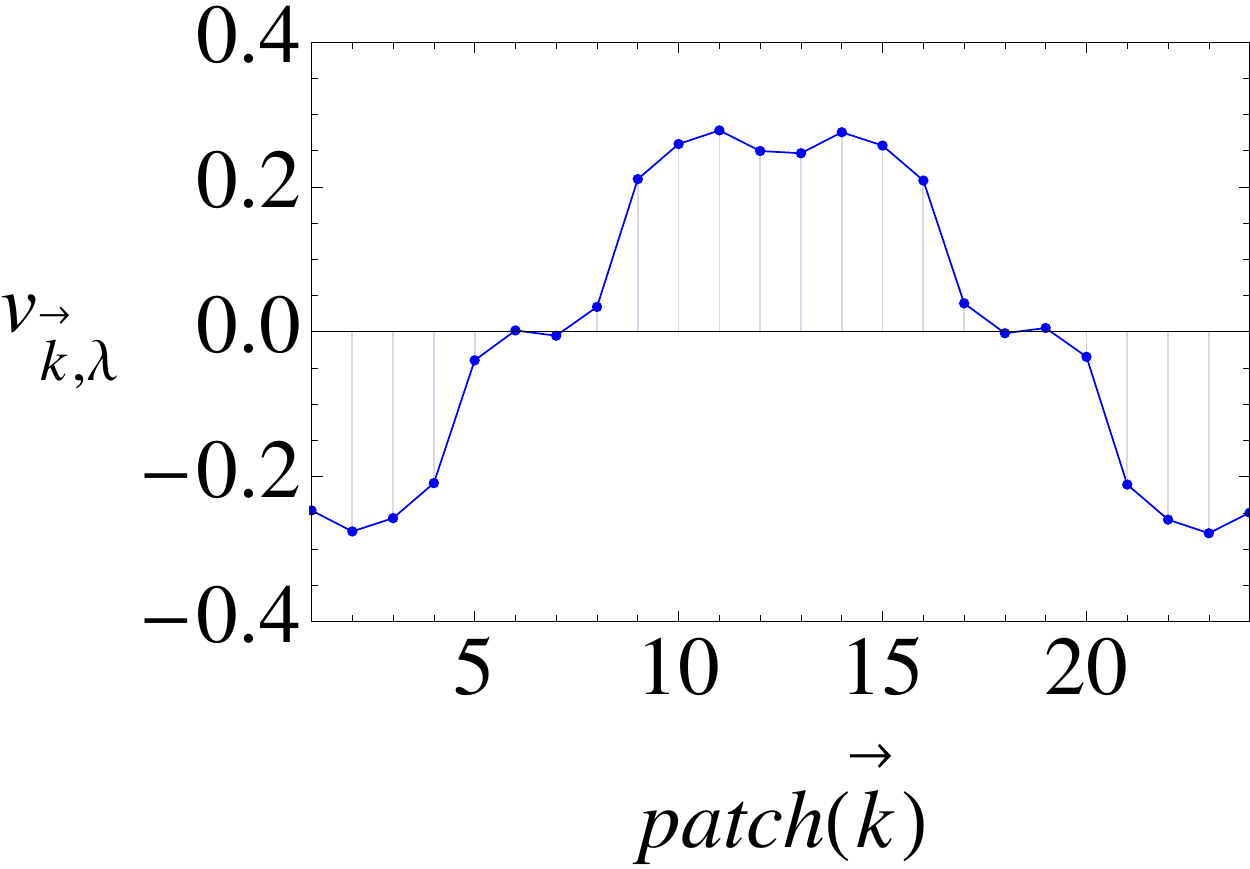}
\caption{The upper left panel shows the divergent part of the interband triplet-$x$ vertex function corresponding
to $d$-wave pairing symmetry for $\delta=0.34$ and $|\JH|/t_{0}=0.525$ with $\JK/t_{0}=1$, where the patch numbers corresponding to patch momentum $\vec{k}_{1}$ are given on the ordinate and $\vec{k}_{2}$ on the abscissa. The remaining free momentum $\vec{k}_{3}$ is fixed to the third patch for better visibility, cf. Fig.~\ref{fig:patching}. The upper right
panel shows the effect of the proximity to the CDW instability in the intraband triplet-$x$ vertex function. The correlations are of
nearest-neighbor $p$-wave type. The lower left panel shows the eigenvector with mixed $s$- and $d$-wave symmetry of the interband triplet-$x$ pair-scattering amplitude along the Fermi surface.  The patch number is enumerated on the abscissa. The lower right panel shows the
corresponding intraband eigenvector with $p$-wave symmetry.}
\label{fig:ff3}
\end{figure}

At small doping $\delta \simeq 0.1$ and for $|\JH|/t_{0} \simeq 0.1 $, the leading instability is of SDW type.
In fact, the divergent momentum structure is the same is in the doped stripy phase, cf. Fig.~\ref{fig:af}. We thus
find a  N$\mathrm{\acute e}$el antiferromagnet in this parameter range driven here by the antiferromagnetic Kitaev exchange. As both $\delta$ and $|\JH|$ are increased, the magnetic
order rather quickly makes way for pairing instabilities and an adjacent CDW instability. The vertex structure corresponding to a CDW instability is shown in Fig.~\ref{fig:ff3}.

The charge density wave is produced by particle-hole fluctuations, in a similar fashion as the antiferromagnetic instability.
A pure particle-particle flow would of course yield a superconducting instability, while a pure particle-hole resummation already
gives us the CDW instability. The origin of the strong CDW ordering tendencies traces back to the repulsive (for $\JH > 0$) nearest-neighbor interaction
between the sublattice charge densities, see \Eqref{eq:hamiltoniancontdH}.  While we thus find the resulting phase diagram as a `competition' of tendencies, the existence of
either one of the instabilities does not hinge on an interplay between different, competing channels. Such behavior would 
manifest itself in the complete absence of a particular instability once either particle-particle or particle-hole bubbles are excluded from the flow.
This observation can be traced back to the $t -\JK - \JH$ model that we take as our starting point. Since important particle-hole fluctuations
of a microscopic model in the Mott insulating phase are already contained in the exchange terms, the subsequent fRG-flow tends to enhance
the `pre-formed' tendencies.
\begin{figure}[t!]
\centering
\includegraphics[height=.45\columnwidth]{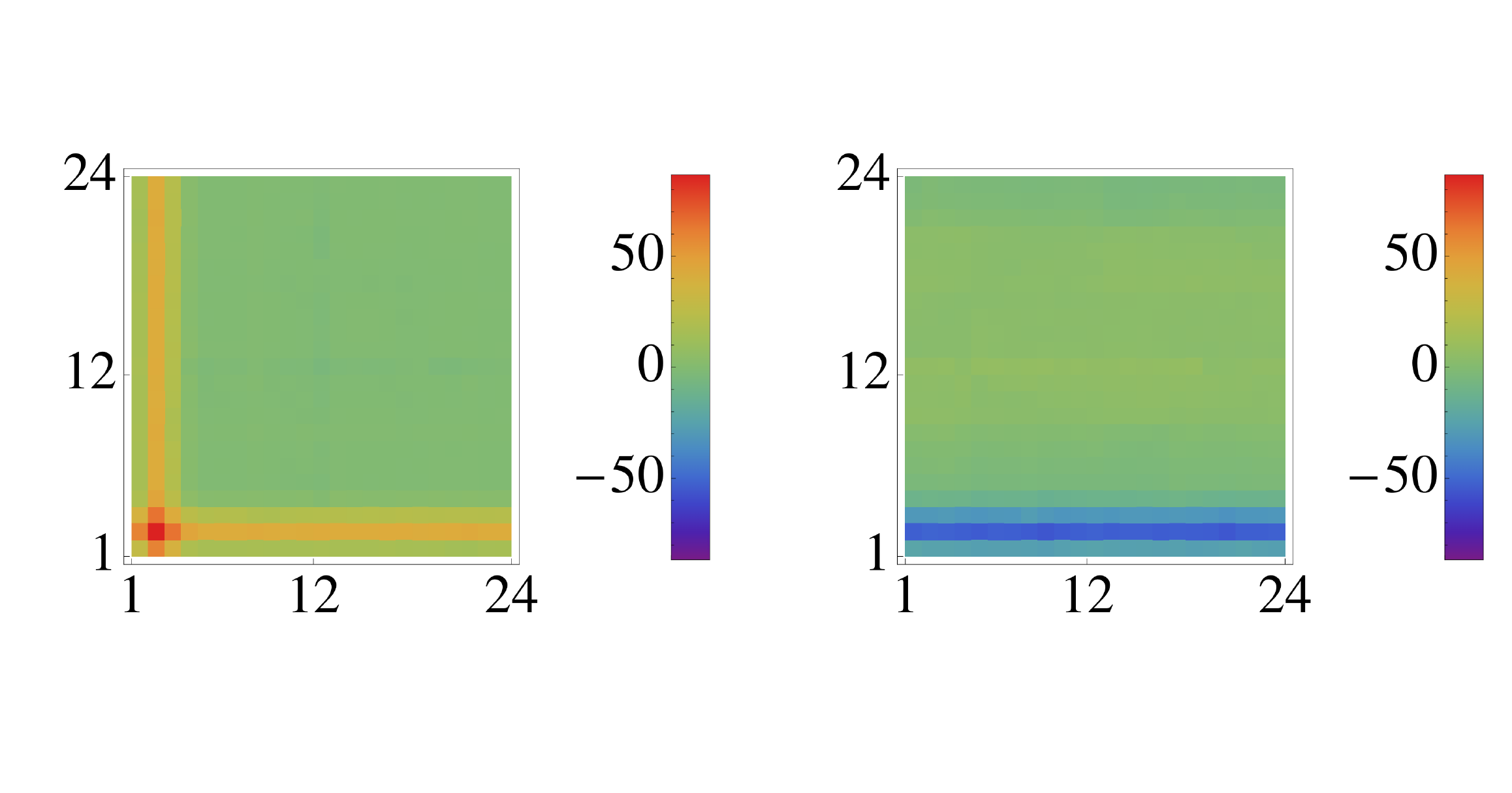}\\
\vspace{-2.5em}
\includegraphics[height=.45\columnwidth]{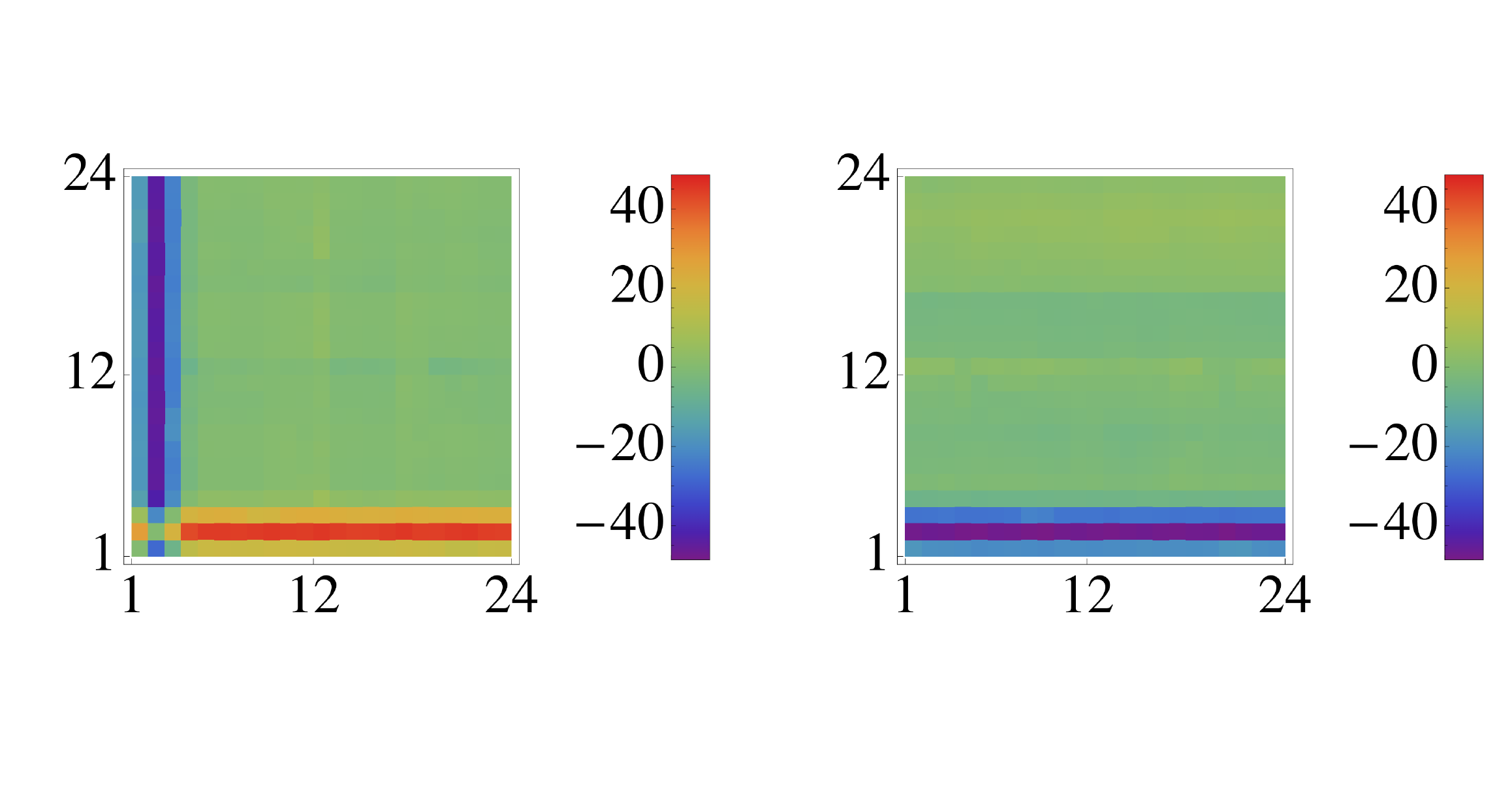}
\caption{The upper panel shows the
intrasublattice (left) and intersublattice (right) component of the divergent singlet vertex function for $\delta=0.14$ and $|\JH|/t_{0}=0.525$ with $\JK/t_{0}=1$, where the patch numbers corresponding to patch momentum $\vec{k}_{1}$ are given on the ordinate and $\vec{k}_{2}$ on the abscissa. The remaining free momentum $\vec{k}_{3}$ is here fixed to the second patch, cf. Fig.~\ref{fig:patching}. With the
same conventions, the lower panel shows the intrasublattice (left) and intersublattice (right) component of the divergent triplet-$x$ vertex function.
The divergent momentum structure corresponds to a CDW instability. Both amplitude ratio of singlet to triplet vertex and the sign structure
conspire to re-combine singlet and triplet pairing interactions into a density-density interaction, cf. \Eqref{eq:cdw}.}
\label{fig:cdw}
\end{figure}
Similar to the case of the antiferromagnet, the low-energy degrees of freedom close to $\Lambda_{c}$ can
be described by a Hamiltonian of the form
\be\label{eq:cdw}
H_{\mathrm{CDW}} \propto - V \sum_{o,o^{\prime}\in A,B} \epsilon_{o,o^{\prime}}\, N_{\vec{q}=0}^{o}\, N_{\vec{q}=0}^{o^{\prime}},\quad V > 0.
\ee
Here, $N_{\vec{q}=0}^{o} = \sum_{\vec{k}} f_{o,\vec{k},\sigma}^{\dagger} [\sigma_{0}]_{\sigma\sigma^{\prime}} f_{o,\vec{k}-\vec{q},\sigma^{\prime}}$ is the sublattice density operator for auxiliary fermions.
It differs from the electron density only by a factor of $\delta$. The system minimizes its energy by having a charge imbalance, e.g. more 
electrons reside on sublattice $A$ than on sublattice $B$, or vice versa. The CDW instability takes up a large part of the phase diagram and
also comes with rather large critical scales. We estimate critical temperatures up to $k_{B}T_{c} \sim 10^{-1}\, t_{0}$. Previous mean-field studies 
did not include CDW order-parameters. In our case, the CDW instability is driven by the density-density term in \Eqref{eq:hamiltoniancontdH} for $\JH < 0$
and outweighs ferromagnetic ordering tendencies. 

We now turn to the superconducting instabilities. The singlet channel determines the leading superconducting instability only
in a rather narrow strip for $|\JH|/t_{0} < 0.2$. For doping up to $\delta = 0.4$,
the intraband pairing symmetry is $d$-wave. Interband correlations are again of $p$-wave type. As the doping level is increased
above $\delta = 0.4$, an $s$-wave pairing symmetry is favored. Also here, the dominating superconducting correlations are of
nearest-neighbor type. Intrasublattice correlations are subleading.

As the strength of the ferromagnetic Heisenberg coupling is increased, at $|\JH|/t_{0} \simeq 0.2$ the leading
instability switches from singlet to triplet. The different ordering tendencies in particle-hole and particle-particle channels
lead to a suppression of intraband-pairing. On moving closer to the triplet-SC/CDW phase boundary, the intraband pairing
correlations of $p$-wave type grow stronger (see Fig.~\ref{fig:ff3}).
Pairing correlations along nearest-neighbor bonds still dominate.  
For the intraband correlations, however, we still observe a \textit{substantial decrease} as compared to ferromagnetic Kitaev and antiferromagnetic Heisenberg exchange (cf. Sects.~\ref{subsec:pp},~\ref{subsec:ppph}), while the interband correlations dominate. The intraband pairing can be described by the following $\vec{d}$-vector
\be\label{eq:dvec2}
\vec{d}_{\vec{k},1} & = & \left[p_{x}(\vec{k}) + 1/\sqrt{3}\, p_{y}(\vec{k})\right]\left(1, 0 , 0\right)^{T}, \nn \\
\vec{d}_{\vec{k},2} & = & \left[p_{x}(\vec{k}) - 1/\sqrt{3}\, p_{y}(\vec{k})\right]\left(0,1 , 0 \right)^{T},  \\
\vec{d}_{\vec{k},3} & = & \left[2/\sqrt{3}\, p_{y}(\vec{k}) \right]\left(0, 0, 1\right)^{T}.  \nn
\ee
As compared to the $\vec{d}$-vector obtained from doping the stripy phase, here the $p$-wave instability is driven
by the ferromagnetic and isotropic Heisenberg exchange.

The interband $\vec{d}$-vector is captured by (see also Fig.~\ref{fig:ff3})
\be\label{eq:dvec3}
\vec{d}_{\vec{k},1} & = & \left[- s(\vec{k}) + 1/3 \, d_{xy}(\vec{k}) + 1/3\sqrt{3} \, d_{x^2-y^2}(\vec{k})\right](1,0,0)^{T}, \nn \\
\vec{d}_{\vec{k},2} & = & \left[ s(\vec{k}) + 1/3 \, d_{xy}(\vec{k}) - 1/3\sqrt{3} \, d_{x^2-y^2}(\vec{k}) \right](0,1,0)^{T}, \nn \\
\vec{d}_{\vec{k},3} & = & \left[ s(\vec{k}) + 2/3\sqrt{3} d_{x^2-y^2}(\vec{k})\right](0,0,1)^{T}. 
\ee
Here, by $s(\vec{k})$ we denote the even parity nearest-neighbor $s$-wave form factor, see~\Eqref{eq:ffsp}. 
The $p$-wave part in~\Eqref{eq:dvec2} was reported previously~\cite{okamoto2013} with dominant intraband pairing.
We here find dominating interband correlations and enhanced $s$-wave contributions close to the CDW phase boundary.

Critical scales/temperatures for the $p$-wave regime decrease upon doping from $k_{B}T_{c} \sim 10^{-4}\, t_{0}$ to $k_{B}T_{c} < 10^{-8}\, t_{0}$. Critical scales
below $10^{-8}\, t_{0}$ could actually not be properly resolved from the fRG-flows, see also Fig.~\ref{fig:critscalePPPH2}. Further, critical scales are not constant along the $\JH$-axis for fixed $\delta$ and $\JK$.

\subsection{Bond-order instabilities at van Hove filling}
\label{sec:bondorder}
%

%
\begin{figure}[t!]
\centering
\includegraphics[height=.5\columnwidth]{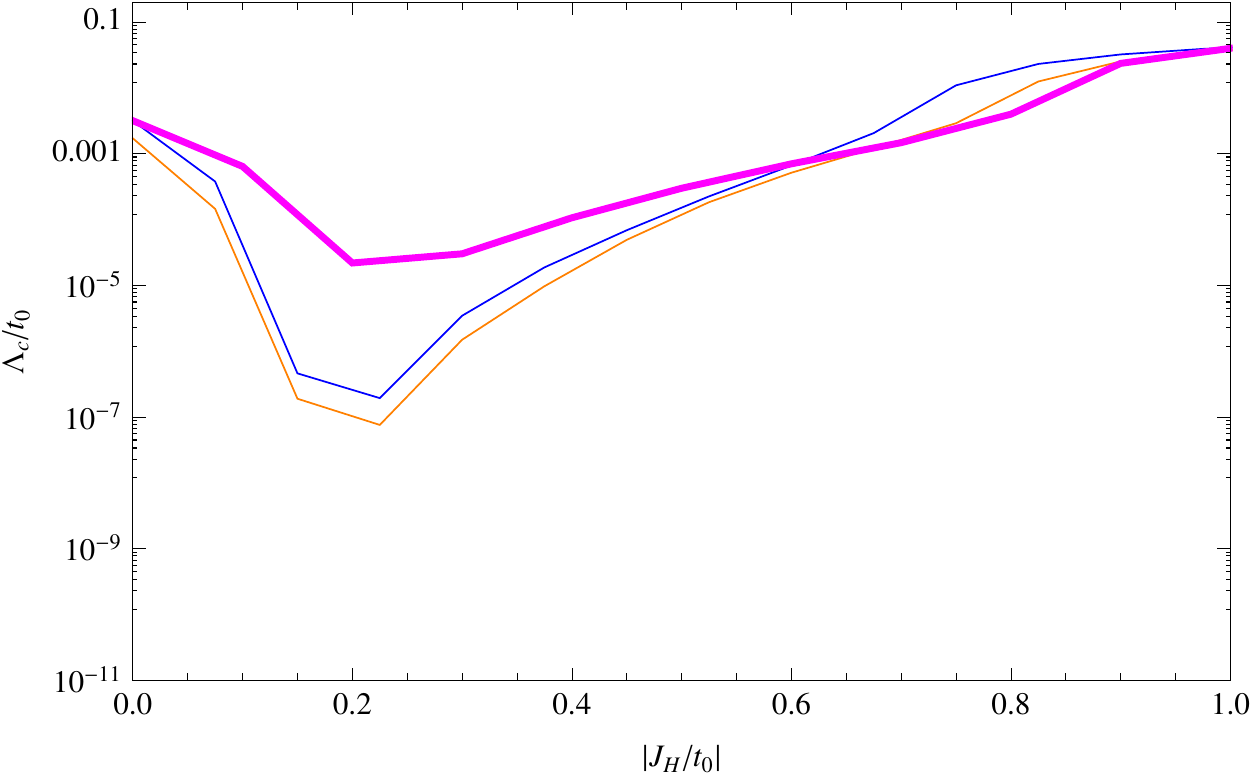}
\caption{Critical scale $\Lambda_{c}$ in units of the bare hopping $t_{0}$ for flows
evaluated at van Hove filling $\delta = 1/4$ (magenta) and fillings $\delta = 1/4 - 0.01$ (blue) and $\delta = 1/4 + 0.01$ (orange) as a function
of ferromagnetic Heisenberg exchange $|\JH|/t_{0}$ and fixed antiferromagnetic Kitaev exchange $\JK/t_{0} = 1$. 
As expected from the enhancement of the single-particle density of states for the nested Fermi surface at van Hove filling, the critical
scale is enhanced by a few orders of magnitude. It drops, however, rather quickly as the doping departs from $\delta = 1/4$. Also,
the effect of DOS enhancement reflected in increased critical scales is rendered ineffective as soon as the CDW instability sets in at $|\JH|/t_{0} \gtrsim 0.7$.}
\label{fig:critscalevHF}
\end{figure}
The filling $\delta = 1/4$ plays a special role in honeycomb lattice models, since perfect Fermi surface nesting and a van Hove singularity coincide. It is thus not surprising, that the effects from nesting
and enhanced density of states (DOS) at the Fermi level lead to a strong impact from the particle-hole
fluctuations on the emerging Fermi surface instability. Since it is the interplay of nesting and
density-of-states enhancement that is important, the ensuing phase at van Hove filling should be considered
as rather fragile with respect to deviations in filling factor. It comes, however, with an 
increased critical scale, i.e., larger critical temperature due to larger Fermi level DOS, cf. Fig,~\ref{fig:critscalevHF}. Additionally we find that for the parameter ranges studied in this work, only for antiferromagnetic Kitaev and ferromagnetic Heisenberg exchange do the particle-hole effects outweigh the pairing instability. 
Consequently, we will focus on $J_{\mathrm{K}} > 0$ and $J_{\mathrm{H}} < 0$ in the following.
As before, we keep the Kitaev interaction fixed at $J_{\mathrm{K}}/t_{0} = 1$ and vary the Heisenberg exchange $|J_{\mathrm{H}}|/t_{0}\in[0,1]$ for fixed doping $\delta = 1/4$. It turns out that
an $N=24$ patching scheme is insufficient to properly capture both DOS enhancement and nesting,
and leads to spurious artefact instabilities throughout the phase diagram. Upon increased angular 
resolution along the Fermi surface, these artefacts disappear at $N=96$ and allow for a clear
identification of the resulting ordering structures. Since $N=24$ patching has proven quite reliable in
interacting honeycomb systems away from van Hove filling~\cite{scherer2012a,lang2012,scherer2012b}, we believe our results for $\delta \neq 1/4$ 
are quite robust. We supported this claim by checking the phase boundaries in Fig.\ref{fig:pd2} with $N=96$. Only the singlet/triplet phase boundary was mildly affected. 

\begin{figure}[t!]
\centering
\includegraphics[height=.25\columnwidth]{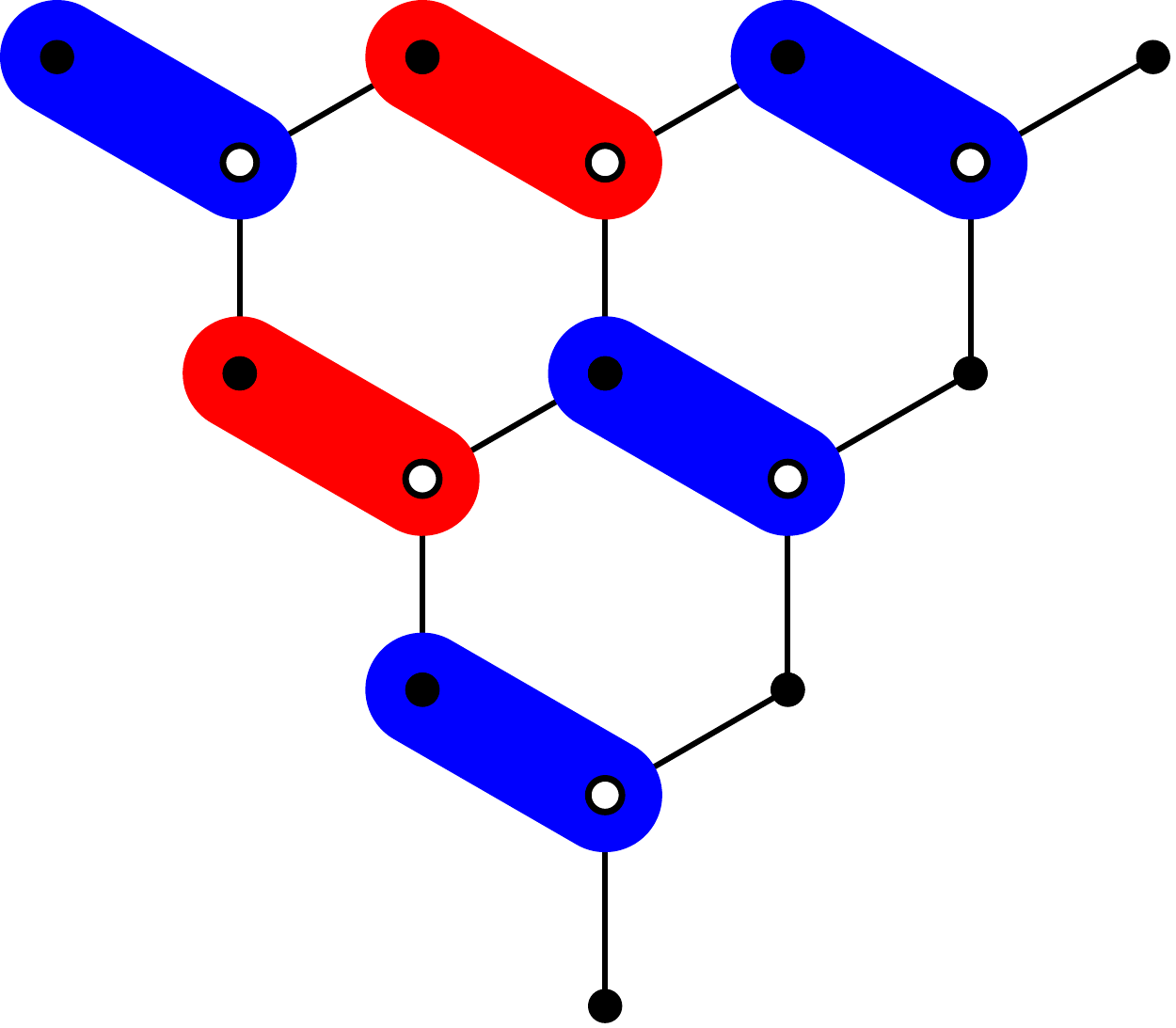}\hspace{0.5em}
\includegraphics[height=.25\columnwidth]{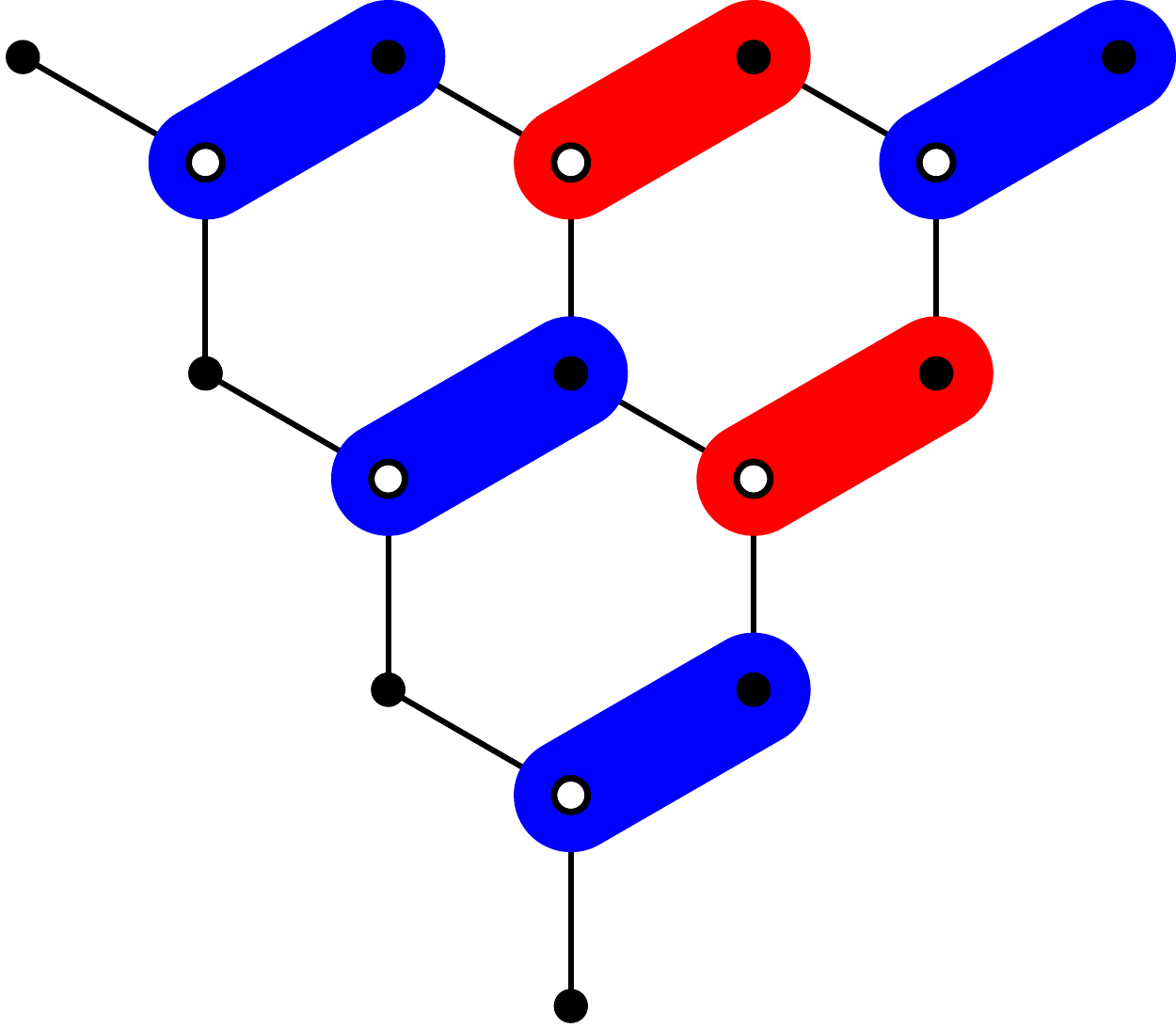}\hspace{0.5em}
\includegraphics[height=.25\columnwidth]{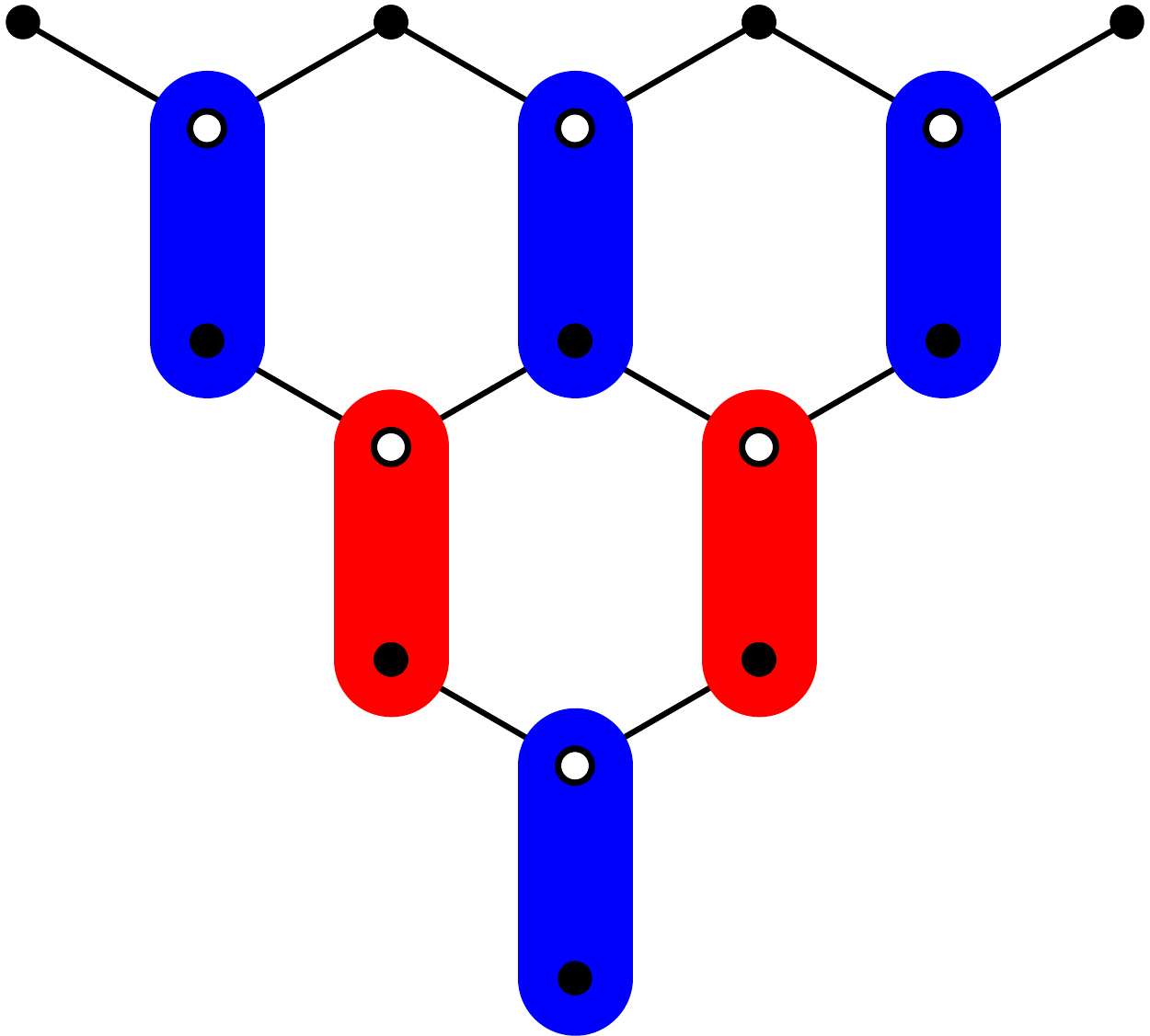}
\caption{Real-space charge bond-order patterns for $|\JH|/t_{0}=0.2$. Red bonds correspond to enhancement
of the hopping amplitude along the given bond, while blue bonds correspond to a decrease of the hopping amplitude. 
White disks mark the $A$-sublattice sites, black disks the $B$-sublattice sites. Shown are the ordering patterns for the
three different ordering wavevectors $Q_{1}$ (left), $Q_{2}$ (middle) and $Q_{3}$ (right). The form-factors
are given by $\cos(\delta_{3}\cdot{\vec{k}})$, $\cos(\vec{\delta}_{1}\cdot\vec{k})$ and $\cos(\vec{\delta}_{3}\cdot\vec{k})$, respectively.}
\label{fig:cBO}
\end{figure}

\begin{figure}[t!]
\centering
\includegraphics[height=.25\columnwidth]{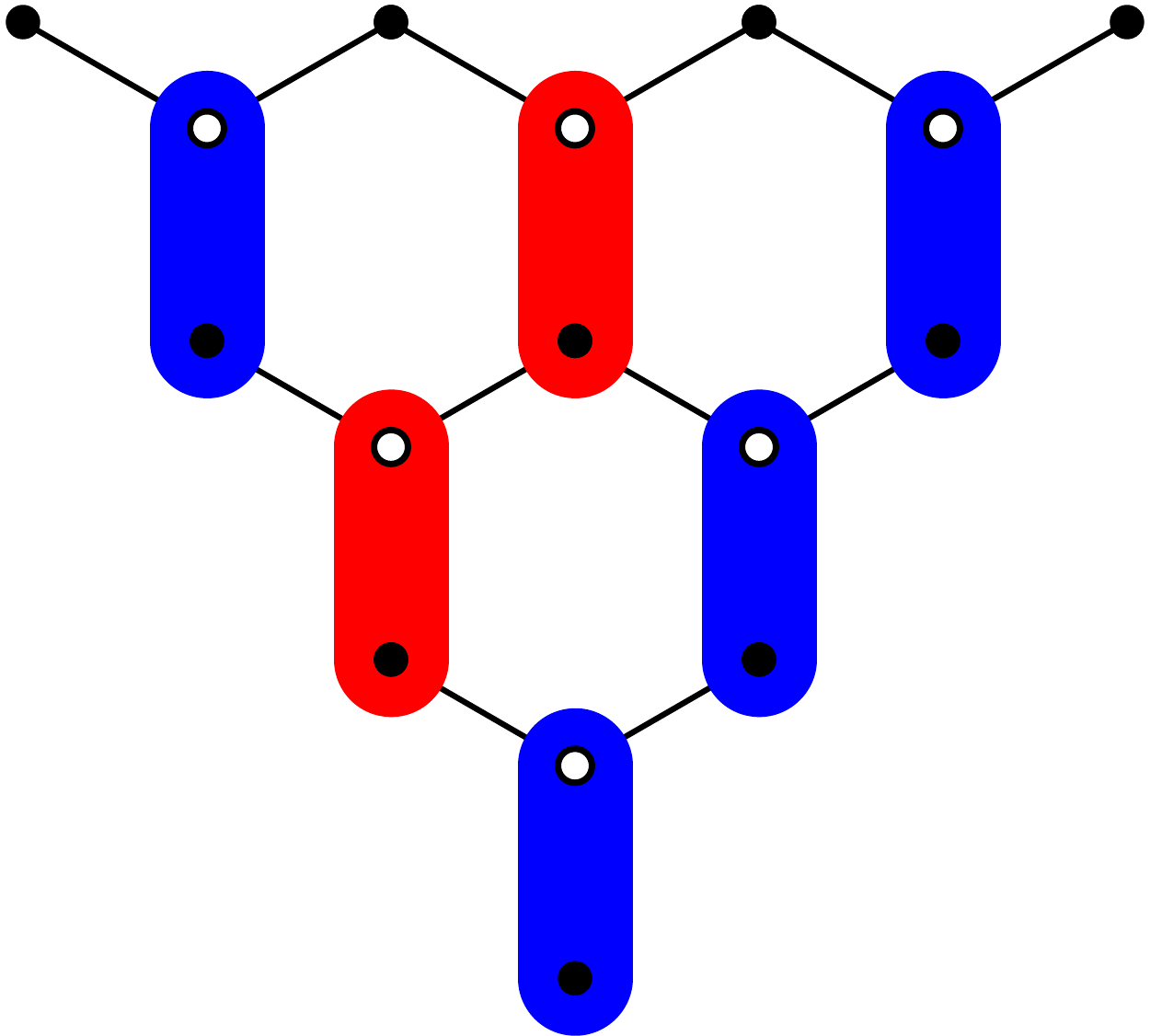}\hspace{0.5em}
\includegraphics[height=.25\columnwidth]{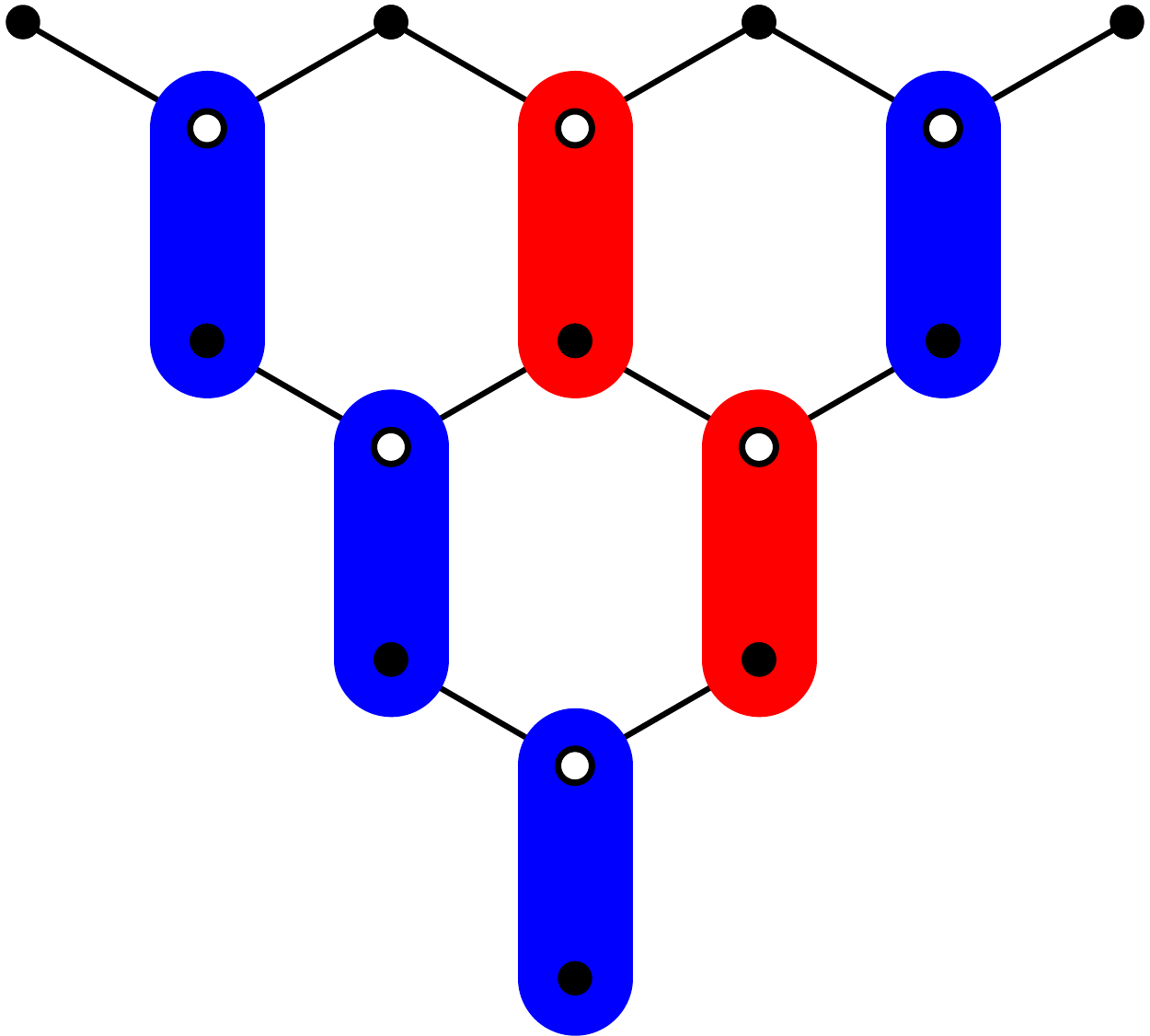}\hspace{0.5em}
\includegraphics[height=.25\columnwidth]{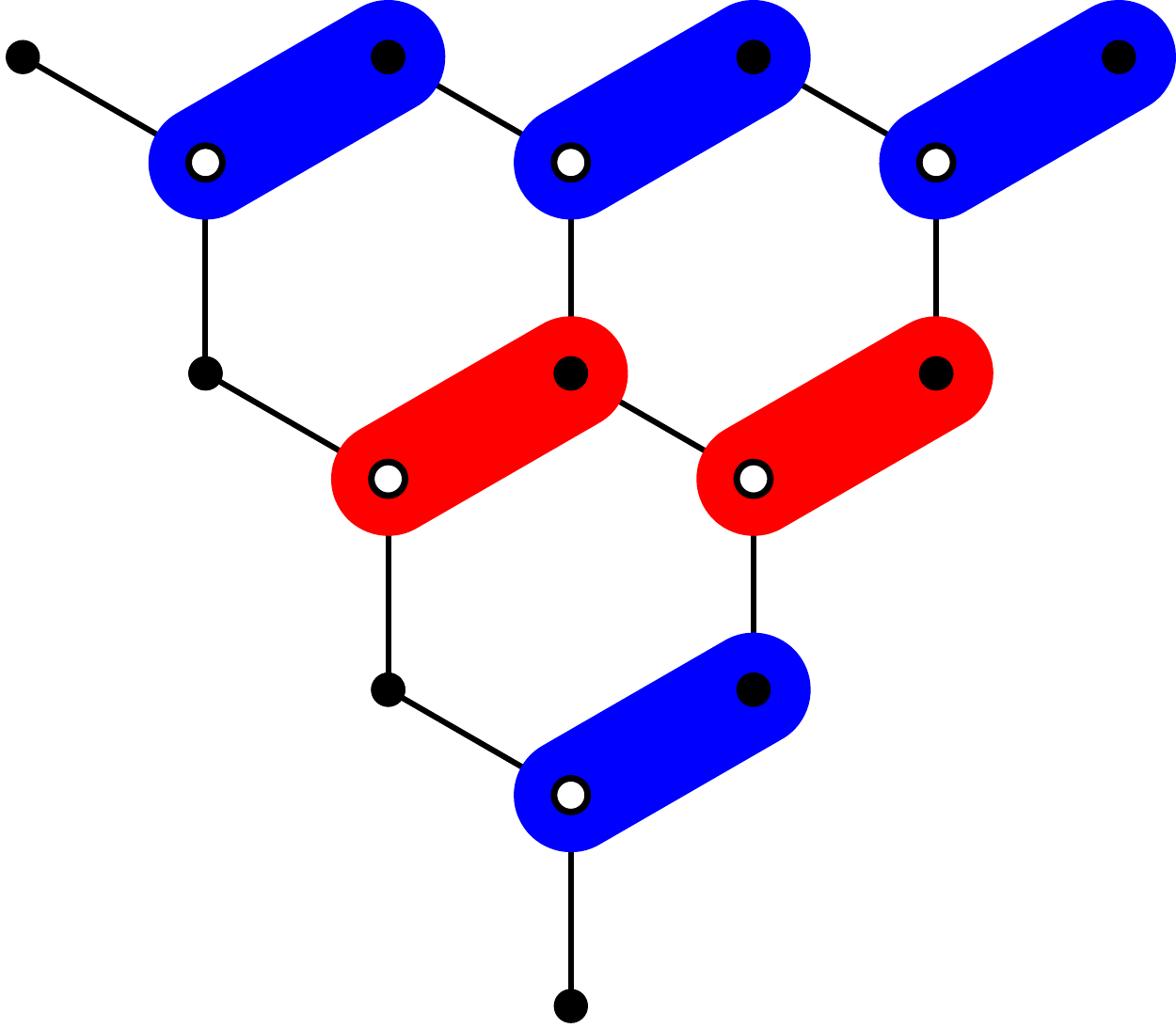}
\caption{Real-space spin bond-order patterns for $|\JH|/t_{0}=0.3$. Red bonds correspond to enhancement
of the hopping amplitude along the given bond, while blue bonds correspond to a decrease of the hopping amplitude. 
White disks mark the $A$-sublattice sites, black disks the $B$-sublattice sites. Shown are the ordering patterns for the
three different ordering wavevectors $Q_{1}$ (left), $Q_{2}$ (middle) and $Q_{3}$ (right). The spin-dependent form-factors (see Tab.~\ref{table:sBO})
are given by $\sigma_{z}\sin(\delta_{3}\cdot{\vec{k}})$, $\sigma_{y}\sin(\vec{\delta}_{3}\cdot\vec{k})$ and $\sigma_{x}\sin(\vec{\delta}_{1}\cdot\vec{k})$, respectively.}
\label{fig:sBO}
\end{figure}

The nesting vectors $\vec{Q}_{i}$, $i=1,2,3$ connect opposite edges of the hexagonal Fermi surface at van Hove filling, see Fig.~\ref{fig:patching}. Modulo reciprocal lattice vectors, these are equivalent to vectors connecting inequivalent, neighboring $M$ points. Explicitly, they are given by 
\be
\vec{Q}_{1} = \pi\left(1/\sqrt{3},1\right)^{T} &, &\quad \vec{Q}_{2} = \pi\left(-1/\sqrt{3},1\right)^{T},\nonumber \\ 
\vec{Q}_{3} & = & \pi\left(-2/\sqrt{3},0\right)^{T}.
\ee
An emergent order parameter with ordering wavevector $\vec{Q}_{i}$ breaks translation invariance
of the underlying lattice and leads to a doubling of the unit cell, i.e., a four atom unit cell in our present case. From analyzing the momentum-space pattern of the renormalized vertex function at the critical scale and employing Fierz identities, we find effective low-energy Hamiltonians of either charge bond-order (cBO) or spin bond-order (sBO) type:
\be
H_{\mathrm{cBO}} & \propto & \sum_{i=1}^{3}  V_{\mathrm{cBO}}^{i}  \Phi_{\vec{Q}_{i}}^{\ast}\Phi_{\vec{Q}_{i}} \\
\label{eq:HsBO}
H_{\mathrm{sBO}} & \propto &  \sum_{i=1}^{3} \sum_{l\in \{x,y,z\}}  V_{\mathrm{sBO}}^{i,l}
\Phi_{l,\vec{Q}_{i}}^{\ast}\Phi_{l,\vec{Q}_{i}} \, ,
\ee
with charge and spin bond-order amplitudes $V_{\mathrm{cBO}}^{i}$ and $V_{\mathrm{sBO}}^{i,l}$, respectively. The fermionic bilinears $\Phi_{\vec{Q}}$ and $\Phi_{l,\vec{Q}}$ are given by
\be
\Phi_{\vec{Q}} \propto \sum_{\vec{k},\sigma} \sum_{o,o^{\prime}} \tilde{\epsilon}_{o,o^{\prime}} t_{\vec{k}}^{o,o^{\prime}} (\vec{Q})  f_{o,\vec{k},\sigma}^{\dagger} f_{o^{\prime},\vec{k}-\vec{Q},\sigma}\, ,
\ee
and
\be
\Phi_{l,\vec{Q}} \propto \sum_{\vec{k},\sigma,\sigma^{\prime}} \sum_{o,o^{\prime}} \tilde{\epsilon}_{o,o^{\prime}}  t_{l,\vec{k}}^{o,o^{\prime}} (\vec{Q})  f_{o,\vec{k},\sigma}^{\dagger} [\sigma_{l}]_{\sigma\sigma^{\prime}} f_{o^{\prime},\vec{k}-\vec{Q},\sigma^{\prime}}\, , \nonumber \\ 
\ee
where $i = 1,\dots,3$ labels the different ordering wavevectors, and $l \in \{x,y,z\}$
labels spin-vector components. Here,
$\tilde{\epsilon}_{o,\mathrm{o}^{\prime}} = +1$ for $o \neq o^{\prime}$ and   
$\tilde{\epsilon}_{o,\mathrm{o}^{\prime}} = 0$ for $o = o^{\prime}$. This particular
form of interlattice correlations corresponds to the dimerization of particle-hole excitations
along a given bond. On a mean-field level,  a finite expectation value $\langle\Phi_{\vec{Q}}\rangle$ of the cBO order-parameter leads to a renormalization of the hopping amplitude and to an enlargement of the unit cell with a corresponding downfolded Brillouin zone and additional bands. From numerical calculations, we find the form factors $t_{\vec{k}}^{o,o^{\prime}}(\vec{Q})$ can be described by $\cos(\vec{\delta}_{j}\cdot\vec{k})$, $j=1,2,3$ form factors for hopping along nearest-neighbor bonds. The resulting real-space patterns are displayed in 
Fig.~\ref{fig:cBO}.
A finite sBO order-parameter $\langle\Phi_{l,\vec{Q}}\rangle$ leads to a renormalized spin-dependent hopping amplitude. For given $\vec{Q}$, the form factors $t_{\vec{k}}^{o,o^{\prime}}(\vec{Q})$ and $t_{l,\vec{k}}^{o,o^{\prime}}(\vec{Q})$, respectively, determine the bond-order pattern within the enlarged unit cell. From our fRG results we infer the leading instability is either
of cBO or sBO type, but the two different instabilities do not coincide. The eigenmodes for different
$\vec{Q}$ extracted from the corresponding reduced vertex functions -- which can again be understood as $N \times N$ matrices -- turn out to be degenerate. 
Further, in the sBO case, there are always two (almost) degenerate eigenmodes with different $l$ for fixed $\vec{Q}$. The association of spin matrices to a given ordering wavevector as obtained from our numerical results is collected in Tab.~\ref{table:sBO}.
For sBO instabilities, the dominant features of the numerically obtained form factors can be described with $\sin(\vec{\delta}_{j}\cdot \vec{k})$, $j=1,2,3$
where $\vec{\delta}_{j}$ are the nearest-neighbor vectors from $A$ to $B$ sublattice, see Fig.~\ref{fig:sBO}. These, of course, can be expressed in terms of
nearest-neighbor $p$-wave form factors. The form factors, however, seem to rotate in the degenerate $p$-wave subspace as 
$\JH$ changes. The modes corresponding to $\cos(\vec{\delta}_{j}\cdot \vec{k})$ form factors turn out
to be subleading for the sBO instability. Fourier transforming the form-factors yields the corresponding modulation of the real-space hopping amplitude. Due to the limitations of our truncation to the exact hierarchy of fRG equations, we cannot determine which linear combination of the different
mean-fields will be realized in the ground state of the system. While some of the sine patterns overlap,
others reside on mutually exclusive bonds. For overlapping patterns, we cannot expect the different ordering patterns to be energetically independent. 
A determination of the lowest-energy configuration, however, is beyond the capabilities of our employed truncation scheme.
As displayed in Fig.~\ref{fig:pd2}, at small $|\JH|/t_{0}$ the singlet pairing instability is leading,
while at $|\JH|/t_{0} = 0.2$ charge bond-order sets in as the leading instability. As $|\JH|/t_{0}$ increases, the leading instability quickly crosses over from charge to spin bond-order with the
aforementioned two degenerate eigenmodes per ordering wavevector. The bond-order instability
is cut off by the CDW for $|\JH|/t_{0} \gtrsim 0.7$

In view of the superconducting neighborhood of the bond-order instabilities at van Hove filling, cf. Fig.~\ref{fig:pd2}, we can infer that while 
the proximity to even-parity singlet pairing also promotes even-parity singlet charge bond-order, odd-parity triplet pairing favors the formation
of odd-parity triplet spin bond-order.

Remarkably, even though we modeled the hopping in the non-interacting Hamiltonian \Eqref{eq:hamiltoniancontd} as spin-independent, the interplay of antiferromagnetic Kitaev and
ferromagnetic Heisenberg exchange with nesting and DOS enhancement lead to dynamical 
re-generation of anisotropic spin-orbit coupling type terms on the level of a mean-field treatment of the
low-energy Hamiltonian \Eqref{eq:HsBO}.
\begin{table}[]
\begin{tabular}{c || c c c }
wavevector & \quad $\sigma_{x}$ \quad & \quad $\sigma_{y}$ \quad & \quad $\sigma_{z}$ \quad \\
\hline
$\vec{Q}_{1}$ & $\checkmark$  & --  & $\checkmark$ \\
$\vec{Q}_{2}$ & --  & $\checkmark$  & $\checkmark$  \\
$\vec{Q}_{3}$ & $\checkmark$  & $\checkmark$  & -- \\
\end{tabular}
\caption{Spin matrices associated with the three different ordering wavevectors as obtained
from our numerical results. For fixed wavevector $\vec{Q}$, emergent spin bond-order yields
spin-dependent hopping described by the checked spin matrices within a mean-field treatment
of the low-energy Hamiltonian \Eqref{eq:HsBO}.}
\label{table:sBO}
\end{table}
 While a detailed analysis of the properties of fermions moving in the background of self-consistently generated bond-order patterns is beyond the scope of this paper, the mean-field Hamiltonian for low-energy fermions with a static bond-order mean-field readily yields a renormalized fermion spectrum. Considering the different ordering wavevectors independently, we obtain a metallic state for each $\vec{Q}$ with a connected Fermi surface in the reduced Brillouin zone. Energetically, a gapless metallic state might seem less favorable for the system than a state with a nodal superconducting gap. But the condensation of bond order seems to occur at critical scales which are well above the critical temperature for the transition to the superconducting state with a nodal gap along the Fermi surface.
 
 We note that even for $\JK = 0$, a finite ferromagnetic Heisenberg coupling $\JH$ is sufficient to drive the system toward a bond-order instability at 
 van Hove filling. A similar observation -- ferromagnetic fluctuations causing a propensity toward bond-order instabilities -- was made for the extended kagome Hubbard model
 with fRG methods~\cite{kiesel2013}. The dimerization pattern corresponds to spin bond-order, due to the restored rotational symmetry, however, all spin components are degenerate. We attribute the fact that we do not observe a magnetically site-ordered state to the dominating role of
the nearest-neighbor density-density interaction term, cf.~\Eqref{eq:hamiltoniancontdH}.
For $\delta \neq 1/4$ and $\JK = 0$ critical scales drop quickly below $10^{-8} t_{0}$. The associated instabilities, if they exist, are thus not observable
 within our current approach.
 
Finally, we comment on the stability of our results in the presence of Fermi surface renormalization. Since in the present truncation
self-energy feedback is completely neglected, the shape of the Fermi surface is fixed during the RG flow. An fRG scheme that takes into account self-energy feedback in principle can modify the Fermi surface and thus might destroy the nesting condition. However, as observed in Ref.~\onlinecite{honerkamp2001} the real part of the flowing self-energy mainly leads to a straightening of the Fermi surface. From this point of view, it seems plausible that effects from nesting and DOS enhancement are stable with respect to
inclusion of self-energy effects. A more complete picture studying the interplay between van Hove singularities and self-energy flow is, however, certainly desirable.
 
\section{Conclusions \& Discussion}
\label{sec:conclusions}

We have analyzed the phase diagram of the doped Kitaev-Heisenberg model on the honeycomb lattice for the situations of
ferromagnetic Kitaev and antiferromagnetic Heisenberg exchange, as well as antiferromagnetic Kitaev and
ferromagnetic Heisenberg exchange. We attacked the problem of describing the correlated, frustrated and spin-orbit
coupled Mott insulator within a slave-boson treatment, and derived functional RG equations for the auxiliary fermionic problem
after the bosonic holon sector was dealt with on a mean-field level. We solved the 
functional flow equations in the static patching approximation, where the patch number for
angular resolution of the Fermi surface ranged from $N=24$ to $N=96$.

While our results corroborate the tendency towards the formation of triplet $p$-wave pairing phases, we demonstrate that other competing orders driven
by particle-hole fluctuations reduce the parameter space where pairing yields the leading instability. We further uncovered instabilities at van Hove filling supporting unconventional
dimerization phases of the electronic liquid. Interestingly, the prediction of emergent topological 
$p$-wave pairing states is unaffected by the inclusion of particle-hole fluctuations. For ferromagnetic Kitaev and antiferromagnetic Heisenberg exchange, the gap-closing transition from trivial to topologically non-trivial $p$-wave is left untouched, although critical temperatures are reduced. Flipping the signs of both exchange terms, a bond-order instability pre-empts the naive pairing mean-field gap-closing transition at van Hove filling. The resulting dimerization state, however, remains gapless. Upon doping beyond van Hove filling, the $p$-wave
phase is restored. Applying the rule of counting the number of time-reversal invariant momenta below the Fermi surface~\cite{qi2009,sato2009a,sato2009b,qi2010,sato2010}, we again obtain a topological $p$-wave state.

While the dimerized state at van Hove filling appears to remain gapless and non-topological, 
the proposal of Ref.~\onlinecite{foyevtsova2013} to include longer-ranged exchange interactions beyond isospins connected by nearest-neighbor bonds to better model the magnetic state of Na$_{2}$IrO$_3$ might also provide a route to dynamically generated topological Mott insulating states at van Hove filling.

Extending the $t - \JK - \JH$ model to a $\JH - V$ model in the Heisenberg sector, where $V$ is now
promoted to an independent coupling for nearest-neighbor density-density interactions (while we restricted our attention to $V=\JH$) provides another route
to generalization. At least for a subset of initial values for $\JH$ and $V$, however,
the fRG-flow will be attracted to the infrared-manifold and the corresponding instabilities we discuss in the present paper. Further, the fRG approach taken in this work might be successfully applied 
to doping induced instabilities in the context of other material-inspired spin-orbit model Hamiltonians~\cite{nussinov2013}.

Before closing the discussion of our results, we briefly comment on the treatment of the $t - \JK - \JH$ model within the fRG framework. 
The fRG approach employed in this work differs from fRG applications to other, weakly correlated electron lattice systems (for a review, see Ref.~\onlinecite{metzner2011}) as its starting point is the renormalized auxiliary fermion Hamiltonian, with a strongly reduced bandwidth. This may cast some doubts on the applicability of a method perturbative in the interactions like the fRG. Here, we do not claim that the results are quantitatively controlled, but we can be confident that qualitatively they capture the right trends. First of all it should be noted that using fRG instead of the common mean-field study of the phase diagram of the auxiliary fermion model is certainly an improvement that removes ambiguities and excludes that important channels may get overlooked. Then we also refer to a number of works with a method dubbed `spin-fRG', cf. Refs.~\onlinecite{reuther,reuther2012,singh2012}. In these works related spin physics is explored in the insulating limit where the kinetic energy is completely quenched. The results obtained there are physically meaningful and give insights into spin physics of frustrated models that are otherwise hard to obtain. In the insulating case, the fermion propagator is purely local and the spin-spin interaction remains of a simpler bilocal form. 
Hence its full frequency dependence can be taken into account. This simplicity is lost in the doped case studied here, as the fermion propagator is non-local and mediates effective interactions different from simple bilocal spin-spin type. Hence, for us it is difficult to treat the frequency dependence of the vertex in addition to the even more important momentum space structure. 
Nevertheless, as our case interpolates between the two extreme cases, insulator and weakly correlated systems, where the approach has been shown to work reasonably, we can be confident that studying the correlated doped case by perturbative fRG is justified.

DDS acknowledges discussions with L.~Kimme, T.~Hyart and M.~Horsdal and technical support by M.~Treffkorn and H.~Nagel. MMS is supported by the grant ERC- AdG-290623.

\newpage

\begin{widetext}

\appendix

\section{Technical supplement}

\subsection{Gamma matrices and superconducting order-parameters}
\label{app:matrix}

The decomposition of the interaction terms in the singlet and triplet pairing channels 
is adapted to analyzing superconducting instabilities. Accordingly, the $\Gamma$ matrices
were chosen following the conventions \cite{sigrist1991} used in the description of unconventional
superconductors. This way, the spin-structure of an emerging superconducting instability is
included automatically in our approach. For convenience, we here give the explicit expressions for the complete set
of $2\times2$ matrices:
\be
\Gamma_{0} & = & \frac{1}{\sqrt{2}}\sigma_{0}\mathrm{i}\sigma_{y} = 
\frac{1}{\sqrt{2}}
\begin{pmatrix}
0 & 1\\
-1 & 0
\end{pmatrix},
\quad 
\Gamma_{x}=\frac{1}{\sqrt{2}}\sigma_{x}\mathrm{i}\sigma_{y} =
\frac{1}{\sqrt{2}}
\begin{pmatrix}
-1 & 0\\
0 & 1
\end{pmatrix}, 
\\
\Gamma_{y} & = & \frac{1}{\sqrt{2}}\sigma_{y}\mathrm{i}\sigma_{y} =
\frac{1}{\sqrt{2}}
\begin{pmatrix}
\mathrm{i} & 0\\
0 & \mathrm{i}
\end{pmatrix},
\quad 
\Gamma_{z}=\frac{1}{\sqrt{2}}\sigma_{z}\mathrm{i}\sigma_{y} = 
\frac{1}{\sqrt{2}}
\begin{pmatrix}
0 & 1\\
1 & 0
\end{pmatrix}, 
\nn
\ee 
\be
\mathrm{Tr}(\Gamma_{\mu}\Gamma_{\nu}^{\dagger})=\delta_{\mu\nu},\quad\mu,\nu\in\{0,x,y,z\}. 
\ee
Here, $\sigma_{x}$, $\sigma_{y}$ and $\sigma_{z}$ are the Pauli matrices, and $\sigma_{0}$ denotes the $2\times2$ unit matrix.
Further, one can derive so-called Fierz identities for this set of matrices in order to re-write quartic pairing terms 
into density-density type interactions. This enables us to obtain CDW and SDW type instabilities from the
singlet and triplet pairing interactions.

The most general superconducting order parameter with singlet and triplet components can now be
compactly written as~\cite{sigrist1991}
\be
\hat{\Delta}_{\vec{k}} = \sqrt{2} \left(\psi_{\vec{k}}\,\Gamma_{0} + \vec{d}_{\vec{k}}\cdot\vec{\Gamma} \right),
\ee
with $\vec{\Gamma} = (\Gamma_{x},\Gamma_{y},\Gamma_{z})^{T}$. The singlet order-parameter is described by a scalar
function $\psi_{\vec{k}}$, while the triplet order-parameter is specified by a three-component vector $\vec{d}_{\vec{k}}=(d_{\vec{k},x},d_{\vec{k},y},d_{\vec{k},z})^{T}$. 
Since in this work we are dealing with a multi-band system, the $2\times2$ order-parameter $\hat{\Delta}_{\vec{k}}$ also carries band 
indices describing intra- ($[\psi_{\vec{k}}]_{b_{1},b_{2}}$, $[\vec{d}_{\vec{k}}]_{b_{1},b_{2}}$ with $b_{1}=b_{2}$) and interband pairing
 ($[\psi_{\vec{k}}]_{b_{1},b_{2}}$, $[\vec{d}_{\vec{k}}]_{b_{1},b_{2}}$ with $b_{1}\neq b_{2}$).
 
\subsection{Form factors on the honeycomb lattice}
\label{app:formfactors}

The form factors that we employ here for the analysis of order-parameter symmetries are obtained
from the irreducible representations of the point group of the hexagonal lattice. For a given representation
in terms of functions defined on the real-space lattice, a momentum space form-factor $f_{\vec{k}}$
can be obtained as $f_{\vec{k}} = \sum_{\vec{r}}\mathrm{e}^{\mathrm{i}\vec{k}\cdot \vec{r}}f_{\vec{r}}$.
As is usually done, the lattice sum is split into nearest-neighbor (NN), next-nearest neighbor (NNN), etc., contributions.
The NN form-factors are thus given by  $f_{\vec{k}} = \sum_{j}\mathrm{e}^{\mathrm{i}\vec{k}\cdot \vec{\delta}_{j}}f_{\vec{r}+\vec{\delta}_{j}}$.
Since the resulting form factors do not come with well-defined parity, we form the appropriate linear combinations yielding
form-factors that are either even or odd with respect to $\vec{k} \to -\vec{k}$. We can proceed accordingly for NNN form factors and
so on. The obtained set of form factors is suitable to analyze order-parameter symmetries for fermions in Bloch/sublattice representation. 
When we switch to the band representation, the NN form-factors pick up a phase factor $\phi_{k}=\sum_{j}\mathrm{e}^{\mathrm{i}\vec{k}\cdot\vec{\delta}_{j}}/|\sum_{j}\mathrm{e}^{\mathrm{i}\vec{k}\cdot\vec{\delta}_{j}}|$ due to the unitary transformation
relating Bloch and band representations. For the NN form-factors in the band representation,
we find the following expressions:
\be
s(\vec{k}) & = & \frac{1}{3} \left(\cos \left(\frac{\sqrt{3} k_{x}}{2}-\frac{k_{y}}{2}+\phi_{\vec{k}} \right)+\cos \left(\frac{\sqrt{3} }{2} k_{x}+\frac{k_{y}}{2}-\phi_{\vec{k}} \right)+\cos (k_{y}+\phi_{\vec{k}} )\right),\label{eq:ffsp}
\\
d_{x^2-y^2}(\vec{k}) & = & -\frac{4}{3}  \left(\cos \left(\frac{\sqrt{3} k_{x}}{2}-\frac{k_{y}}{2}+\phi_{\vec{k}} \right)+\cos \left(\frac{\sqrt{3} }{2} k_{x}+\frac{k_{y}}{2}-\phi_{\vec{k}} \right)-2 \cos (k_{y}+\phi_{\vec{k}} )\right),\label{eq:ffdx2y2}
\\
d_{xy}(\vec{k}) & = & \frac{8 \sin \left(\frac{\sqrt{3} k_{x}}{2}\right) \sin \left(\frac{k_{y}}{2}-\phi_{\vec{k}}\right)}{\sqrt{3}},\label{eq:ffdxy}
\\
p_{x}(\vec{k}) & = & \frac{2 \sin \left(\frac{\sqrt{3} k_{x}}{2}\right) \cos \left(\frac{k_{y}}{2}-\phi_{\vec{k}}\right)}{\sqrt{3}},\label{eq:ffpx}
\\
p_{y}(\vec{k}) & = & \frac{1}{3} \left(-\sin \left(\frac{\sqrt{3} k_{x}}{2}-\frac{k_{y}}{2}+\phi_{\vec{k}} \right)+\sin \left(\frac{\sqrt{3} }{2} k_{x}+\frac{k_{y}}{2}-\phi_{\vec{k}} \right)+2 \sin (k_{y}+\phi_{\vec{k}} )\right).\label{eq:ffpy}
\ee
In the limit $\phi_{\vec{k}} \to 0$, we recover the NN form-factors in the Bloch/sublattice representation.

\subsection{Flow equations}
\label{app:flow}

In this section we summarize the RG-contributions to the right hand sides of \Eqref{eq:singletflow} and \Eqref{eq:tripletflow}, respectively. We here stick to the conventions of Ref.~\onlinecite{salmhofer2001}. 
We define the shorthand $\int\!\!d\teta$ to represent integration/summation over
loop variables. The spin projection $\sigma = \uparrow, \downarrow$ is not included and has already been traced over in going from $\xi$ to $\txi$ and 
$\eta$ to $\teta$.
Carrying out the projections onto singlet and triplet channels and defining the loop kernel $L = S^{\Lambda} G_{0}^{\Lambda} + G_{0}^{\Lambda} S^{\Lambda}$
with the single-scale propagator $S^{\Lambda} = d/d\Lambda\, G_{0}^{\Lambda}$, we find for the singlet case the particle-particle 
contribution
\be
\phi_{\mathrm{pp}}^{(\mathrm{s})}(\txi_1,\txi_2,\txi_3,\txi_4) = \frac{1}{2}\prod_{\nu=1}^{4}\!\int\!\!d\teta_{\nu}\, L(\teta_2,\teta_1,\teta_3,\teta_4)
V^{(\mathrm{s})}(\txi_2,\txi_1,\teta_2,\teta_3)V^{(\mathrm{s})}(\teta_4,\teta_1,\txi_3,\txi_4).
\ee
The particle-particle bubble-contribution to the triplet channel is given by
\be
\phi_{\mathrm{pp};l}^{(\mathrm{t})}(\txi_1,\txi_2,\txi_3,\txi_4) = \frac{1}{2}\prod_{\nu=1}^{4}\!\int\!\!d\teta_{\nu}\, L(\teta_2,\teta_1,\teta_3,\teta_4)
V^{(\mathrm{t})}_{l}(\txi_2,\txi_1,\teta_2,\teta_3)V^{(\mathrm{t})}_{l}(\teta_4,\teta_1,\txi_3,\txi_4).
\ee
Obviously, particle-particle fluctuations do not couple singlet and triplet vertex-funtions.
The singlet particle-hole fluctuations read
\be
\phi_{\mathrm{ph}}^{(\mathrm{s})}(\txi_1,\txi_2,\txi_3,\txi_4) & = &  - \frac{1}{4}\prod_{\nu=1}^{4}\!\int\!\!d\teta_{\nu}\, L(\teta_1,\teta_2\teta_3,\teta_4)\times \\ 
& & \Bigl[
V^{(\mathrm{s})}(\teta_4,\txi_2,\txi_3,\teta_1)V^{(\mathrm{s})}(\txi_1,\teta_2,\teta_3,\txi_4) + 
\sum_{i,j}V^{(\mathrm{t})}_{i}(\teta_4,\txi_2,\txi_3,\teta_1)V^{(\mathrm{t})}_{j}(\txi_1,\teta_2,\teta_3,\txi_4) + \nn \\
& & \sum_{i}V^{(\mathrm{s})}(\teta_4,\txi_2,\txi_3,\teta_1)V^{(\mathrm{t})}_{i}(\txi_1,\teta_2,\teta_3,\txi_4) + 
\sum_{i}V^{(\mathrm{t})}_{i}(\teta_4,\txi_2,\txi_3,\teta_1)V^{(\mathrm{s})}(\txi_1,\teta_2,\teta_3,\txi_4) 
\Bigr], \nn
\ee
while the triplet contribution is given by
\be
\phi_{\mathrm{ph};l}^{(\mathrm{t})}(\txi_1,\txi_2,\txi_3,\txi_4)  & = & - \frac{1}{4}\prod_{\nu=1}^{4}\!\int\!\!d\teta_{\nu}\, L(\teta_1,\teta_2,\teta_3,\teta_4)\times \\
& & \Bigl[
\,V^{(\mathrm{s})}(\teta_4,\txi_2,\txi_3,\teta_1)V^{(\mathrm{s})}(\txi_1,\teta_2,\teta_3,\txi_4) + 
\sum_{i,j}c_{ij}^{l}\,V^{(\mathrm{t})}_{i}(\teta_4,\txi_2,\txi_3,\teta_1)V^{(\mathrm{t})}_{j}(\txi_1,\teta_2,\teta_3,\txi_4) \nn \\
& &\sum_{i}c_{il}\,V^{(\mathrm{s})}(\teta_4,\txi_2,\txi_3,\teta_1)V^{(\mathrm{t})}_{i}(\txi_1,\teta_2,\teta_3,\txi_4) + 
\sum_{i}c_{il}\,V^{(\mathrm{t})}_{i}(\teta_4,\txi_2,\txi_3,\teta_1)V^{(\mathrm{s})}(\txi_1,\teta_2,\teta_3,\txi_4)
\Bigr]. \nn
\ee
The four coefficient matrices $c_{il}$ and $c_{ij}^{l}$ that result from performing internal spin summations encode a specific sign structure,
\be
c_{il} =
\begin{pmatrix}
+1 & -1 & -1 \\
-1 & +1 & -1 \\
-1 & -1 & +1
\end{pmatrix}_{il},\quad 
c_{ij}^{x}=
\begin{pmatrix}
+1 & +1 & +1 \\
+1 & +1 & -1 \\
+1 & -1 & +1
\end{pmatrix}_{ij}
\quad
c_{ij}^{y}=
\begin{pmatrix}
+1 & +1 & -1 \\
+1 & +1 & +1 \\
-1  & +1 & +1
\end{pmatrix}_{ij},\quad
c_{ij}^{z}=
\begin{pmatrix}
+1 & -1 & +1 \\
-1 & +1 & +1 \\
+1 & +1 & +1 
\end{pmatrix}_{ij},
\ee
and $i,l\in \{x,y,z\}$. The crossed and direct particle-hole contributions entering the flow \Eqref{eq:singletflow}, \Eqref{eq:tripletflow}  are defined
as $\phi_{\mathrm{ph,cr}}^{(\mathrm{s})}(\txi_1,\txi_2,\txi_3,\txi_4)\equiv\phi_{\mathrm{ph}}^{(\mathrm{s})}(\txi_1,\txi_2,\txi_3,\txi_4)$ and
$\phi_{\mathrm{ph,d}}^{(\mathrm{s})}(\txi_1,\txi_2,\txi_3,\txi_4)\equiv\phi_{\mathrm{ph}}^{(\mathrm{s})}(\txi_1,\txi_2,\txi_4,\txi_3)$
for the singlet case. An analogous definition holds for the triplet case. 

Further we note that a delta function taking care of global momentum conservation can be factored out from the flow equations, which leaves only three independent momenta. The final flow equations are formulated and implemented in terms of reduced vertex functions with three independent momenta. For the sake of convenience, we denote full and reduced vertex functions with the same symbol.

\subsection{Vertex reconstruction from Ward identity}
\label{app:ward}

The symmetries of the Hamiltonian \Eqref{eq:slaveboson} can be efficiently described by embedding the two-dimensional honeycomb lattice
into a three-dimensional cubic lattice~\cite{you2012}. Then a rotation around the $\hat{n}=\frac{1}{\sqrt{3}}(1,1,1)^{T}$ axis
by $\pm2\pi /3$ corresponds to the $C_{3}$ or $C_{3}^{-1}$ element, repsectively, of the point group acting on a site in the honeycomb lattice.
This rotation also preserves the sublattice index, i.e., both $A$ and $B$ sublattices are mapped onto themselves. We note
that the coordinate system is adapted to an embedding of the honeycomb lattice in a 3D cubic lattice~\cite{you2012}. Under a rotation by $-2\pi/3$ the spin components along the bonds are mapped as $S_{x} \to S_{y}$, $S_{y} \to S_{z}$ and $S_{z} \to S_{x}$. 
As expected from the strong spin-orbit coupling scenario realized in the iridates, the Kitaev term is only invariant under simultaneous transformations
of spin and lattice (orbital) degrees of freedom. The corresponding transformation on the lattice is a rotation $R_{\hat{n}}(\theta)$ with $\theta = 2\pi /3$. 
This operation can also be represented through combinations of reflections or rotations and reflections.
The two transformations (spin and lattice rotation) taken together leave the Hamiltonian \Eqref{eq:slaveboson} invariant. The $SU(2)$-transformation matrix acting on the fermionic degrees of freedom is given by $S_{\hat{n}}(-\theta)=\exp(\mathrm{i}/2\, \theta\, \hat{n}\cdot\vec{\sigma})$. Point group transformations acting on the real
space lattice also induce a representation on the Brillouin zone. BZ momenta accordingly transform as $\vec{k} \rightarrow \vec{k}^{\prime} = R_{\hat{n}}^{T}(\theta)\vec{k}$. 

Moving to a functional formulation and replacing operator valued fields ($f^{\dagger}$, $f$) with Grassmann variables ($\bar{f}$, $f$), the symmetry of
the system is expressed as the invariance of the generating functional. For the generating functional of 1PI vertices $\Gamma[\bar{f},f]$,
this statement reads as~\cite{kopietz} 
\be\label{eq:ward}
\Gamma[\bar{f}^{\prime},f^{\prime}]=\Gamma[\bar{f},f],
\ee
where the prime denotes transformed fields. We note that since the kinetic part of the Hamiltonian is also invariant under this 
rotational symmetry, so are both the bare and the regularized bare propagator in band representation. This in turn implies that the identity \Eqref{eq:ward} is
valid also for the scale-dependent generating functional $\Gamma_{\Lambda}$ for $\Lambda > 0$. We thus obtain
\be\label{eq:scaleward}
\Gamma^{\Lambda}[\bar{f}^{\prime},f^{\prime}]=\Gamma^{\Lambda}[\bar{f},f],
\ee
For the discrete rotational symmetry discussed above, we arrive at the following transformation rules for the fermion fields in momentum space:
\be
f_{o,\sigma,\vec{k}} \quad & \rightarrow & \quad f_{o,\sigma^{\prime},\vec{k}^{\prime}}^{\prime} = \sum_{\sigma,\vec{k}}\,[S_{\hat{n}}(\theta)]_{\sigma^{\prime} \sigma} \, \delta(\vec{k}^{\prime}-R_{\hat{n}}^{T}(\theta)\vec{k}) \, f_{o,\sigma,\vec{k}}\, , \\ 
\bar{f}_{o,\sigma,\vec{k}} \quad & \rightarrow &\quad \bar{f}_{o,\sigma^{\prime},\vec{k}^{\prime}}^{\prime} = \sum_{\sigma,\vec{k}}\,[S^{-1}_{\hat{n}}(\theta)]_{\sigma\sigma^{\prime}} \, \delta(\vec{k}^{\prime}-R_{\hat{n}}^{T}(\theta)\vec{k}) \, \bar{f}_{o,\sigma^{\prime},\vec{k}^{\prime}}\, .
\ee
Expanding both sides of~\Eqref{eq:scaleward} in transformed and original fields and using the explicit representation of the transformation rules,
we find the following set of Ward identities for the (scale-dependent) singlet and triplet vertex functions:
\be
V_{o_1,o_2,o_3,o_4}^{(\mathrm{s})}(\vec{k}_1,\vec{k}_{2},\vec{k}_{3},\vec{k}_{4}) & = & 
V_{o_1,o_2,o_3,o_4}^{(\mathrm{s})}(\vec{k}_1^{\prime},\vec{k}_{2}^{\prime},\vec{k}_{3}^{\prime},\vec{k}_{4}^{\prime}), \\
V_{x;o_1,o_2,o_3,o_4}^{(\mathrm{t})}(\vec{k}_1,\vec{k}_{2},\vec{k}_{3},\vec{k}_{4}) & = &
V_{y;o_1,o_2,o_3,o_4}^{(\mathrm{t})}(\vec{k}_1^{\prime},\vec{k}_{2}^{\prime},\vec{k}_{3}^{\prime},\vec{k}_{4}^{\prime}), \\
V_{y;o_1,o_2,o_3,o_4}^{(\mathrm{t})}(\vec{k}_1,\vec{k}_{2},\vec{k}_{3},\vec{k}_{4}) & = & 
V_{z;o_1,o_2,o_3,o_4}^{(\mathrm{t})}(\vec{k}_1^{\prime},\vec{k}_{2}^{\prime},\vec{k}_{3}^{\prime},\vec{k}_{4}^{\prime}), \\
V_{z;o_1,o_2,o_3,o_4}^{(\mathrm{t})}(\vec{k}_1,\vec{k}_{2},\vec{k}_{3},\vec{k}_{4}) & = & 
V_{x;o_1,o_2,o_3,o_4}^{(\mathrm{t})}(\vec{k}_1^{\prime},\vec{k}_{2}^{\prime},\vec{k}_{3}^{\prime},\vec{k}_{4}^{\prime}), 
\ee
where we made the orbital indices explicit, i.e., $V_{o_1,o_2,o_3,o_4}^{(\mathrm{s})}(\vec{k}_1,\vec{k}_{2},\vec{k}_{3},\vec{k}_{4}) \equiv V^{(\mathrm{s})}(\txi_1,\txi_2,\txi_3,\txi_4)$
and $V_{l;o_1,o_2,o_3,o_4}^{(\mathrm{t})}(\vec{k}_1,\vec{k}_{2},\vec{k}_{3},\vec{k}_{4}) \equiv V_{l}^{(\mathrm{t})}(\txi_1,\txi_2,\txi_3,\txi_4)$ with $\txi = (\omega,o,\vec{k})$ in the
static approximation $\omega = 0$. These Ward identities give us the important information, that for one given triplet vertex function, 
the other remaining two vertex functions can be reconstructed. This fact was exploited in designing an efficient numerical implementation
of the flow \Eqref{eq:singletflow} and \Eqref{eq:tripletflow}. The Ward identity for the singlet vertex function was not directly employed in the numerical implementation. 
Besides fermionic exchange symmetry, however,  it serves as an important consistency check for the numerical solution of the flow equation. 

\end{widetext}


\end{document}